\documentclass[superscriptaddress, reprint, amsmath, amssymb, aps, prx, floatfix]{revtex4-2}

\usepackage{bm}

\usepackage{mathtools}
\usepackage{amsfonts}
\usepackage{amssymb}
\usepackage{dsfont}
\usepackage{amsthm}
\usepackage[english]{babel}
\usepackage{graphicx}
\usepackage{amsmath}
\usepackage{lipsum}
\usepackage{wrapfig}
\usepackage{float}
\usepackage{enumitem} 
\usepackage{color} 
\usepackage{hyperref}
\usepackage[normalem]{ulem}
\usepackage{url}

\useunder{\uline}{\ul}{}

\hypersetup{colorlinks = true, pdfauthor = , pdftitle = A QOC perspective on VQAs}

\def\ket#1{\ensuremath{\mathinner{|{#1}\rangle}}}

\newcommand{\needcite}[1]{\textcolor{red}{[Ref needed]}}
\usepackage{qcircuit}

\begin{document}

\title{From pulses to circuits and back again:\\
A quantum optimal control perspective on variational quantum algorithms}

\author{Alicia B. Magann}\thanks{These two authors contributed equally.}
\affiliation{Department of Chemical \& Biological Engineering, Princeton University, Princeton, New Jersey 08544, USA}
\affiliation{Extreme-Scale Data Science \& Analytics, Sandia National Laboratories, Livermore, California 94550, USA}

\author{Christian Arenz}\thanks{These two authors contributed equally.}
\affiliation{Department of Chemistry, Princeton University, Princeton, New Jersey 08544, USA}

\author{Matthew D. Grace}
\affiliation{Extreme-Scale Data Science \& Analytics, Sandia National Laboratories, Livermore, California 94550, USA}

\author{Tak-San Ho}
\affiliation{Department of Chemistry, Princeton University, Princeton, New Jersey 08544, USA}

\author{Robert L. Kosut}
\affiliation{SC Solutions, Sunnyvale CA, 94085}
\affiliation{Department of Chemistry, Princeton University, Princeton, New Jersey 08544, USA}

\author{Jarrod R. McClean}
\affiliation{Google Research, 340 Main Street, Venice, CA 90291, USA}

\author{Herschel A. Rabitz}
\affiliation{Department of Chemistry, Princeton University, Princeton, New Jersey 08544, USA}

\author{Mohan Sarovar}
\affiliation{Extreme-Scale Data Science \& Analytics, Sandia National Laboratories, Livermore, California 94550, USA}

\date{\today}

\begin{abstract}

The last decade has witnessed remarkable progress in the development of quantum technologies. Although fault-tolerant devices likely remain years away, the noisy intermediate-scale quantum devices of today may be leveraged for other purposes. Leading candidates are variational quantum algorithms (VQAs), which have been developed for applications including chemistry, optimization, and machine learning, but whose implementations on quantum devices have yet to demonstrate improvements over classical capabilities. In this Perspective, we propose a variety of ways that the performance of VQAs could be informed by quantum optimal control theory. To set the stage, we identify VQAs and quantum optimal control as formulations of variational optimization at the circuit level and pulse level, respectively, where these represent just two levels in a broader hierarchy of abstractions that we consider. In this unified picture, we suggest several ways that the different levels of abstraction may be connected, in order to facilitate the application of quantum optimal control theory to VQA challenges associated with ansatz selection, optimization landscapes, noise, and robustness. A major theme throughout is the need for sufficient control resources in VQA implementations; we discuss different ways this need can manifest, outline a variety of open questions, and conclude with a look to the future. 

\end{abstract}

\maketitle

\section{Introduction}

The development of large scale, fault-tolerant quantum computers would enable diverse and disruptive applications, such as the ability to break RSA encryption protocols using Shor's factoring algorithm \cite{10.1109/SFCS.1994.365700} and to efficiently simulate the dynamics of complex quantum systems \cite{1996Sci...273.1073L}. Although significant progress has been made \cite{egan2020faulttolerant}, persistent technological challenges in current quantum devices means they cannot yet serve as platforms for implementing these landmark algorithms at scale. As such, a major goal is to identify classically difficult problems that could be solved with these noisy intermediate-scale quantum (NISQ) \cite{preskill_quantum_2018} devices. 

This goal has motivated the development of variational quantum algorithms (VQAs) for a variety of applications including ground state chemistry \cite{peruzzo_variational_2013}, optimization \cite{2014arXiv1411.4028F}, and machine learning \cite{Dunjko2020nonreviewofquantum}. In VQAs, the problem to be solved is reformulated as an optimization problem, whose solution is sought using quantum hardware and classical optimization in concert \cite{McClean_2016}. The quantum device is used to evaluate the objective function, which is accomplished via a relatively shallow, parametrized quantum circuit applied to an appropriately initialized register of qubits, after which the value of the objective function can be determined by measuring the register. Meanwhile, the classical co-processor iteratively optimizes the parameters of the shallow quantum circuit. To-date, hardware implementations of VQAs have not yet demonstrated improvements over the capabilities of classical computers, and the aim of this Perspective is to examine how progress can be made towards meeting this milestone in the future. In particular, this Perspective will consider VQAs, their associated challenges, and potential paths forward, through the lens of quantum optimal control. To motivate this choice, we first look back and review certain aspects of the research efforts that have led us to the NISQ era of today. 

We begin by recalling early efforts to create quantum computers, which focused on developing methods to control their components, and involved one or a few qubits \cite{nakamura1999coherent, yamamoto2003demonstration}. In these experiments, control was typically realized using electromagnetic fields or ``pulses'' designed to drive the dynamics of the qubits in a desired fashion. Techniques for qubit control were studied extensively, especially in the context of implementing high-fidelity entangling gates between qubits, as this is a necessary ingredient for quantum computation in the gate model. One method, quantum optimal control (QOC), stands out for its ability to improve gate fidelities beyond what other techniques could offer \cite{waldherr2014quantum,dolde2014high}. In QOC, an objective functional is defined that quantifies how well a desired control task is achieved; then, the pulses to minimize the objective functional are sought using iterative optimization methods \footnote{We note that we use the term \emph{quantum optimal control} to refer solely to the closed-loop control scenario just described, and do not consider other quantum control settings such as real-time feedback control in this paper.}. 

Following these early demonstrations of qubit control, devices began to scale up to higher qubit counts, which has led to the advent of the NISQ era today, and in tandem, to the development of VQAs. These concurrent developments have inspired significant research on circuit compilation and optimization \cite{venturelli2018compiling,khatri2019quantum,murali2019noise}, similar in spirit to many of the earlier efforts that studied pulse optimization in the context of QOC. Thus, the reach of technology and the focus of the community have broadened \emph{``from pulses to circuits''}. In this Perspective, we explore how the development of VQAs can be informed by going from the circuit level \emph{``back again''} to the pulse level to strengthen ties to QOC and leverage results and tools from this well-developed field. 

\begin{figure}[t]
\centering
\includegraphics[width=0.9\columnwidth]{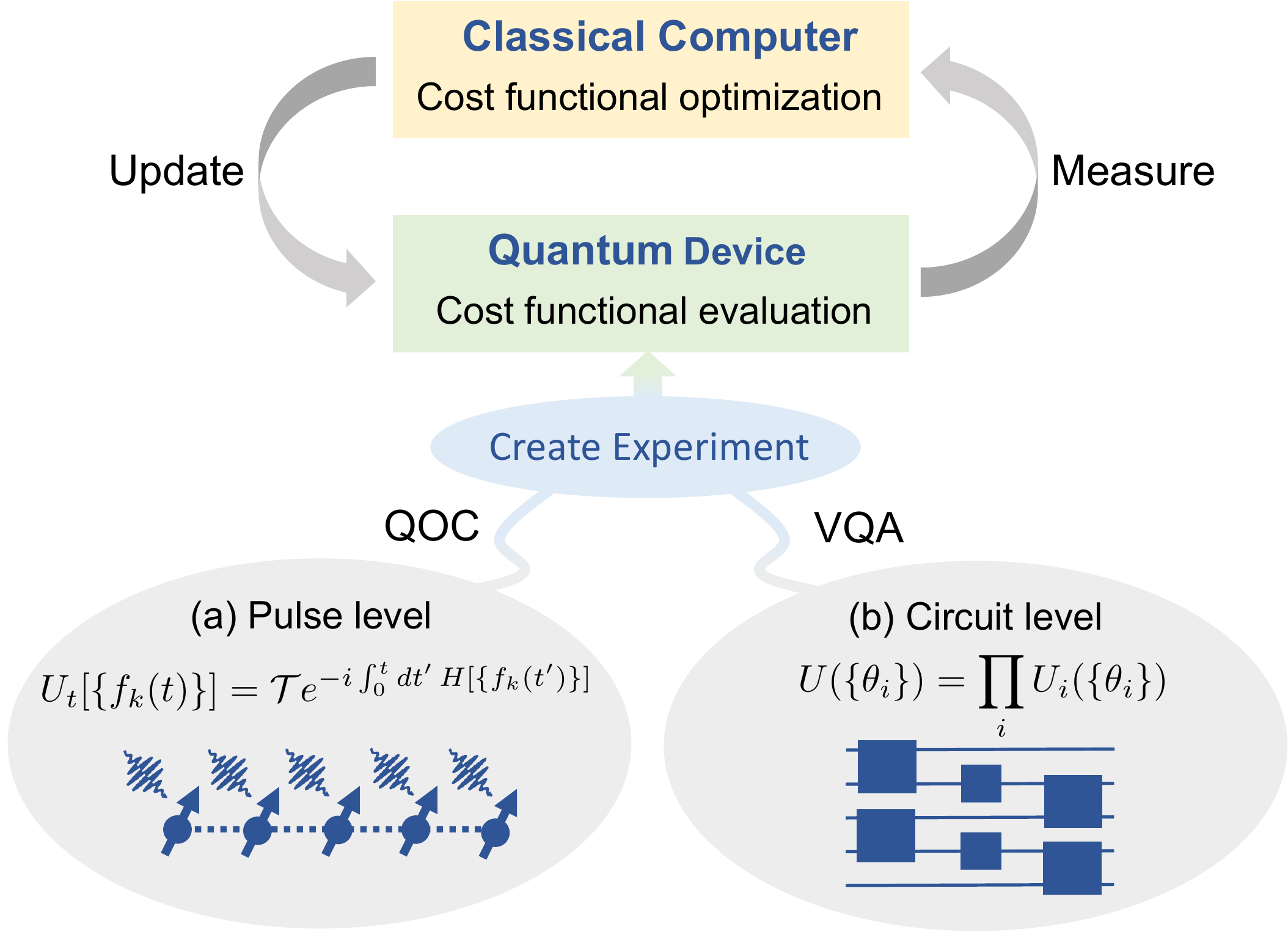}
\caption{The quantum-classical optimization loop used in VQA and QOC experiments is shown. The quantum device (green) evaluates the objective functional, whose value is determined via measurements and input into a classical computer (yellow), which updates the optimization parameters. In conventional QOC experiments, these parameters are defined at the pulse level (a), and serve to parametrize a set of control fields $\{f_{k}(t)\}$ entering in a Hamiltonian describing the physics of the system. In VQAs, the parameters typically enter at the circuit level (b), via a set of gates parametrized through a set of e.g., gate angles $\{\theta_{i}\}$. In this Perspective, we explore how VQAs may be informed by returning to a more physical, pulse-level description inspired by QOC.}
\label{Scheme}
\end{figure}

As illustrated in Fig. 1, we explore this prospect by framing VQA implementations and QOC experiments as quantum-classical optimization loops; in the former (b), the optimization is done over quantum circuit elements, in contrast to a conventional QOC experiment (a), where the optimization is performed over a set of continuous pulses. We remark now that although the parametrized quantum circuits in (b) are formed by gates, which are in turn implemented using pulses, a user seeking to implement a VQA typically has no need for any knowledge of what happens at the pulse level, which offers a layer of abstraction separating the user from the underlying hardware physics.

With this unifying picture in mind, we now describe the remainder the this article. We begin by introducing the concept of VQAs and discussing their current state in Section \ref{sec:VQA}. We then discuss QOC theory in Section \ref{sec:QOC} and review connections that have been made to VQAs to-date in Section \ref{sec:PriorWork}. In Section \ref{sec:futuredirections} we present a hierarchy of abstractions in variational optimization that serves to provide a common framework for VQA and QOC experiments. This is followed by an in-depth discussion of QOC-motivated future research directions aimed at addressing four important challenges associated with VQAs: ansatz selection, optimization landscapes, noise, and robustness. In each of these cases, we emphasize the need for appropriate control resources to enhance the performance of VQAs. Finally, we conclude in Section \ref{sec:outlook} with a look ahead.

\section{Variational quantum algorithms}

\label{sec:VQA}

Variational quantum algorithms seek to solve problems by leveraging the dynamical and representational power of quantum computers in conjunction with classical computers. They do so by reformulating problems of interest as the minimization of some objective or loss function $J[\{\theta_i\}]$ over a set of parameters $\{\theta_i\}$, as
\begin{align}
\label{eq:vqaoptimizationproblem}
 \min_{\{\theta_i\}} J[\{\theta_i\}],
\end{align}
where $\{\theta_i\}$ parametrizes a quantum circuit
\begin{align}
U(\{\theta_i\}) = \prod_{i}U_{i}(\theta_{i}),
\end{align}
on $n$ qubits. The objective function can be nonlinear in the general case, such as cross-entropy for machine learning. However, for reasons of simplicity, it is often formulated as the minimization of the expectation of a linear operator $H_{\mathrm{p}}$, referred to as the ``problem'' Hamiltonian \footnote{The term ``problem Hamiltonian'' for this quantity is standard in optimization applications of VQAs. It is often called ``system Hamiltonian'' in other applications.}:
\begin{equation}
J[\{\theta_i\}] = \langle \psi(\{\theta_i\})|H_{\mathrm{p}}|\psi(\{\theta_i\})\rangle,
\label{eq:loss}
\end{equation}
where the state of the qubits at the culmination of the circuit is $|\psi(\{\theta_i\})\rangle = U(\{\theta_i\})|\psi_{0}\rangle$, with $|\psi_{0}\rangle$ denoting their fixed initial state. Exact minimization of $J$ corresponds to the preparation of the ground state of $H_{\mathrm{p}}$. 

The set of variational parameters $\{\theta_i\}$ that minimize $J[\{\theta_i\}]$ are sought iteratively, where at every iteration, $J[\{\theta_i\}]$ is evaluated by preparing the qubits in the state $|\psi_{0}\rangle$, applying a circuit $U(\{\theta_i\})$ with a particular parametrization $\{\theta_i\}$, and then measuring a set of qubit observables to estimate $J[\{\theta_i\}]$. This can be accomplished by expanding $H_{\mathrm{p}}$ in the Pauli operator basis as per $H_{\mathrm{p}} = \sum_{j=1}^N \alpha_j P_j$, where $\alpha_j$ are scalar coefficients, $P_j$ are tensor products of Pauli operators that are easy to measure on a quantum device, and $N = \mathcal{O}({\rm poly}(n))$\footnote{We remark here that this is a common property of optimization problems, like the ones we consider in this Perspective, but does not always hold.}, and then measuring the expectations of each of the $N$ Pauli basis operators in the expansion. Due to the stochastic nature of measurement in quantum mechanics, repeated measurements on an identically prepared state $\ket{\psi(\{\theta_i\})}$ are needed to estimate these expectations. Then, the expectation value of $H_{\mathrm{p}}$ can be computed to determine $J$ by classically evaluating the weighted sum $\langle \psi(\{\theta_i\})|H_{\mathrm{p}}|\psi(\{\theta_j\})\rangle = \sum_{j=1}^N \alpha_j \langle\psi(\{\theta_j\})| P_j|\psi(\{\theta_i\})\rangle$. After the value of $J[\{\theta_j\}]$ has been evaluated in this manner, a classical optimization routine is then used to iteratively update the values of $\{\theta_i\}$ until convergence. Using this method to estimate the value of $J[\{\theta_j\}]$ to a specified precision $\epsilon$ requires a number of re-preparations and measurements of the state that scales as $N_s \propto \lambda^2/\epsilon^2$, where $\lambda=\sum_j |\alpha_j|$~\cite{rubin2018application}. Recent work has shown that clever grouping of terms and other techniques can be used to reduce the naive scaling of these measurements by orders of magnitude, even with techniques in the near-term~\cite{huggins2019efficient}. As quantum computers advance, it is possible to improve the scaling of this estimation even further using techniques that leverage phase estimation, but the increased resource costs for such approaches can be prohibitive.

For practical reasons, the circuit represented by $U(\{\theta_i\})$ is usually assumed to be formed by a sequence of elementary one- and two-qubit gates drawn from a specified gate set, which is constructed and parametrized according to a particular ansatz. However, we remark here that knowledge of how the parameters $\{\theta_i\}$ explicitly enter into the circuit is not always required, as $J[\{\theta_i\}]$ is evaluated via measurements using the quantum device. That is, the method is robust to many types of labeling errors, since properties of the quantum system rather than specific parameter values are of interest. This means that VQAs possess some degree of robustness to drifts, crosstalk, and other systematic errors that can occur during the implementation of the circuit \cite{OMalley2016}, which is a primary reason why VQAs are believed to be a way to derive practical algorithmic use from NISQ devices.

Furthermore, we note that the choice of ansatz for any VQA is a crucial step. Although there is no general approach for developing good ans\"atze, they are often derived from physical intuition (e.g., the QAOA ansatz, see below), knowledge of states generated by a particular ansatz (e.g., the coupled-cluster ansatz, see below), or practical convenience (e.g., hardware-efficient ans\"atze \cite{peruzzo_variational_2013, kandala2017hardware}). A critical feature of an ansatz is that it should be scalable, i.e., having a circuit depth that scales as $\mathcal{O}({\rm poly}(n))$ for an $n$-qubit quantum computer. Furthermore, the number of variational parameters $\{\theta_{i}\}$ should also scale as $\mathcal{O}({\rm poly}(n))$.

VQAs have been developed for numerous application areas including machine learning \cite{Dunjko2020nonreviewofquantum}, linear systems \cite{2019arXiv190905820B}, and the compilation of quantum circuits \cite{khatri_quantum-assisted_2019}. However, the first application area for VQAs was the ground state problem in quantum chemistry \cite{mcclean2018barren,McClean_2016}. In this context, the variational quantum eigensolver (VQE) was developed as a VQA for seeking the electronic ground state of a chemical system in a field of fixed nuclear charges. The solution of this chemistry problem has a variety of applications, including in chemical reaction prediction, the determination of molecular properties, etc. One common ansatz is the unitary coupled-cluster ansatz~\cite{peruzzo_variational_2013,taube2006new}, which is a norm-preserving variant of the common coupled-cluster ansatz used in quantum chemistry, which constructs a size-extensive ansatz through an exponential parametrization. Due to its unitary formulation, it naturally preserves physical properties of the state, but it is not efficient to evaluate classically. It represents an example of a structured ansatz that carries the fermionic structure into its translation into gates after the use of a Jordan-Wigner transformation and Suzuki-Trotter splitting. Strictly speaking, this formulation represents a slight deviation from the formal construction \cite{grimsley2019trotterized}, but the construction remains unitary, independent of the choice of parameters, and upon repetition it can be used to express arbitrary states within the manifold of fixed particle number.

With the evidence that NISQ devices can achieve classically intractable, but perhaps not useful, tasks \cite{arute_quantum_2019}, there is a belief that variational preparation of ground states of correlated systems may represent one of the first classically intractable and useful roles for NISQ devices. As such, the study of correlated quantum systems remains a major objective for NISQ devices, and is a focus of many experimental and theoretical efforts at present. VQE demonstrations have been shown experimentally on a variety of photonic, ion trap, and superconducting qubit setups in combination with error mitigation techniques~\cite{OMalley2016, kandala2017hardware, PhysRevX.8.031022, kandala2019error, Siddiqi2017, PhysRevA.100.022517, PhysRevA.100.010302}. Indeed the discrepancy between available qubits in current quantum devices ($\gtrapprox$ 50 qubits) and the number used in VQE experiments ($\lesssim$ 10 qubits) is that the impact of noise makes the experiments incompatible with the high accuracy necessary to claim an application advantage. With the help of error mitigation from symmetries in the reduced subspace, the largest variational calculation performed to chemical accuracy on a quantum computer utilized 12 qubits and approximately 200 quantum gates to simulate a 12-atom hydrogen chain~\cite{google2020hartree}. Using more qubits will require advances in both hardware and error reduction techniques.

Another major application of VQAs is combinatorial optimization. Here, the quantum approximate optimization algorithm (QAOA) was developed as a variational method for determining approximate solutions to combinatorial optimization problems, by encoding them into diagonal Ising Hamiltonians, such that the solution of the problem is encoded in the ground state of the Hamiltonian \cite{2014arXiv1411.4028F}. QAOA seeks to find the solution by variationally minimizing the expectation value of the Ising Hamiltonian. Unlike the VQE, the QAOA ansatz is typically fixed, and consists of an alternating sequence of unitary operations generated by the problem Hamiltonian $H_{\mathrm{p}}$, and a so-called ``driver'' Hamiltonian, which does not commute with the problem Hamiltonian, and is denoted by $H_{\mathrm{d}}$. Explicitly, the QAOA ansatz is formed by $p$ rounds of alternating applications of these two Hamiltonians,
\begin{equation}
\begin{aligned}
\label{eq:qaoa}
U(\{\theta_i\}) = \prod_{j = 1}^{p} \exp{\left( - i\beta_j H_{\mathrm{d}} \right)} \exp{\left( - i\gamma_j H_{\mathrm{p}} \right)},
\end{aligned}
\end{equation}
where $\{\theta_i\}$ is the set of variational parameters $\{\beta_j,\gamma_j\}$, and the true optimum for the original combinatorial optimization problem can be achieved as $p\rightarrow \infty$ \cite{2014arXiv1411.4028F}. 

QAOA has been implemented experimentally using superconducting circuits for up to 23 qubits \cite{otterbach2017unsupervised,willsch2020benchmarking,abrams2019implementation,bengtsson2019quantum,arute2020quantum}, a photonic system with 2 qubits \cite{qiang2018large}, and trapped ions with up to 40 qubits \cite{pagano2019quantum}. Due to limited qubit coherence times, most of these implementations (with some exceptions \cite{abrams2019implementation,arute2020quantum}) considered combinatorial optimization problems defined on the connectivity graph of the hardware only, keeping circuit depths at a minimum. Recently, it has been shown that success in quantum optimization can be related to increased flexibility in the training landscape of a continuous extension of the original problem~\cite{mcclean2020low}, which is connected to the condition of full controllability in quantum control theory. Furthermore, although in theory, ans\"atze with higher $p$ should improve on the quality of the solutions achievable at lower $p$, they also require deeper circuits. So in practice, the effects of noise and decoherence may negate any improvements in quality. For this reason, most experiments implemented a single round of QAOA only, i.e., $p = 1$. These cases involved only two parameters whose optimal values could be obtained analytically, meaning that a variational optimization loop was not necessary. However, in recent superconducting circuit experiments it was shown that going to $p=2$ \cite{bengtsson2019quantum} or $p=3$ \cite{arute2020quantum} and performing the optimization could improve the solution quality beyond $p=1$. 

Although these recent hardware demonstrations show a promising trend, the ultimate goal of implementing VQAs to solve problems that are intractable classically has yet to be reached. The path towards meeting this goal will involve a confluence of theoretical and experimental progress. In the following, we propose a few ways that this progress could be informed by QOC. In particular, we focus in on how methods and results from QOC could be leveraged to address VQA challenges associated with ansatz selection, optimization landscapes, noise and robustness. Before getting to this, we first introduce QOC and discuss its connections to VQAs. 

\section{Quantum optimal control}
\label{sec:QOC}

The aim of QOC is to design one or more electromagnetic fields or ``pulses'' to steer the dynamics of a quantum system towards a desired control target, which can be a state, observable expectation value, or evolution operator, at some terminal time $T$. A standard formulation in QOC seeks to minimize a control objective functional $J[\{f_{k}(t)\}]$ over $\{f_{k}(t)\}$, as \cite{kosloff1989wavepacket, shi1990quantum, sundermann1999extensions, dong2010quantum, lloyd2014information}
\begin{align}
\label{eq:optimizationproblem}
 \min_{\{f_{k}(t)\}} J[\{f_{k}(t)\}],
\end{align}
where $J[\{f_{k}(t)\}]$ includes the control target and physical constraints, often along with other criteria, which can be defined to represent available laboratory resources or quantify robustness to errors or uncertainties \cite{PhysRevA.37.4950}. 

The set of pulses $\{f_{k}(t)\}$ used in QOC are typically considered to be classical fields in the semiclassical approximation \cite{Brif2011}, in contrast to fully quantized fields. In addition, the wavelengths of the fields are typically assumed to be much greater than the size of the controlled quantum system in the dipole approximation \cite{Brif2011}, such that the control fields $\{f_{k}(t)\}$ typically enter the time-dependent Hamiltonian $H[\{f_{k}(t)\}]$, as follows \cite{elliott2009bilinear, d2007introduction}:
\begin{align}
\label{eq:Hamiltoniancontinuous}
H[\{f_{k}(t)\}] = H_{0} + \sum_{k} f_{k}(t) H_{k},
\end{align} 
where $H_0$ is the drift Hamiltonian describing the time-independent system and $\{H_{k}\}$ is the set of control Hamiltonians that couple the fields to the system, e.g., via dipole interactions. The dynamical equation for the system time-evolution operator $U_{t}$ is given by the Schr\"odinger equation $\dot{U}_{t} = -i H[\{f_{k}(t)\}]U_{t}$,  with  $U_{0} = \mathds{1}$ (throughout this article, we set $\hbar=1$). This is a bilinear control system, making an analytical formulation of QOC solutions intractable in general \cite{jurdjevic1972control, elliott2009bilinear, d2007introduction}. Its formal solution reads 
\begin{align}
U_{t}[\{f_{k}(t)\}]=\mathcal T e^{-i\int_{0}^{t}dt^{\prime}\, H[\{f_{k}(t^{\prime})\}]},
\label{Eq:Ut}
\end{align}
where $\mathcal T$ indicates time ordering, such that the system state at time $t$ is given by $|\psi(t)\rangle = U_{t}|\psi_{0}\rangle$ where $|\psi_0\rangle$ is the initial state. 

In analogy to the formulation of VQAs in Eqs. (\ref{eq:vqaoptimizationproblem}) and (\ref{eq:loss}), we now turn our attention to the QOC problem in Eq. (\ref{eq:optimizationproblem}), where we define 
\begin{equation}
    J[\{f_{k}(t)\}] = \langle\psi(T)|H_{\mathrm{p}}|\psi(T)\rangle,
\end{equation} 
such that (unconstrained) $\{f_{k}(t)\}$ are sought to minimize the expectation value of $H_{\mathrm{p}}$ at a designated time $T$. Solutions of this QOC problem, subject to the dynamical constraint that $|\psi(t)\rangle$ evolves according to the Schr\"odinger equation, are stationary points of the control objective functional $J$, given the initial condition $|\psi(0)\rangle = |\psi_0\rangle$. Optimal control fields can be constructed iteratively via the corresponding Euler-Lagrange equations \cite{stengel_optimal_1994}. To this end, a plethora of methods have been developed for updating the QOC solutions, including GRAPE \cite{khaneja2005optimal}, Krotov \cite{palao2003optimal, reich2012monotonically, goerz2019krotov}, TBQCP \cite{ho2010accelerated, ho2011general, liao2011fast}, D-MORPH \cite{rothman2005observable, rothman2005quantum, rothman2006exploring}, and SCP \cite{Kosut2013_Robust}, until specified optimality or convergence criteria are satisfied.

In practice, the minimization of $J$ is usually accomplished by first parametrizing one or more continuous control fields by a set of variables $\{c_{i}\}$, such that the Hamiltonian describing the controlled system becomes 
\begin{equation}
H(t,\{c_i\}) = H_0 + \sum_k f_k(t,\{c_{i,k}\}) H_k.
\end{equation}
Common parametrizations include setting $\{c_i\}$ to be the amplitudes and phases of a set of frequency components of the field, or setting $\{c_i\}$ to be piecewise-constant field amplitudes in the time domain. Then, the objective is to optimize $\{c_i\}$ to generate $U_T(\{c_i\}) = \mathcal{T}e^{-i\int_0^T H(t,\{c_i\})dt}$ such that the control objective functional $J[\{c_i\}]$ is minimized at the terminal time $T$. In practice, this minimization is typically performed in an iterative fashion. If a tractable and accurate model is available, this iteration can be performed numerically. However, it can also be carried out experimentally via learning control \cite{PhysRevLett.68.1500, egger2014adaptive, lu2017enhancing, li2017hybrid, chen2020combining, 2020arXiv200803874Y}, which does not require knowledge of the underlying system model. Instead, at each iteration of such QOC experiments, $J[\{c_i\}]$ is evaluated by first preparing the system in a specified initial state, then evolving it in the presence of applied fields with parametrization $\{c_i\}$, and finally measuring the observable expectation value(s) needed to estimate $J[\{c_i\}]$. A classical optimization routine is used to update the values of $\{c_i\}$ from one iteration to the next, until $J[\{c_i\}]$ converges \cite{PhysRevLett.68.1500}. In this manner, objective functional evaluations are performed by the quantum system directly, inherently accounting for parameter uncertainties and other systematic errors, limitations, etc.

This quantum-classical optimization loop associated with QOC at the pulse level is directly analogous to the procedure used in VQA implementations at the circuit level, as shown in Fig.~\ref{Scheme}. In fact, if one has access to the quantum circuit at the pulse level, then Eq. (\ref{eq:vqaoptimizationproblem}) can instead be solved by optimizing over the set of continuous pulses $\{f_{k}(t)\}$ that are available. As such, we consider VQA implementations to be a form of \emph{digital} QOC experiments on qubits, where the quantum circuit generating the unitary transformation $U(\{\theta_i\})$ is designed directly, through the selection and optimization of a parametrized unitary ansatz. In general, the depth, dimension, and structure of the ansatz, as well as the continuous and discrete parameters within it, are all tunable, giving the resulting optimization space both continuous and discrete degrees of freedom.

In the past few years, this fundamental relationship between QOC and VQA has been exploited to derive a deeper understanding of VQAs and novel variational strategies for quantum computing problems. In the following section, we review some of this work at the intersection of VQAs and QOC.

\section{Prior work connecting QOC with VQAs} \label{sec:PriorWork}

As argued above, standard parametrized quantum circuit ans\"atze can be viewed as examples of digitized QOC implementations. There are many benefits to relaxing, or embedding, such digitized ans\"{a}tze into a continuously (in time) parametrized framework, similar to the typical setting in QOC. Such a relaxation often allows one to eliminate any discrete optimization component of the problem, and more importantly, by formulating VQAs within a standard control theory setting, this enables one to apply many powerful methods and results of optimal control theory. Of course, any relaxation of a parametrized ansatz into a continuous-in-time ansatz must be done in a way that is consistent with the capabilities of the available control hardware -- e.g., since most quantum computing platforms have native one- and two-qubit gates, the natural continuously parametrized ans\"atze will be composed of pulses that address individual qubits or the coupling between two qubits.

An early example of work with this reformulation is by Yang et al., \cite{PhysRevX.7.021027}, who used a continuously-parametrized formulation of variational quantum optimization to demonstrate that a bang-bang approach (similar to the the alternating structure of QAOA) is optimal for preparing the state encoding the optimization solution, given amplitude-constrained control fields and a finite time to solution. This is achieved by applying Pontryagin's minimum principle \cite{stengel_optimal_1994} to the continuously-parametrized formulation of the problem. Similarly, Lin et al.  apply Pontryagin's minimum principle associated with time-optimal quantum control to Grover's quantum search problem, and find that the time-optimal control solution has a bang-singular-bang structure \cite{PhysRevA.100.022327}.

Another direction has considered connections between the QOC concept of \emph{controllability} and the quantum computing concept of \emph{computational universality} \cite{Ramakrishna96a}. In brief, controllability is the study of which control objectives can be realized with a given set of controls and constraints. For unconstrained control fields, the dynamical Lie algebra $\mathfrak{L}$, formed by iterated commutators of the drift and the control Hamiltonians and their real linear combinations, is a powerful tool for deciding these matters. In particular, the dynamical Lie algebra gives rise to the Lie rank criterion \cite{elliott2009bilinear, d2007introduction}, which states that if $\mathfrak{L}$ spans the full space (i.e., the special unitary algebra $\mathfrak{su}(2^{n})$ for a $n$ qubit system), then every unitary transformation $V\in\text{SU}(2^{n})$ can be created to arbitrary precision in finite time by shaping the control fields, and the system is said to be \emph{fully operator controllable}. A vast literature characterizing the controllability of quantum systems has been developed in recent decades \cite{altafini2002controllability, albertini2002lie, schirmer2001complete, turinici2000controllability, fu2001complete, turinici2003wavefunction, burgarth2013zero, heule2010local, zimboras2014dynamic, genoni2012dynamical, dirr2009lie, burgarth2014exponential, arenz2016universal}. More recently, the relationship between controllability and computational universality was utilized in refs.~\cite{2018arXiv181211075L, 2019arXiv190903123M} to show that the QAOA ansatz is universal for quantum computing for specific choices of the problem Hamiltonian. In addition, Mbeng et al. made connections between digitized quantum annealing, QAOA, and QOC, e.g., analyzing the number of angles that are needed for controllability \cite{2019arXiv190608948B}, while Akshay et al. examined reachability deficits in QAOA, providing strategies for improving reachability \cite{PhysRevLett.124.090504}.

Some groups have investigated using the QAOA ans\"{a}tze for bang-bang control of state transitions in quantum spin systems, e.g., Refs.~\cite{2019arXiv191100789D, Niu2019_Optimizing, 2020arXiv200201068Y}, exploring robustness and reachability as a function of the ansatz depth. In addition, Bapat and Jordan analyzed the performance of bang-bang control protocols for optimization algorithms, showing that on certain problem instances, these protocols can yield an exponential speedup for both classical and quantum optimization, compared with quasistatic scheduling \cite{Bapat2019_Bang}. Using QOC, Brady et al. \cite{2020arXiv200308952B} investigate the optimal continuous-time control solution for preparing an approximation of a ground state with a time-dependent Hamiltonian that is the sum of the driver and problem Hamiltonians found in QAOA, and the total evolution time is fixed. They demonstrate that the optimal solution can in fact be more general than the bang-bang solution identified by Yang et al. \cite{PhysRevX.7.021027}, because an assumption made in that work (the absence of ``singular'' intervals) is not generally valid. With a careful variational analysis, along with numerical simulations of various tranverse-field Ising models, Brady et al. argue that the more general form of the optimal QAOA variational schedule is a \emph{bang-anneal-bang} schedule, whereby the control parameters take extreme values at the beginning and end of the time interval, but take on intermediate values that smoothly vary during intermediate times. With a fermionic representation, Wang et al., \cite{PhysRevA.97.022304} show that the evolution of a quantum system implementing QAOA on the so-called ``ring of disagrees'' problem translates into QOC of an ensemble of independent spins, thereby simplifying the determination of the optimal angle vectors. On a related note, Wu et al. propose a scheme to machine learning tasks into corresponding QOC problems on NISQ devices \cite{2020arXiv200313658W}.

Recently, several groups have proposed adaptive,  variational, or QOC-inspired approaches to design improved VQA ans\"{a}tze \cite{Grimsley2019_Adaptive, Tang2020_Qubit, Zhu2020_Adaptive, Choquette2020_QOCA, Meitei2020_Gate, Zhang2020_Mutual}. For example, the approach developed in Refs.~\cite{Grimsley2019_Adaptive, Tang2020_Qubit, Zhu2020_Adaptive} uses derivative information to adaptively modify the circuit depth and ansatz structure using a ``pool'' of predetermined single- or multi-qubit Hamiltonian operators. Whereas, in the context of quantum chemistry, the approach presented in Ref.~\cite{Choquette2020_QOCA} uses a set of QOC-informed driving Hamiltonians to generate VQA ans\"{a}tze with symmetry-breaking features that can decrease the circuit depth required for convergence. Similarly, ref.~\cite{Meitei2020_Gate} proposes ans\"{a}tze determined by QOC at the device level, rather than parametrized quantum circuits, to perform VQE simulations. These last works especially strengthen the connections between QOC and VQAs, and provide a natural segue to the next section, where we outline some promising new directions of research at this intersection.

\section{New directions for VQAs informed by QOC} \label{sec:futuredirections}

\begin{figure}[t] 
\centering
\includegraphics[width=1\columnwidth]{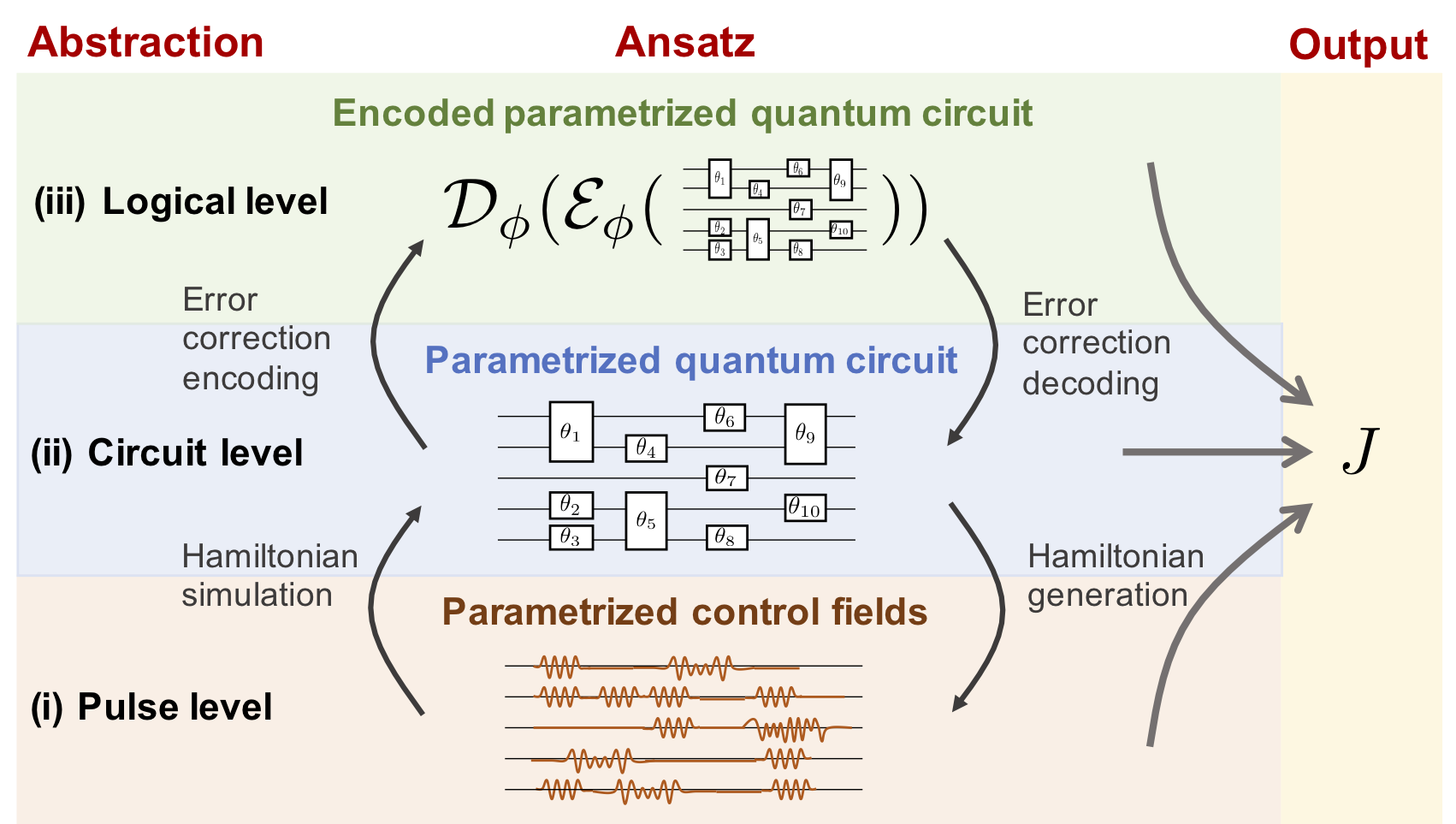}
\caption{A representation of the hierarchy of abstractions used to model quantum hardware, and how variational optimization enters into each level of the hierarchy. Output produced by the hardware is the desired objective function, $J$, which is parametrized in different ways by each level of the hierarchy -- by control field parameters $c_i$ at the pulse level (i), by circuit parameters (usually angles $\theta_i$) at the quantum circuit level (ii), and by the structure of the circuit and the encoding and decoding maps ($\mathcal{E}_\phi$ and $\mathcal{D}_\phi$) at the logical circuit level (iii). By understanding the relationship between the natural variational parametrizations at each of the levels, we can develop a rich family of variational ans\"{a}tze.
\label{fig:AnsatzFamily}}
\end{figure}

A unifying view of VQAs and QOC can be obtained by viewing both as formulations of variational optimization at different levels within an abstraction hierarchy. This is illustrated in Fig.~\ref{fig:AnsatzFamily}, which we now discuss. We first assume that the objective function $J$, determined by the application, is shared between the two approaches. Then, one can think of experimental ways to evaluate $J$ within a hierarchy of abstractions modeling the experimental hardware. At each level (i)-(iii) of this hierarchy, there is a natural parametrization of the control one has over the hardware and this defines a natural variational ansatz at that level. 

At the bottom of this hierarchy is a pulse-level abstraction (i), where we are closest to a first-principles model of the hardware and think in terms of Hamiltonians for the localized computing elements (e.g., qubits, qudits) and fields coupling them. At this level, the parametrization of control is in terms of a continuously-parametrized control field or control Hamiltonian (with parameters $c_i$) that is often realized by a set of electromagnetic fields, which are coupled to the computing elements. QOC typically operates at this level of the modeling hierarchy.

In the middle of the hierarchy (ii), we abstract away first-principles descriptions of the hardware and think in terms of universal circuit elements, or gates, that (ideally) perform well-defined maps on the computing elements. Although there might be a discrete set of types of gates, they can be continuously-parametrized by some set of gate parameters $\{\theta_i\}$. VQAs typically operate at this level of the modeling hierarchy.

Finally, at the highest level of the modeling hierarchy, the logical circuit level (iii), we think in terms of circuits operating on quantum states encoded within an error-correcting code. Any physical circuit at the circuit level of abstraction can be converted into a logical circuit given an error-correcting code and its associated logical gates, with an error rate that is determined by the code and the hardware. Strictly speaking, with conventional error-correcting codes, only random errors in discrete gates are thought to be error correctable to arbitrary precision, thus, arbitrary rotation gates depending on $\theta_i$ are synthesized as a sequence of discrete gates, which performs this rotation to a specified precision. Hence, the rotation angle $\theta_i$ is still present in the logical circuit, but only up to the precision that is given by the synthesis and code procedure. In practice, one may optimize the angle as if it is continuous, so long as the synthesis map is performed after, and the precision is great enough to impact the optimization in practice. Error-corrected quantum computing experiments operate at this level of the modeling hierarchy, and benefit from decreased susceptibility to hardware noise due to the encoding and careful implementation of fault-tolerant operations.

Importantly, an error-correction procedure could be parametrized and considered as variational parameters (these are denoted by the collective parameter $\phi$ in Fig. \ref{fig:AnsatzFamily}). This is not done conventionally, but is a new avenue that opens up when one considers performing VQAs at the logical level of abstraction. For example, if (even partial) characterization of the noise sources in a NISQ device is possible, as shown in \cite{ReimpellW:05,FletcherSW:06,KosutSL:08a,KosutL:06,TaghaviKL:10}, efficient quantum error correction (QEC) which can significantly outperform standard schemes is possible. Unitary encoding and recovery operations tuned to the range and ``character'' of noise-induced uncertainties can be calculated via a bi-convex optimization. The circuit representations of these channel optimized unitaries can then be parametrized. By adding these parameters to the VQA parameters, the QEC system becomes another part of the ``learning control'' environment. Furthermore, continuous parameters like error likelihoods in decoding could be adjusted to attempt to improve performance, and even more sophisticated schemes that allow discrete code modifications could be another potential area of research. 
While maintaining fault tolerance under variation of the error correction procedure can be challenging, doing such variational optimization of small encodings could be useful in the intermediate term where fully fault-tolerant operation is not feasible, but access to limited error correction is possible.

\subsection{Ansatz selection}

With this hierarchy of modeling abstractions and a description of variational optimization at each level, we can exploit the connections between the levels in the hierarchy to define new and richer variational ans\"{a}tze, and in fact, a family of ans\"{a}tze, built from paths on this diagram. To see this, we explore a few sample paths across this diagram and understand the implications of following particular trajectories.

As a first example, we consider a path from the pulse level (i) to the circuit level (ii). We begin at the pulse level (i), with a continuous control perspective of fields acting on $n$ qubits, which is then discretized as a parametrized control. We now have a time-dependent Hamiltonian $H(t, \{c_i\})$ acting on a system of $n$ qubits. The field of quantum algorithms of simulation of time-dependent Hamiltonians is well developed, and a range of methods exist for simulating the time evolution generated by $H(t)$ over some desired time interval $[0,T]$ to arbitrary accuracy. Among the simplest is an operator splitting, also known as ``Trotter factorization'' \cite{suzuki1991general}, but a host of methods with more accurate implementations without direct classical simulation analogs, including quantum walk and so-called linear combinations of unitary approaches, exist that may also be used \cite{berry2009black, berry2015simulating, low2017optimal}. Hence, we may understand this step as implying that the parametrized control can be combined with Hamiltonian simulation to yield a quantum circuit, whose parametrization is naturally understood at the level of $H(t)$. As such, optimizations proceed on a landscape determined by a parametrization at the pulse level (i), despite being implemented at the circuit level (ii). That is, while arbitrary rotations depending on some $\theta_i$ exist in the circuit, they are entirely determined by the composition $\theta_i (\{c_i\})$ in conjunction with the chosen Hamiltonian simulation map. This has the advantage that this circuit can be converted to a logical circuit by discretization into error-correctable gates, mitigating the effects of decoherence while retaining the essential prescription of control. In this case, we retain all of the power of QOC machinery in manipulating and understanding the ansatz, while leveraging the power of quantum algorithms and error correction to ensure theoretical guarantees of implementation.

We can take this same path farther, by appending to it additional gate parametrizations. That is, if Hamiltonian simulation also maps parametrized control to a circuit at the quantum circuit level (ii), e.g., ref.~\cite{Magann2020_Digital}, additional parameters may also be added to the generated circuits, creating a hybrid ansatz. To explore the full power of these connections though, let us explore the other direction that one can take in this perspective. 

Consider starting at the quantum circuit level (ii), with descriptions of hardware in terms of quantum circuits. To make these circuits realizable, a structure and parametrization is selected. Once this circuit has been determined, one can, in principle, map this back to a QOC problem at the pulse level (i) in a number of ways, where now the control parametrization depends on the angles from the circuit such that $c_i(\{\theta_i\})$ is determined through a map that we denote as ``Hamiltonian generation''. This mapping is typically non-unique in a more severe way than in the other direction, especially if one considers mapping the entire circuit to a generating Hamiltonian, whether it be time independent or time dependent. However, much better formed mappings can be used and are related to existing strategies for gate design. For example, one map can perform a gate-wise mapping for each parametrized gate back to the Hamiltonian control parameters $c_i(\{\theta_i\})$ that are used to accurately implement this gate. Similar to before, a hybrid control ansatz can then be created by dropping the dependence on $\theta_i$ and allowing free variation of the parameter $c_i$. This can also be done for fixed gates without parametrizations in the gate model, increasing the overall expressiveness of the resulting circuit in a systematic and new way. We see then, that the gate formulation also naturally provides a family of control ans\"{a}tze.

To give a specific example, we consider a system and ansatz that has received considerable attention in the VQA literature, which is the preparation of the H$_2$ ground state in a minimal, molecular orbital basis. Without symmetry reduction, the ground state of this problem is given by $\ket{\Psi_{\mathrm{GS}}} = \cos{\theta} \ket{0011} + \sin{\theta} \ket{1100}$. A simplified version of a circuit that contains this state within its parametrization is given by

\begin{figure}[h!]
\footnotesize
\centering
\[
  \Qcircuit @C=1em @R=.7em {
 \lstick{\ket{0}} & \gate{R_y(\theta)} & \ctrl{1} & \qw & \qw & \qw & \qw & \qw \\
 \lstick{\ket{0}} & \gate{H} & \ctrl{-1} & \gate{H} & \ctrl{1} & \qw & \qw & \qw \\
 \lstick{\ket{0}} & \gate{X} &  \qw & \gate{H} & \ctrl{-1} & \gate{H} & \ctrl{1} & \qw \\
 \lstick{\ket{0}} & \gate{X} & \qw & \qw & \qw &  \gate{H} & \ctrl{-1} & \gate{H}
  }
\]
\end{figure}
Where, in this diagram, $H$ is the Hadamard gate, $R_y$ is a rotation about the Pauli $Y$ axis, and the connecting line is a two-qubit controlled-Z gate, $CZ=\text{diag}(1,1,1,-1)$. In this case, variation over the parameter $\theta$ provides the variational freedom required to prepare the ground state. In addition, as outlined above, we can map back to control implementations of gates, including those that don't yet have parameters. For example, a controlled-phase gates may be written as a time-independent evolution under a Hamiltonian of the form $H_z = c_{12} Z_1 Z_2 + c_1 Z_2 + c_2 Z_2$ for a time $\tau$, up to an unimportant global phase. If this is how the gate is implemented in our physical system, then the parameter $\tau$ can be added to the list of parameters to create a hybrid control ansatz. The ansatz in this case includes the beginning of the circuit, as well as analog evolution under $H_z$ which may be implemented by physical means.

Finally, taking the mapping in the other direction, the evolution under the Hamiltonian $H$ for a time $\tau$ in this case, could be re-digitized into a Trotter factorization of $\exp(-i H_z \tau) = \exp(-i \tau c_{12} Z_1 Z_2) \exp(-i \tau c_{1} Z_1) \exp(-i \tau c_{2} Z_2)$, where each term in the product can be translated back into a digital sequence, and now we can vary the set of parameters $\{\theta, c_1, c_2, c_{12}\}$, where we have retained the $\theta$ parametrization and the rest of the circuit, but added more controls. Note that for this simple problem, if the implementation is error-free, this additional freedom may be unnecessary. However, if some errors occur in the gates, this freedom can be used to improve the result. Moreover, problems that are similar, but not identical to the hydrogen molecule, can then leverage this base ansatz and natural extensions from it. This simple example is designed to illustrate the ways that one can move from the circuit level (ii) to a hybrid ansatz that combines tools from the circuit and pulse levels (ii) and (i) and back again, while changing the nature of the ansatz.

Finally, we see that beginning at any level in the hierarchy in Fig.~\ref{fig:AnsatzFamily}, it is possible to iterate on the connections between these approaches, choosing to carry forward parametrizations or develop new ones, and map back to the other corresponding formalism. For example, if one takes a variational parametrization at the control-field level, maps it to a quantum circuit with a Hamiltonian simulation map, then selects a particular gate and maps it back to a control parametrization, and uses the combined circuit, we get a new hybrid parametrization that can be manipulated closer to a device-level description. It is easy to see that these choices can be mixed, matched, and iterated on to create an entire family of parametrizations. Here, we envision that the tools of QOC are bolstered by concepts from digital simulation, including error correction and advanced time-propagation algorithms, and vice versa.

In addition to expanding the set of options and tools of both areas, we hope that this formulation and perspective of connections allows more direct cross-pollination of ideas. For example, by formulating a circuit ansatz based on control theory, we more directly understand the consequences of controllability in a circuit model. In turn, by developing a control-theoretic ansatz from a circuit family known to be universal, we may find new insights in relation to entanglement-generating power or expressiveness of quantum circuits. 

\subsection{Optimization landscapes} \label{Sec:OptLand}

A critical consideration in VQAs is the difficulty of optimizing over the circuit parameters. The ease of finding the global minimum of $J$ during this optimization process is dictated by the structure of the underlying optimization landscape as well as available prior information about the location of good optima. A central theme of QOC is the study of such landscapes, and translating these tools into the context of VQAs will allow for new strategies in ansatz and problem design.

At their core, VQAs represent the mapping of a convex optimization problem in an exponentially-large linear space, to a non-convex optimization problem over parametrized quantum circuits. This can be seen by expanding the variational state $\ket{\psi(\{\theta_{i}\})}$ in the eigenbasis $\{\ket{n}\}$ of $H_{\mathrm{p}}$. The objective functional then takes the form $J[\{\theta_{i}\}] = \sum_{n=1}^{d} \lambda_{n}(\{\theta_{i}\}) E_{n}$, where $E_{n}$ are the eigenvalues of $H_{\mathrm{p}}$ and $\lambda_{n}=|\langle n\ket{\psi(\{\theta_{i}\})}|^{2}$. When $J$ is optimized over the set $\{\lambda_{n}\}$, the optimization problem is convex, which implies that the corresponding optimization landscape is free from local optima. In contrast, the parameters $\theta_{i}$ typically enter in a non-linear fashion in $\lambda_{n}$, often via an exponential map in $U_i(\theta_{i})$. As such, when $J$ is optimized over the set $\{\theta_{i}\}$ the optimization problem is non-convex and in general, local optima can appear. 

In the context of QOC, to address these and other landscape considerations, a sizeable body of research has analyzed the topological features and properties of the (dynamic) QOC landscape, defined by $J$ as a functional of the control fields \cite{rabitz2004quantum, rabitz2006topology, HO2006226, ho2009landscape, pechen2011there, de2013closer}. To characterize these landscapes, recall that $J$ depends on the time-evolution operator up to time $T$, such that we have $J = J(U_{T}[f(t)])$. Then, the functional derivative of $J$ with respect to a single field $\frac{\delta J}{\delta f(t)}$ is given by the composition $\frac{\delta J}{\delta f(t)} = \frac{\partial J}{\partial U_{T}} \circ \frac{\delta U_{T}}{\delta f(t)}$. The first term in the composition captures the properties of the kinematic control landscape, i.e., $J$ defined as a function of $U_{T}$. The composition above implies that if the variation of $U_{T}$ with respect to the control field $\frac{\delta U_{T}}{\delta f(t)}$ is assumed to be full rank, i.e., equal to $d^2$, then the dynamic and kinematic critical points coincide. This result is referred to as \emph{local surjectivity}. In this scenario, the topology of the dynamic control landscape is fully characterized by the critical point structure of the kinematic control landscape. Results from QOC theory have shown that the kinematic control landscapes of typical objective functionals consist of global extrema and saddles \cite{HO2006226, ho2009landscape, chakrabarti2007quantum, rabitz2006optimal, hsieh2008optimal}. As such, if we are able to create every unitary transformation such that $J$ as a function of $U_{T}$ can be varied arbitrarily, and if $U_{T}$ can be varied arbitrarily also by varying the control field, then the landscape of $J$ consists of saddles and global extrema. More precisely, under the assumptions of (1) a full dynamical Lie algebra, (2) unconstrained control fields of arbitrary length and shape, and (3) local surjectivity, the control landscapes for commonly-employed objective functionals are free from local extrema. Although debated in the literature \cite{pechen2011there, PhysRevLett.108.198901, PhysRevLett.108.198902, Moore-Tibbetts2012Exploring, de2013closer,riviello2014searching,russell2017control, Zhdanov_2018, Russell_2018}, extensive QOC simulation and laboratory studies indicate the relative ease of satisfying all three assumptions \cite{Brif2011}. Furthermore, it has been shown that assumptions (1) and (3) are almost always satisfied in a measure theoretic sense \cite{altafini2002controllability,russell2017control,2020arXiv200402729A}, thereby implying that QOC landscapes are almost always free of local minima under the premise of sufficient control field resources \cite{russell2017control}. While the precise meaning of ``sufficient'' is application-dependent and remains an open challenge to systematically assess, it has recently been shown that local surjectivity is almost always met when the control fields allow for approximating Haar random evolutions \cite{banchi2017driven,2020arXiv200402729A} within the time interval of interest: $[0, T]$. As such, even though performing the optimization over a set of parametrized control fields does not avoid the non-convexity of the problem, provided the assumptions (1)-(3) hold, the interior of the optimization landscape provably contains saddles only. 

Looking ahead, we hope that the theory of QOC landscapes can provide insight into optimization landscapes of VQAs, which could include better tools to understand different ans\"{a}tze. Most directly, the optimization landscape of a variational ansatz formulated at the control-field level in Fig. \ref{fig:AnsatzFamily} could be analyzed in terms of existing QOC landscape theory. Here, in addition to the theoretical foundations, numerical tools, such as D-MORPH (a flexible continuation-method developed in the context of QOC) \cite{moore2011exploring, Hocker2014_Characterization} and FLACCO (software for feature-based landscape analysis of continuous and constrained optimization problems) \cite{Kerschke2017_FLACCO}, could be employed to explore these landscapes. However, there are a number of caveats to consider for this direct approach; that is, although QOC has many well developed tools for the characterization of QOC landscapes, there remain challenges for applying these tools to scalable and practical implementations of VQAs. For instance, characterizing the landscape topology in the presence of constraints is challenging; although these landscapes are expected to be free from local extrema when sufficiently many unconstrained variables $\{c_i\}$ are used to parametrize the control fields, there is evidence that an insufficient number of parameters (i.e., less than $d^{2}$), leading to a violation of the surjectivity assumption (3), and control field constraints are among the reasons for local extrema to appear \cite{Riviello2015_Searching}. In general, further research is needed when the control resources do not scale with the dimension $d$ of the quantum system being controlled, but rather only scale as $\mathcal O(\mathrm{poly}(\mathrm{log}(d)))$. To this end, it may be interesting to study systems that are not fully controllable, but where all states within a certain subspace defined by the symmetries of the (controlled) system are reachable. One may then ask whether a (significantly) smaller number of control parameters could be used to obtain a trap-free landscape when moving only in a restricted subspace (containing the ground state) whose dimension does not scale exponentially. 

We also remark that as the dimension $d$ of the quantum system increases the landscape flattens out, such that for certain objective functionals, gradients become exponentially small as a function of the number of qubits, making these regions difficult to leave with local optimization algorithms. These so-called barren plateaus \cite{mcclean2018barren} are a consequence of a concentration of measure phenomenon, and occur in both VQA implementations and QOC \cite{2020arXiv200402729A}. Understanding how to avoid them in each context independently could yield useful and transferable techniques benefiting both areas of study, and will likely involve careful design of ansatz, initialization, and training methods, e.g., ref.~\cite{Volkoff2020_Large}.

\subsection{Noise and time-optimal control} \label{Sec:Noise}

Today, errors in NISQ devices severely restrict the circuit depths that are achievable for VQAs. For example, certain errors can arise from stochastic fluctuations in the gate implementations, leading to errors that accumulate as circuit depth increases. In QOC, robust control strategies have been developed to suppress random errors like this, by seeking pulses that are robust in the presence of finite control noise; the condition of the Hessian of $J$ can be used to determine such properties \cite{Hocker2014_Characterization}. We believe more direct translation of robust control tools from QOC into variational algorithms will help improve their robustness against general errors.

Other, often more insidious, errors stem from interactions of the qubits with their surrounding environment that can cause the system to decohere over time. Thus, the timescale associated with performing a quantum circuit should ideally be restricted to the coherence time of the system. When coherence times are limited and gates are dominated by stochastic errors, it is desirable to seek VQA circuits that drive the qubits to the minimum of $J$ as quickly as possible, i.e., using circuits with minimum depth. In the context of QOC, the theory of time-optimal control, and the associated theory of quantum speed limits, provide a powerful framework for considering this issue at the pulse level, and may offer valuable tools to enhance VQA performance. In general, identifying time-optimal fields is a challenging task, as it corresponds to finding the associated geodesic on the unitary group \cite{carlini2006time, nielsen2006quantum, nielsen2006optimal, carlini2007time, wang2015quantum, Moore-Tibbetts2012Exploring}. Exact results are only known for certain systems consisting of 1-3 qubits \cite{doi:10.1063/1.4906137, PhysRevA.90.062302, PhysRevLett.111.260501,  khaneja2001time,khaneja2002sub}. However, upper and lower bounds on $T$ can be found \cite{arenz2017roles, lee2018dependence, arenz2018controlling, lee2020upper}. In addition, for $n$ qubit networks, the upper-bounds in \cite{arenz2018controlling, lee2020upper} allow for characterizing the unitary transformations that can be created efficiently with $2n$ local fields, i.e., where $T = \mathcal{O}(\textrm{poly}(n))$. While some progress has recently been made to characterize efficiently-controllable qubit graphs \cite{gokler2017efficiently}, in general, it remains an open challenge to systematically determine the set of unitary transformations that are reachable in polynomial time with fewer controls. Finally, we remark that additional control resources not only offer faster control strategies and in general, richer ans\"atze, but can also allow for counteracting decoherence by enabling decoupling schemes \cite{viola2002quantum, west2010high, arenz2014control,  kosut2019quantum} that can suppress interactions with the environment. 

\subsection{Robust control through digitization} \label{Sec:quantdata}
All the directions covered by the previous subsections are fairly natural in the VQA setting. However, if we think more broadly, the scope of interactions between VQAs and QOC could be even more fruitful. In particular, consider how VQAs could impact traditional QOC experiments. Until this point, we have considered a setting natural to VQA's, where one typically starts from a known initial state, e.g., $\ket{00..0}$ and builds up the state of interest using a set of parametrized gates or controls. However, many QOC experiment deals with situations where one is given an unknown quantum state $\ket{\psi}$, and wishes to manipulate it. For example, this quantum state may be the state of reactants in a chemical system and the control goal is to steer the reaction products. Alternatively, in a quantum sensing context, the quantum state may be the state of the sensed environment and the control goal is to distinguish features in the environment.

If we now consider such a QOC experiment from the point of view of VQAs, and think of the applied controls as a parametrized unitary transformation, one can translate these controls into a parametrized circuit ansatz \cite{Magann2020_Digital} and further, into an error-corrected circuit. This would have the effect of ensuring that the controls are applied to an arbitrarily high degree of accuracy, an advantage of taking the digital point of view. Of course, to be fully compatible with quantum error correction, the input quantum state may also need to be encoded, and effective digital encoding of quantum states from nature is an experimental challenge that has not yet been realized, but this perspective has the potential to enable QOC experiments of arbitrary accuracy on many-body systems.

The increasing difficulty of precisely controlling a quantum system with non-ideal analog controls as its dimension increases can be understood in various ways. Conceptually, this is a consequence of the orthogonality catastrophe -- where small perturbations from the ideal Hamiltonian (in this setting, caused by small errors in the analog controls) can lead to an exponential decay (with system dimension) of fidelity with the state under ideal evolution \cite{dowdall_fast_2017,fogarty_orthogonality_2020}. More operationally, consider an $N$-body quantum system with state vector of dimension $d\sim \exp(N)$. 
We can estimate the requirements on control precision with a simple model. Assume we want to prepare the state with 1 in the first entry of the state vector, and 0 on the remaining $(d-1)$ entries. Now for each dimension we are able to achieve a precision $\delta$, and consider for simplicity the normalized erred state that is off by $\delta$ in each of the entries except the first. In this case, the fidelity with the target state is given by $1 / [1 + (d-1) \delta^2]$, which means to maintain a constant fidelity, one must have $\delta \propto d^{-1/2}$. This is an exponential (in $N$) precision requirement for the control.

In contrast, if the state of this $N$-body system can be \emph{transduced} into $n = \log_2(d)$ qubits, the control of the state can be encoded in a VQA circuit \cite{Magann2020_Digital}. Then, assuming the fidelity of qubit gates are sufficiently high and using the tools of quantum error correction, the fidelity of controlling this state can be brought arbitrarily close to $1$ with logarithmic overhead in physical resources.

At a broader application level, the manipulation of unknown quantum states in a VQA manner encompasses what is sometimes called ``quantum machine learning'' for training data provided in the form of quantum states rather than classical inputs, or quantum data. This idea allows for the tools QOC to come to bear in this community. Moreover, this permits a closer connection to query based proofs in quantum computer science, where stronger proofs are possible in the setting where only a limited number of quantum states are available \cite{knill1996quantum, cleve2004consequences, Harrow:2017a, bravyi2018quantum, centrone2020experimental}.

\section{Outlook}
\label{sec:outlook}

In this perspective, we have explored the connections between VQAs and the theory of QOC. While VQA applications constitute some of the most promising applications of near-term quantum computers, more progress bridging the gap between theoretical VQAs and NISQ computing hardware is necessary to realize this promise. We have argued that concepts from QOC, and the more general framework outlined in Sec. \ref{sec:futuredirections}, which integrates traditional models of QOC with VQAs and extends both, are critical to bridging this gap. Both fields have something to gain from exploring this fertile area of overlap. From the perspective of VQAs, the mature theory and powerful tools of QOC can provide richer variational structures and offer a deeper understanding of variational experiments. Conversely, from the perspective of QOC, VQAs present an exciting frontier of many-body quantum control, pushing this established field in new directions. For these reasons, we expect bountiful fruits at the intersection of these two areas of research in the years to come.

\begin{acknowledgments}

A.B.M., M.D.G., and M.S. were supported by the U.S. Department of Energy, Office of Science, Office of Advanced Scientific Computing Research, under the Quantum Computing Application Teams program. A.B.M. also acknowledges support from the U.S. Department of Energy, Office of Science, Office of Advanced Scientific Computing Research, Department of Energy Computational Science Graduate Fellowship under Award Number DE-FG02-97ER25308. M.D.G. also acknowledges support from the U.S. Department of Energy, Office of Advanced Scientific Computing Research, under the Quantum Algorithm Teams program. C.A. and H.A.R. were supported by the U.S. Army Research Office, grant numbers W911NF-16-1-0014 and W911NF-19-1-0382, respectively. T.S.H. acknowledges support from the U.S. Department of Energy, grant number DE-FG02-02ER15344. 

Sandia National Laboratories is a multimission laboratory managed and operated by National Technology \& Engineering Solutions of Sandia, LLC, a wholly owned subsidiary of Honeywell International Inc., for the U.S. Department of Energy's National Nuclear Security Administration under contract DE-NA0003525. This paper describes objective technical results and analysis. Any subjective views or opinions that might be expressed in the paper do not necessarily represent the views of the U.S. Department of Energy or the United States Government. 

This report was prepared as an account of work sponsored by an agency of the United States Government. Neither the United States Government nor any agency thereof, nor any of their employees, makes any warranty, express or implied, or assumes any legal liability or responsibility for the accuracy, completeness, or usefulness of any information, apparatus, product, or process disclosed, or represents that its use would not infringe privately owned rights. Reference herein to any specific commercial product, process, or service by trade name, trademark, manufacturer, or otherwise does not necessarily constitute or imply its endorsement, recommendation, or favoring by the United States Government or any agency thereof. The views and opinions of authors expressed herein do not necessarily state or reflect those of the United States Government or any agency thereof. 

\end{acknowledgments}

\bibliography{bib}

\begin{thebibliography}{162}%
\makeatletter
\providecommand \@ifxundefined [1]{%
 \@ifx{#1\undefined}
}%
\providecommand \@ifnum [1]{%
 \ifnum #1\expandafter \@firstoftwo
 \else \expandafter \@secondoftwo
 \fi
}%
\providecommand \@ifx [1]{%
 \ifx #1\expandafter \@firstoftwo
 \else \expandafter \@secondoftwo
 \fi
}%
\providecommand \natexlab [1]{#1}%
\providecommand \enquote  [1]{``#1''}%
\providecommand \bibnamefont  [1]{#1}%
\providecommand \bibfnamefont [1]{#1}%
\providecommand \citenamefont [1]{#1}%
\providecommand \href@noop [0]{\@secondoftwo}%
\providecommand \href [0]{\begingroup \@sanitize@url \@href}%
\providecommand \@href[1]{\@@startlink{#1}\@@href}%
\providecommand \@@href[1]{\endgroup#1\@@endlink}%
\providecommand \@sanitize@url [0]{\catcode `\\12\catcode `\$12\catcode
  `\&12\catcode `\#12\catcode `\^12\catcode `\_12\catcode `\%12\relax}%
\providecommand \@@startlink[1]{}%
\providecommand \@@endlink[0]{}%
\providecommand \url  [0]{\begingroup\@sanitize@url \@url }%
\providecommand \@url [1]{\endgroup\@href {#1}{\urlprefix }}%
\providecommand \urlprefix  [0]{URL }%
\providecommand \Eprint [0]{\href }%
\providecommand \doibase [0]{http://dx.doi.org/}%
\providecommand \selectlanguage [0]{\@gobble}%
\providecommand \bibinfo  [0]{\@secondoftwo}%
\providecommand \bibfield  [0]{\@secondoftwo}%
\providecommand \translation [1]{[#1]}%
\providecommand \BibitemOpen [0]{}%
\providecommand \bibitemStop [0]{}%
\providecommand \bibitemNoStop [0]{.\EOS\space}%
\providecommand \EOS [0]{\spacefactor3000\relax}%
\providecommand \BibitemShut  [1]{\csname bibitem#1\endcsname}%
\let\auto@bib@innerbib\@empty
\bibitem [{\citenamefont {Shor}(1994)}]{10.1109/SFCS.1994.365700}%
  \BibitemOpen
  \bibfield  {author} {\bibinfo {author} {\bibfnamefont {P.~W.}\ \bibnamefont
  {Shor}},\ }in\ \href {\doibase 10.1109/SFCS.1994.365700} {\emph {\bibinfo
  {booktitle} {Proceedings of the 35th Annual Symposium on Foundations of
  Computer Science}}},\ \bibinfo {series and number} {SFCS ’94}\ (\bibinfo
  {publisher} {IEEE Computer Society},\ \bibinfo {address} {USA},\ \bibinfo
  {year} {1994})\ p.\ \bibinfo {pages} {124–134}\BibitemShut {NoStop}%
\bibitem [{\citenamefont {{Lloyd}}(1996)}]{1996Sci...273.1073L}%
  \BibitemOpen
  \bibfield  {author} {\bibinfo {author} {\bibfnamefont {S.}~\bibnamefont
  {{Lloyd}}},\ }\href {\doibase 10.1126/science.273.5278.1073} {\bibfield
  {journal} {\bibinfo  {journal} {Science}\ }\textbf {\bibinfo {volume}
  {273}},\ \bibinfo {pages} {1073} (\bibinfo {year} {1996})}\BibitemShut
  {NoStop}%
\bibitem [{\citenamefont {Egan}\ \emph {et~al.}(2020)\citenamefont {Egan},
  \citenamefont {Debroy}, \citenamefont {Noel}, \citenamefont {Risinger},
  \citenamefont {Zhu}, \citenamefont {Biswas}, \citenamefont {Newman},
  \citenamefont {Li}, \citenamefont {Brown}, \citenamefont {Cetina},\ and\
  \citenamefont {Monroe}}]{egan2020faulttolerant}%
  \BibitemOpen
  \bibfield  {author} {\bibinfo {author} {\bibfnamefont {L.}~\bibnamefont
  {Egan}}, \bibinfo {author} {\bibfnamefont {D.~M.}\ \bibnamefont {Debroy}},
  \bibinfo {author} {\bibfnamefont {C.}~\bibnamefont {Noel}}, \bibinfo {author}
  {\bibfnamefont {A.}~\bibnamefont {Risinger}}, \bibinfo {author}
  {\bibfnamefont {D.}~\bibnamefont {Zhu}}, \bibinfo {author} {\bibfnamefont
  {D.}~\bibnamefont {Biswas}}, \bibinfo {author} {\bibfnamefont
  {M.}~\bibnamefont {Newman}}, \bibinfo {author} {\bibfnamefont
  {M.}~\bibnamefont {Li}}, \bibinfo {author} {\bibfnamefont {K.~R.}\
  \bibnamefont {Brown}}, \bibinfo {author} {\bibfnamefont {M.}~\bibnamefont
  {Cetina}}, \ and\ \bibinfo {author} {\bibfnamefont {C.}~\bibnamefont
  {Monroe}},\ }\href {https://arxiv.org/abs/2009.11482} {\bibfield  {journal}
  {\bibinfo  {journal} {arXiv preprint arXiv:1906.02700}\ } (\bibinfo {year}
  {2020})}\BibitemShut {NoStop}%
\bibitem [{\citenamefont {Preskill}(2018)}]{preskill_quantum_2018}%
  \BibitemOpen
  \bibfield  {author} {\bibinfo {author} {\bibfnamefont {J.}~\bibnamefont
  {Preskill}},\ }\href {\doibase 10.22331/q-2018-08-06-79} {\bibfield
  {journal} {\bibinfo  {journal} {Quantum}\ }\textbf {\bibinfo {volume} {2}},\
  \bibinfo {pages} {79} (\bibinfo {year} {2018})}\BibitemShut {NoStop}%
\bibitem [{\citenamefont {Peruzzo}\ \emph {et~al.}(2013)\citenamefont
  {Peruzzo}, \citenamefont {McClean}, \citenamefont {Shadbolt}, \citenamefont
  {Yung}, \citenamefont {Zhou}, \citenamefont {Love}, \citenamefont
  {Aspuru-Guzik},\ and\ \citenamefont {O'Brien}}]{peruzzo_variational_2013}%
  \BibitemOpen
  \bibfield  {author} {\bibinfo {author} {\bibfnamefont {A.}~\bibnamefont
  {Peruzzo}}, \bibinfo {author} {\bibfnamefont {J.}~\bibnamefont {McClean}},
  \bibinfo {author} {\bibfnamefont {P.}~\bibnamefont {Shadbolt}}, \bibinfo
  {author} {\bibfnamefont {M.-H.}\ \bibnamefont {Yung}}, \bibinfo {author}
  {\bibfnamefont {X.-Q.}\ \bibnamefont {Zhou}}, \bibinfo {author}
  {\bibfnamefont {P.~J.}\ \bibnamefont {Love}}, \bibinfo {author}
  {\bibfnamefont {A.}~\bibnamefont {Aspuru-Guzik}}, \ and\ \bibinfo {author}
  {\bibfnamefont {J.~L.}\ \bibnamefont {O'Brien}},\ }\href {\doibase
  10.1038/ncomms5213} {\bibfield  {journal} {\bibinfo  {journal} {Nat.
  Commun.}\ }\textbf {\bibinfo {volume} {5}},\ \bibinfo {pages} {4213}
  (\bibinfo {year} {2013})}\BibitemShut {NoStop}%
\bibitem [{\citenamefont {{Farhi}}\ \emph {et~al.}(2014)\citenamefont
  {{Farhi}}, \citenamefont {{Goldstone}},\ and\ \citenamefont
  {{Gutmann}}}]{2014arXiv1411.4028F}%
  \BibitemOpen
  \bibfield  {author} {\bibinfo {author} {\bibfnamefont {E.}~\bibnamefont
  {{Farhi}}}, \bibinfo {author} {\bibfnamefont {J.}~\bibnamefont
  {{Goldstone}}}, \ and\ \bibinfo {author} {\bibfnamefont {S.}~\bibnamefont
  {{Gutmann}}},\ }\href@noop {} {\bibfield  {journal} {\bibinfo  {journal}
  {arXiv e-prints}\ ,\ \bibinfo {eid} {arXiv:1411.4028}} (\bibinfo {year}
  {2014})},\ \Eprint {http://arxiv.org/abs/1411.4028} {arXiv:1411.4028
  [quant-ph]} \BibitemShut {NoStop}%
\bibitem [{\citenamefont {Dunjko}\ and\ \citenamefont
  {Wittek}(2020)}]{Dunjko2020nonreviewofquantum}%
  \BibitemOpen
  \bibfield  {author} {\bibinfo {author} {\bibfnamefont {V.}~\bibnamefont
  {Dunjko}}\ and\ \bibinfo {author} {\bibfnamefont {P.}~\bibnamefont
  {Wittek}},\ }\href {\doibase 10.22331/qv-2020-03-17-32} {\bibfield  {journal}
  {\bibinfo  {journal} {{Quantum Views}}\ }\textbf {\bibinfo {volume} {4}},\
  \bibinfo {pages} {32} (\bibinfo {year} {2020})}\BibitemShut {NoStop}%
\bibitem [{\citenamefont {McClean}\ \emph {et~al.}(2016)\citenamefont
  {McClean}, \citenamefont {Romero}, \citenamefont {Babbush},\ and\
  \citenamefont {Aspuru-Guzik}}]{McClean_2016}%
  \BibitemOpen
  \bibfield  {author} {\bibinfo {author} {\bibfnamefont {J.~R.}\ \bibnamefont
  {McClean}}, \bibinfo {author} {\bibfnamefont {J.}~\bibnamefont {Romero}},
  \bibinfo {author} {\bibfnamefont {R.}~\bibnamefont {Babbush}}, \ and\
  \bibinfo {author} {\bibfnamefont {A.}~\bibnamefont {Aspuru-Guzik}},\ }\href
  {\doibase 10.1088/1367-2630/18/2/023023} {\bibfield  {journal} {\bibinfo
  {journal} {New J. Phys.}\ }\textbf {\bibinfo {volume} {18}},\ \bibinfo
  {pages} {023023} (\bibinfo {year} {2016})}\BibitemShut {NoStop}%
\bibitem [{\citenamefont {Nakamura}\ \emph {et~al.}(1999)\citenamefont
  {Nakamura}, \citenamefont {Pashkin},\ and\ \citenamefont
  {Tsai}}]{nakamura1999coherent}%
  \BibitemOpen
  \bibfield  {author} {\bibinfo {author} {\bibfnamefont {Y.}~\bibnamefont
  {Nakamura}}, \bibinfo {author} {\bibfnamefont {Y.~A.}\ \bibnamefont
  {Pashkin}}, \ and\ \bibinfo {author} {\bibfnamefont {J.~S.}\ \bibnamefont
  {Tsai}},\ }\href {https://www.nature.com/articles/19718} {\bibfield
  {journal} {\bibinfo  {journal} {Nature}\ }\textbf {\bibinfo {volume} {398}},\
  \bibinfo {pages} {786} (\bibinfo {year} {1999})}\BibitemShut {NoStop}%
\bibitem [{\citenamefont {Yamamoto}\ \emph {et~al.}(2003)\citenamefont
  {Yamamoto}, \citenamefont {Pashkin}, \citenamefont {Astafiev}, \citenamefont
  {Nakamura},\ and\ \citenamefont {Tsai}}]{yamamoto2003demonstration}%
  \BibitemOpen
  \bibfield  {author} {\bibinfo {author} {\bibfnamefont {T.}~\bibnamefont
  {Yamamoto}}, \bibinfo {author} {\bibfnamefont {Y.~A.}\ \bibnamefont
  {Pashkin}}, \bibinfo {author} {\bibfnamefont {O.}~\bibnamefont {Astafiev}},
  \bibinfo {author} {\bibfnamefont {Y.}~\bibnamefont {Nakamura}}, \ and\
  \bibinfo {author} {\bibfnamefont {J.-S.}\ \bibnamefont {Tsai}},\ }\href
  {https://www.nature.com/articles/nature02015} {\bibfield  {journal} {\bibinfo
   {journal} {Nature}\ }\textbf {\bibinfo {volume} {425}},\ \bibinfo {pages}
  {941} (\bibinfo {year} {2003})}\BibitemShut {NoStop}%
\bibitem [{\citenamefont {Waldherr}\ \emph {et~al.}(2014)\citenamefont
  {Waldherr}, \citenamefont {Wang}, \citenamefont {Zaiser}, \citenamefont
  {Jamali}, \citenamefont {Schulte-Herbr{\"u}ggen}, \citenamefont {Abe},
  \citenamefont {Ohshima}, \citenamefont {Isoya}, \citenamefont {Du},
  \citenamefont {Neumann} \emph {et~al.}}]{waldherr2014quantum}%
  \BibitemOpen
  \bibfield  {author} {\bibinfo {author} {\bibfnamefont {G.}~\bibnamefont
  {Waldherr}}, \bibinfo {author} {\bibfnamefont {Y.}~\bibnamefont {Wang}},
  \bibinfo {author} {\bibfnamefont {S.}~\bibnamefont {Zaiser}}, \bibinfo
  {author} {\bibfnamefont {M.}~\bibnamefont {Jamali}}, \bibinfo {author}
  {\bibfnamefont {T.}~\bibnamefont {Schulte-Herbr{\"u}ggen}}, \bibinfo {author}
  {\bibfnamefont {H.}~\bibnamefont {Abe}}, \bibinfo {author} {\bibfnamefont
  {T.}~\bibnamefont {Ohshima}}, \bibinfo {author} {\bibfnamefont
  {J.}~\bibnamefont {Isoya}}, \bibinfo {author} {\bibfnamefont
  {J.}~\bibnamefont {Du}}, \bibinfo {author} {\bibfnamefont {P.}~\bibnamefont
  {Neumann}},  \emph {et~al.},\ }\href
  {https://www.nature.com/articles/nature12919} {\bibfield  {journal} {\bibinfo
   {journal} {Nature}\ }\textbf {\bibinfo {volume} {506}},\ \bibinfo {pages}
  {204} (\bibinfo {year} {2014})}\BibitemShut {NoStop}%
\bibitem [{\citenamefont {Dolde}\ \emph {et~al.}(2014)\citenamefont {Dolde},
  \citenamefont {Bergholm}, \citenamefont {Wang}, \citenamefont {Jakobi},
  \citenamefont {Naydenov}, \citenamefont {Pezzagna}, \citenamefont {Meijer},
  \citenamefont {Jelezko}, \citenamefont {Neumann}, \citenamefont
  {Schulte-Herbr{\"u}ggen} \emph {et~al.}}]{dolde2014high}%
  \BibitemOpen
  \bibfield  {author} {\bibinfo {author} {\bibfnamefont {F.}~\bibnamefont
  {Dolde}}, \bibinfo {author} {\bibfnamefont {V.}~\bibnamefont {Bergholm}},
  \bibinfo {author} {\bibfnamefont {Y.}~\bibnamefont {Wang}}, \bibinfo {author}
  {\bibfnamefont {I.}~\bibnamefont {Jakobi}}, \bibinfo {author} {\bibfnamefont
  {B.}~\bibnamefont {Naydenov}}, \bibinfo {author} {\bibfnamefont
  {S.}~\bibnamefont {Pezzagna}}, \bibinfo {author} {\bibfnamefont
  {J.}~\bibnamefont {Meijer}}, \bibinfo {author} {\bibfnamefont
  {F.}~\bibnamefont {Jelezko}}, \bibinfo {author} {\bibfnamefont
  {P.}~\bibnamefont {Neumann}}, \bibinfo {author} {\bibfnamefont
  {T.}~\bibnamefont {Schulte-Herbr{\"u}ggen}},  \emph {et~al.},\ }\href
  {https://www.nature.com/articles/ncomms4371} {\bibfield  {journal} {\bibinfo
  {journal} {Nat. Commun.}\ }\textbf {\bibinfo {volume} {5}},\ \bibinfo {pages}
  {3371} (\bibinfo {year} {2014})}\BibitemShut {NoStop}%
\bibitem [{Note1()}]{Note1}%
  \BibitemOpen
  \bibinfo {note} {We note that we use the term \protect \emph {quantum optimal
  control} to refer solely to the closed-loop control scenario just described,
  and do not consider other quantum control settings such as real-time feedback
  control in this paper.}\BibitemShut {Stop}%
\bibitem [{\citenamefont {Venturelli}\ \emph {et~al.}(2018)\citenamefont
  {Venturelli}, \citenamefont {Do}, \citenamefont {Rieffel},\ and\
  \citenamefont {Frank}}]{venturelli2018compiling}%
  \BibitemOpen
  \bibfield  {author} {\bibinfo {author} {\bibfnamefont {D.}~\bibnamefont
  {Venturelli}}, \bibinfo {author} {\bibfnamefont {M.}~\bibnamefont {Do}},
  \bibinfo {author} {\bibfnamefont {E.}~\bibnamefont {Rieffel}}, \ and\
  \bibinfo {author} {\bibfnamefont {J.}~\bibnamefont {Frank}},\ }\href
  {https://iopscience.iop.org/article/10.1088/2058-9565/aaa331/meta?casa_token=3MT3WoW0cJQAAAAA:z4iYXGD79TEsGVjKH7PIo161iVIlKgnMulgEux3405DtOAy-IciAEl-BjzXQt4JGRxxz2tny924}
  {\bibfield  {journal} {\bibinfo  {journal} {Quantum Sci. Technol.}\ }\textbf
  {\bibinfo {volume} {3}},\ \bibinfo {pages} {025004} (\bibinfo {year}
  {2018})}\BibitemShut {NoStop}%
\bibitem [{\citenamefont {Khatri}\ \emph
  {et~al.}(2019{\natexlab{a}})\citenamefont {Khatri}, \citenamefont {LaRose},
  \citenamefont {Poremba}, \citenamefont {Cincio}, \citenamefont {Sornborger},\
  and\ \citenamefont {Coles}}]{khatri2019quantum}%
  \BibitemOpen
  \bibfield  {author} {\bibinfo {author} {\bibfnamefont {S.}~\bibnamefont
  {Khatri}}, \bibinfo {author} {\bibfnamefont {R.}~\bibnamefont {LaRose}},
  \bibinfo {author} {\bibfnamefont {A.}~\bibnamefont {Poremba}}, \bibinfo
  {author} {\bibfnamefont {L.}~\bibnamefont {Cincio}}, \bibinfo {author}
  {\bibfnamefont {A.~T.}\ \bibnamefont {Sornborger}}, \ and\ \bibinfo {author}
  {\bibfnamefont {P.~J.}\ \bibnamefont {Coles}},\ }\href
  {https://quantum-journal.org/papers/q-2019-05-13-140/} {\bibfield  {journal}
  {\bibinfo  {journal} {Quantum}\ }\textbf {\bibinfo {volume} {3}},\ \bibinfo
  {pages} {140} (\bibinfo {year} {2019}{\natexlab{a}})}\BibitemShut {NoStop}%
\bibitem [{\citenamefont {Murali}\ \emph {et~al.}(2019)\citenamefont {Murali},
  \citenamefont {Baker}, \citenamefont {Javadi-Abhari}, \citenamefont {Chong},\
  and\ \citenamefont {Martonosi}}]{murali2019noise}%
  \BibitemOpen
  \bibfield  {author} {\bibinfo {author} {\bibfnamefont {P.}~\bibnamefont
  {Murali}}, \bibinfo {author} {\bibfnamefont {J.~M.}\ \bibnamefont {Baker}},
  \bibinfo {author} {\bibfnamefont {A.}~\bibnamefont {Javadi-Abhari}}, \bibinfo
  {author} {\bibfnamefont {F.~T.}\ \bibnamefont {Chong}}, \ and\ \bibinfo
  {author} {\bibfnamefont {M.}~\bibnamefont {Martonosi}},\ }in\ \href
  {https://dl.acm.org/doi/abs/10.1145/3297858.3304075?casa_token=Q72jf4ncc4UAAAAA:lJU2hEzD5kvb4JudB1KDeGWx1ZPJcmm0AVaBVSJMoV1jQImJr1dKAfJuvWpYq1-9tTIGOmth5BcLSA}
  {\emph {\bibinfo {booktitle} {Proceedings of the Twenty-Fourth International
  Conference on Architectural Support for Programming Languages and Operating
  Systems}}}\ (\bibinfo {year} {2019})\ pp.\ \bibinfo {pages}
  {1015--1029}\BibitemShut {NoStop}%
\bibitem [{Note2()}]{Note2}%
  \BibitemOpen
  \bibinfo {note} {The term ``problem Hamiltonian'' for this quantity is
  standard in optimization applications of VQAs. It is often called ``system
  Hamiltonian'' in other applications.}\BibitemShut {Stop}%
\bibitem [{Note3()}]{Note3}%
  \BibitemOpen
  \bibinfo {note} {We remark here that this is a common property of
  optimization problems, like the ones we consider in this Perspective, but
  does not always hold.}\BibitemShut {Stop}%
\bibitem [{\citenamefont {Rubin}\ \emph {et~al.}(2018)\citenamefont {Rubin},
  \citenamefont {Babbush},\ and\ \citenamefont
  {McClean}}]{rubin2018application}%
  \BibitemOpen
  \bibfield  {author} {\bibinfo {author} {\bibfnamefont {N.~C.}\ \bibnamefont
  {Rubin}}, \bibinfo {author} {\bibfnamefont {R.}~\bibnamefont {Babbush}}, \
  and\ \bibinfo {author} {\bibfnamefont {J.}~\bibnamefont {McClean}},\ }\href
  {https://iopscience.iop.org/article/10.1088/1367-2630/aab919/meta} {\bibfield
   {journal} {\bibinfo  {journal} {New J. Phys.}\ }\textbf {\bibinfo {volume}
  {20}},\ \bibinfo {pages} {053020} (\bibinfo {year} {2018})}\BibitemShut
  {NoStop}%
\bibitem [{\citenamefont {Huggins}\ \emph {et~al.}(2019)\citenamefont
  {Huggins}, \citenamefont {McClean}, \citenamefont {Rubin}, \citenamefont
  {Jiang}, \citenamefont {Wiebe}, \citenamefont {Whaley},\ and\ \citenamefont
  {Babbush}}]{huggins2019efficient}%
  \BibitemOpen
  \bibfield  {author} {\bibinfo {author} {\bibfnamefont {W.~J.}\ \bibnamefont
  {Huggins}}, \bibinfo {author} {\bibfnamefont {J.}~\bibnamefont {McClean}},
  \bibinfo {author} {\bibfnamefont {N.}~\bibnamefont {Rubin}}, \bibinfo
  {author} {\bibfnamefont {Z.}~\bibnamefont {Jiang}}, \bibinfo {author}
  {\bibfnamefont {N.}~\bibnamefont {Wiebe}}, \bibinfo {author} {\bibfnamefont
  {K.~B.}\ \bibnamefont {Whaley}}, \ and\ \bibinfo {author} {\bibfnamefont
  {R.}~\bibnamefont {Babbush}},\ }\href {https://arxiv.org/abs/1907.13117}
  {\bibfield  {journal} {\bibinfo  {journal} {arXiv preprint arXiv:1907.13117}\
  } (\bibinfo {year} {2019})}\BibitemShut {NoStop}%
\bibitem [{\citenamefont {O'Malley}\ \emph {et~al.}(2016)\citenamefont
  {O'Malley}, \citenamefont {Babbush}, \citenamefont {Kivlichan}, \citenamefont
  {Romero}, \citenamefont {McClean}, \citenamefont {Barends}, \citenamefont
  {Kelly}, \citenamefont {Roushan}, \citenamefont {Tranter}, \citenamefont
  {Ding}, \citenamefont {Campbell}, \citenamefont {Chen}, \citenamefont {Chen},
  \citenamefont {Chiaro}, \citenamefont {Dunsworth}, \citenamefont {Fowler},
  \citenamefont {Jeffrey}, \citenamefont {Megrant}, \citenamefont {Mutus},
  \citenamefont {Neill}, \citenamefont {Quintana}, \citenamefont {Sank},
  \citenamefont {Vainsencher}, \citenamefont {Wenner}, \citenamefont {White},
  \citenamefont {Coveney}, \citenamefont {Love}, \citenamefont {Neven},
  \citenamefont {Aspuru-Guzik},\ and\ \citenamefont {Martinis}}]{OMalley2016}%
  \BibitemOpen
  \bibfield  {author} {\bibinfo {author} {\bibfnamefont {P.~J.~J.}\
  \bibnamefont {O'Malley}}, \bibinfo {author} {\bibfnamefont {R.}~\bibnamefont
  {Babbush}}, \bibinfo {author} {\bibfnamefont {I.~D.}\ \bibnamefont
  {Kivlichan}}, \bibinfo {author} {\bibfnamefont {J.}~\bibnamefont {Romero}},
  \bibinfo {author} {\bibfnamefont {J.~R.}\ \bibnamefont {McClean}}, \bibinfo
  {author} {\bibfnamefont {R.}~\bibnamefont {Barends}}, \bibinfo {author}
  {\bibfnamefont {J.}~\bibnamefont {Kelly}}, \bibinfo {author} {\bibfnamefont
  {P.}~\bibnamefont {Roushan}}, \bibinfo {author} {\bibfnamefont
  {A.}~\bibnamefont {Tranter}}, \bibinfo {author} {\bibfnamefont
  {N.}~\bibnamefont {Ding}}, \bibinfo {author} {\bibfnamefont {B.}~\bibnamefont
  {Campbell}}, \bibinfo {author} {\bibfnamefont {Y.}~\bibnamefont {Chen}},
  \bibinfo {author} {\bibfnamefont {Z.}~\bibnamefont {Chen}}, \bibinfo {author}
  {\bibfnamefont {B.}~\bibnamefont {Chiaro}}, \bibinfo {author} {\bibfnamefont
  {A.}~\bibnamefont {Dunsworth}}, \bibinfo {author} {\bibfnamefont {A.~G.}\
  \bibnamefont {Fowler}}, \bibinfo {author} {\bibfnamefont {E.}~\bibnamefont
  {Jeffrey}}, \bibinfo {author} {\bibfnamefont {A.}~\bibnamefont {Megrant}},
  \bibinfo {author} {\bibfnamefont {J.~Y.}\ \bibnamefont {Mutus}}, \bibinfo
  {author} {\bibfnamefont {C.}~\bibnamefont {Neill}}, \bibinfo {author}
  {\bibfnamefont {C.}~\bibnamefont {Quintana}}, \bibinfo {author}
  {\bibfnamefont {D.}~\bibnamefont {Sank}}, \bibinfo {author} {\bibfnamefont
  {A.}~\bibnamefont {Vainsencher}}, \bibinfo {author} {\bibfnamefont
  {J.}~\bibnamefont {Wenner}}, \bibinfo {author} {\bibfnamefont {T.~C.}\
  \bibnamefont {White}}, \bibinfo {author} {\bibfnamefont {P.~V.}\ \bibnamefont
  {Coveney}}, \bibinfo {author} {\bibfnamefont {P.~J.}\ \bibnamefont {Love}},
  \bibinfo {author} {\bibfnamefont {H.}~\bibnamefont {Neven}}, \bibinfo
  {author} {\bibfnamefont {A.}~\bibnamefont {Aspuru-Guzik}}, \ and\ \bibinfo
  {author} {\bibfnamefont {J.~M.}\ \bibnamefont {Martinis}},\ }\href {\doibase
  http://dx.doi.org/10.1103/PhysRevX.6.031007} {\bibfield  {journal} {\bibinfo
  {journal} {Phys. Rev. X}\ }\textbf {\bibinfo {volume} {6}},\ \bibinfo {pages}
  {31007} (\bibinfo {year} {2016})}\BibitemShut {NoStop}%
\bibitem [{\citenamefont {Kandala}\ \emph {et~al.}(2017)\citenamefont
  {Kandala}, \citenamefont {Mezzacapo}, \citenamefont {Temme}, \citenamefont
  {Takita}, \citenamefont {Brink}, \citenamefont {Chow},\ and\ \citenamefont
  {Gambetta}}]{kandala2017hardware}%
  \BibitemOpen
  \bibfield  {author} {\bibinfo {author} {\bibfnamefont {A.}~\bibnamefont
  {Kandala}}, \bibinfo {author} {\bibfnamefont {A.}~\bibnamefont {Mezzacapo}},
  \bibinfo {author} {\bibfnamefont {K.}~\bibnamefont {Temme}}, \bibinfo
  {author} {\bibfnamefont {M.}~\bibnamefont {Takita}}, \bibinfo {author}
  {\bibfnamefont {M.}~\bibnamefont {Brink}}, \bibinfo {author} {\bibfnamefont
  {J.~M.}\ \bibnamefont {Chow}}, \ and\ \bibinfo {author} {\bibfnamefont
  {J.~M.}\ \bibnamefont {Gambetta}},\ }\href
  {https://www.nature.com/articles/nature23879} {\bibfield  {journal} {\bibinfo
   {journal} {Nature}\ }\textbf {\bibinfo {volume} {549}},\ \bibinfo {pages}
  {242} (\bibinfo {year} {2017})}\BibitemShut {NoStop}%
\bibitem [{\citenamefont {{Bravo-Prieto}}\ \emph {et~al.}(2019)\citenamefont
  {{Bravo-Prieto}}, \citenamefont {{LaRose}}, \citenamefont {{Cerezo}},
  \citenamefont {{Subasi}}, \citenamefont {{Cincio}},\ and\ \citenamefont
  {{Coles}}}]{2019arXiv190905820B}%
  \BibitemOpen
  \bibfield  {author} {\bibinfo {author} {\bibfnamefont {C.}~\bibnamefont
  {{Bravo-Prieto}}}, \bibinfo {author} {\bibfnamefont {R.}~\bibnamefont
  {{LaRose}}}, \bibinfo {author} {\bibfnamefont {M.}~\bibnamefont {{Cerezo}}},
  \bibinfo {author} {\bibfnamefont {Y.}~\bibnamefont {{Subasi}}}, \bibinfo
  {author} {\bibfnamefont {L.}~\bibnamefont {{Cincio}}}, \ and\ \bibinfo
  {author} {\bibfnamefont {P.~J.}\ \bibnamefont {{Coles}}},\ }\href@noop {}
  {\bibfield  {journal} {\bibinfo  {journal} {arXiv e-prints}\ ,\ \bibinfo
  {eid} {arXiv:1909.05820}} (\bibinfo {year} {2019})},\ \Eprint
  {http://arxiv.org/abs/1909.05820} {arXiv:1909.05820 [quant-ph]} \BibitemShut
  {NoStop}%
\bibitem [{\citenamefont {Khatri}\ \emph
  {et~al.}(2019{\natexlab{b}})\citenamefont {Khatri}, \citenamefont {LaRose},
  \citenamefont {Poremba}, \citenamefont {Cincio}, \citenamefont {Sornborger},\
  and\ \citenamefont {Coles}}]{khatri_quantum-assisted_2019}%
  \BibitemOpen
  \bibfield  {author} {\bibinfo {author} {\bibfnamefont {S.}~\bibnamefont
  {Khatri}}, \bibinfo {author} {\bibfnamefont {R.}~\bibnamefont {LaRose}},
  \bibinfo {author} {\bibfnamefont {A.}~\bibnamefont {Poremba}}, \bibinfo
  {author} {\bibfnamefont {L.}~\bibnamefont {Cincio}}, \bibinfo {author}
  {\bibfnamefont {A.~T.}\ \bibnamefont {Sornborger}}, \ and\ \bibinfo {author}
  {\bibfnamefont {P.~J.}\ \bibnamefont {Coles}},\ }\href {\doibase
  10.22331/q-2019-05-13-140} {\bibfield  {journal} {\bibinfo  {journal}
  {Quantum}\ }\textbf {\bibinfo {volume} {3}},\ \bibinfo {pages} {140}
  (\bibinfo {year} {2019}{\natexlab{b}})}\BibitemShut {NoStop}%
\bibitem [{\citenamefont {McClean}\ \emph {et~al.}(2018)\citenamefont
  {McClean}, \citenamefont {Boixo}, \citenamefont {Smelyanskiy}, \citenamefont
  {Babbush},\ and\ \citenamefont {Neven}}]{mcclean2018barren}%
  \BibitemOpen
  \bibfield  {author} {\bibinfo {author} {\bibfnamefont {J.~R.}\ \bibnamefont
  {McClean}}, \bibinfo {author} {\bibfnamefont {S.}~\bibnamefont {Boixo}},
  \bibinfo {author} {\bibfnamefont {V.~N.}\ \bibnamefont {Smelyanskiy}},
  \bibinfo {author} {\bibfnamefont {R.}~\bibnamefont {Babbush}}, \ and\
  \bibinfo {author} {\bibfnamefont {H.}~\bibnamefont {Neven}},\ }\href
  {https://www.nature.com/articles/s41467-018-07090-4} {\bibfield  {journal}
  {\bibinfo  {journal} {Nat. Commun.}\ }\textbf {\bibinfo {volume} {9}},\
  \bibinfo {pages} {1} (\bibinfo {year} {2018})}\BibitemShut {NoStop}%
\bibitem [{\citenamefont {Taube}\ and\ \citenamefont
  {Bartlett}(2006)}]{taube2006new}%
  \BibitemOpen
  \bibfield  {author} {\bibinfo {author} {\bibfnamefont {A.~G.}\ \bibnamefont
  {Taube}}\ and\ \bibinfo {author} {\bibfnamefont {R.~J.}\ \bibnamefont
  {Bartlett}},\ }\href
  {https://onlinelibrary.wiley.com/doi/abs/10.1002/qua.21198} {\bibfield
  {journal} {\bibinfo  {journal} {Int. J. Quantum Chem.}\ }\textbf {\bibinfo
  {volume} {106}},\ \bibinfo {pages} {3393} (\bibinfo {year}
  {2006})}\BibitemShut {NoStop}%
\bibitem [{\citenamefont {Grimsley}\ \emph
  {et~al.}(2019{\natexlab{a}})\citenamefont {Grimsley}, \citenamefont
  {Claudino}, \citenamefont {Economou}, \citenamefont {Barnes},\ and\
  \citenamefont {Mayhall}}]{grimsley2019trotterized}%
  \BibitemOpen
  \bibfield  {author} {\bibinfo {author} {\bibfnamefont {H.~R.}\ \bibnamefont
  {Grimsley}}, \bibinfo {author} {\bibfnamefont {D.}~\bibnamefont {Claudino}},
  \bibinfo {author} {\bibfnamefont {S.~E.}\ \bibnamefont {Economou}}, \bibinfo
  {author} {\bibfnamefont {E.}~\bibnamefont {Barnes}}, \ and\ \bibinfo {author}
  {\bibfnamefont {N.~J.}\ \bibnamefont {Mayhall}},\ }\href
  {https://pubs.acs.org/doi/10.1021/acs.jctc.9b01083} {\bibfield  {journal}
  {\bibinfo  {journal} {J. Chem. Theory Comput.}\ }\textbf {\bibinfo {volume}
  {16}},\ \bibinfo {pages} {1} (\bibinfo {year}
  {2019}{\natexlab{a}})}\BibitemShut {NoStop}%
\bibitem [{\citenamefont {Arute}\ \emph {et~al.}(2019)\citenamefont {Arute},
  \citenamefont {Arya}, \citenamefont {Babbush}, \citenamefont {Bacon},
  \citenamefont {Bardin}, \citenamefont {Barends}, \citenamefont {Biswas},
  \citenamefont {Boixo}, \citenamefont {Brandao}, \citenamefont {Buell},
  \citenamefont {Burkett}, \citenamefont {Chen}, \citenamefont {Chen},
  \citenamefont {Chiaro}, \citenamefont {Collins}, \citenamefont {Courtney},
  \citenamefont {Dunsworth}, \citenamefont {Farhi}, \citenamefont {Foxen},
  \citenamefont {Fowler}, \citenamefont {Gidney}, \citenamefont {Giustina},
  \citenamefont {Graff}, \citenamefont {Guerin}, \citenamefont {Habegger},
  \citenamefont {Harrigan}, \citenamefont {Hartmann}, \citenamefont {Ho},
  \citenamefont {Hoffmann}, \citenamefont {Huang}, \citenamefont {Humble},
  \citenamefont {Isakov}, \citenamefont {Jeffrey}, \citenamefont {Jiang},
  \citenamefont {Kafri}, \citenamefont {Kechedzhi}, \citenamefont {Kelly},
  \citenamefont {Klimov}, \citenamefont {Knysh}, \citenamefont {Korotkov},
  \citenamefont {Kostritsa}, \citenamefont {Landhuis}, \citenamefont
  {Lindmark}, \citenamefont {Lucero}, \citenamefont {Lyakh}, \citenamefont
  {Mandrà}, \citenamefont {McClean}, \citenamefont {McEwen}, \citenamefont
  {Megrant}, \citenamefont {Mi}, \citenamefont {Michielsen}, \citenamefont
  {Mohseni}, \citenamefont {Mutus}, \citenamefont {Naaman}, \citenamefont
  {Neeley}, \citenamefont {Neill}, \citenamefont {Niu}, \citenamefont {Ostby},
  \citenamefont {Petukhov}, \citenamefont {Platt}, \citenamefont {Quintana},
  \citenamefont {Rieffel}, \citenamefont {Roushan}, \citenamefont {Rubin},
  \citenamefont {Sank}, \citenamefont {Satzinger}, \citenamefont {Smelyanskiy},
  \citenamefont {Sung}, \citenamefont {Trevithick}, \citenamefont
  {Vainsencher}, \citenamefont {Villalonga}, \citenamefont {White},
  \citenamefont {Yao}, \citenamefont {Yeh}, \citenamefont {Zalcman},
  \citenamefont {Neven},\ and\ \citenamefont {Martinis}}]{arute_quantum_2019}%
  \BibitemOpen
  \bibfield  {author} {\bibinfo {author} {\bibfnamefont {F.}~\bibnamefont
  {Arute}}, \bibinfo {author} {\bibfnamefont {K.}~\bibnamefont {Arya}},
  \bibinfo {author} {\bibfnamefont {R.}~\bibnamefont {Babbush}}, \bibinfo
  {author} {\bibfnamefont {D.}~\bibnamefont {Bacon}}, \bibinfo {author}
  {\bibfnamefont {J.~C.}\ \bibnamefont {Bardin}}, \bibinfo {author}
  {\bibfnamefont {R.}~\bibnamefont {Barends}}, \bibinfo {author} {\bibfnamefont
  {R.}~\bibnamefont {Biswas}}, \bibinfo {author} {\bibfnamefont
  {S.}~\bibnamefont {Boixo}}, \bibinfo {author} {\bibfnamefont {F.~G. S.~L.}\
  \bibnamefont {Brandao}}, \bibinfo {author} {\bibfnamefont {D.~A.}\
  \bibnamefont {Buell}}, \bibinfo {author} {\bibfnamefont {B.}~\bibnamefont
  {Burkett}}, \bibinfo {author} {\bibfnamefont {Y.}~\bibnamefont {Chen}},
  \bibinfo {author} {\bibfnamefont {Z.}~\bibnamefont {Chen}}, \bibinfo {author}
  {\bibfnamefont {B.}~\bibnamefont {Chiaro}}, \bibinfo {author} {\bibfnamefont
  {R.}~\bibnamefont {Collins}}, \bibinfo {author} {\bibfnamefont
  {W.}~\bibnamefont {Courtney}}, \bibinfo {author} {\bibfnamefont
  {A.}~\bibnamefont {Dunsworth}}, \bibinfo {author} {\bibfnamefont
  {E.}~\bibnamefont {Farhi}}, \bibinfo {author} {\bibfnamefont
  {B.}~\bibnamefont {Foxen}}, \bibinfo {author} {\bibfnamefont
  {A.}~\bibnamefont {Fowler}}, \bibinfo {author} {\bibfnamefont
  {C.}~\bibnamefont {Gidney}}, \bibinfo {author} {\bibfnamefont
  {M.}~\bibnamefont {Giustina}}, \bibinfo {author} {\bibfnamefont
  {R.}~\bibnamefont {Graff}}, \bibinfo {author} {\bibfnamefont
  {K.}~\bibnamefont {Guerin}}, \bibinfo {author} {\bibfnamefont
  {S.}~\bibnamefont {Habegger}}, \bibinfo {author} {\bibfnamefont {M.~P.}\
  \bibnamefont {Harrigan}}, \bibinfo {author} {\bibfnamefont {M.~J.}\
  \bibnamefont {Hartmann}}, \bibinfo {author} {\bibfnamefont {A.}~\bibnamefont
  {Ho}}, \bibinfo {author} {\bibfnamefont {M.}~\bibnamefont {Hoffmann}},
  \bibinfo {author} {\bibfnamefont {T.}~\bibnamefont {Huang}}, \bibinfo
  {author} {\bibfnamefont {T.~S.}\ \bibnamefont {Humble}}, \bibinfo {author}
  {\bibfnamefont {S.~V.}\ \bibnamefont {Isakov}}, \bibinfo {author}
  {\bibfnamefont {E.}~\bibnamefont {Jeffrey}}, \bibinfo {author} {\bibfnamefont
  {Z.}~\bibnamefont {Jiang}}, \bibinfo {author} {\bibfnamefont
  {D.}~\bibnamefont {Kafri}}, \bibinfo {author} {\bibfnamefont
  {K.}~\bibnamefont {Kechedzhi}}, \bibinfo {author} {\bibfnamefont
  {J.}~\bibnamefont {Kelly}}, \bibinfo {author} {\bibfnamefont {P.~V.}\
  \bibnamefont {Klimov}}, \bibinfo {author} {\bibfnamefont {S.}~\bibnamefont
  {Knysh}}, \bibinfo {author} {\bibfnamefont {A.}~\bibnamefont {Korotkov}},
  \bibinfo {author} {\bibfnamefont {F.}~\bibnamefont {Kostritsa}}, \bibinfo
  {author} {\bibfnamefont {D.}~\bibnamefont {Landhuis}}, \bibinfo {author}
  {\bibfnamefont {M.}~\bibnamefont {Lindmark}}, \bibinfo {author}
  {\bibfnamefont {E.}~\bibnamefont {Lucero}}, \bibinfo {author} {\bibfnamefont
  {D.}~\bibnamefont {Lyakh}}, \bibinfo {author} {\bibfnamefont
  {S.}~\bibnamefont {Mandrà}}, \bibinfo {author} {\bibfnamefont {J.~R.}\
  \bibnamefont {McClean}}, \bibinfo {author} {\bibfnamefont {M.}~\bibnamefont
  {McEwen}}, \bibinfo {author} {\bibfnamefont {A.}~\bibnamefont {Megrant}},
  \bibinfo {author} {\bibfnamefont {X.}~\bibnamefont {Mi}}, \bibinfo {author}
  {\bibfnamefont {K.}~\bibnamefont {Michielsen}}, \bibinfo {author}
  {\bibfnamefont {M.}~\bibnamefont {Mohseni}}, \bibinfo {author} {\bibfnamefont
  {J.}~\bibnamefont {Mutus}}, \bibinfo {author} {\bibfnamefont
  {O.}~\bibnamefont {Naaman}}, \bibinfo {author} {\bibfnamefont
  {M.}~\bibnamefont {Neeley}}, \bibinfo {author} {\bibfnamefont
  {C.}~\bibnamefont {Neill}}, \bibinfo {author} {\bibfnamefont {M.~Y.}\
  \bibnamefont {Niu}}, \bibinfo {author} {\bibfnamefont {E.}~\bibnamefont
  {Ostby}}, \bibinfo {author} {\bibfnamefont {A.}~\bibnamefont {Petukhov}},
  \bibinfo {author} {\bibfnamefont {J.~C.}\ \bibnamefont {Platt}}, \bibinfo
  {author} {\bibfnamefont {C.}~\bibnamefont {Quintana}}, \bibinfo {author}
  {\bibfnamefont {E.~G.}\ \bibnamefont {Rieffel}}, \bibinfo {author}
  {\bibfnamefont {P.}~\bibnamefont {Roushan}}, \bibinfo {author} {\bibfnamefont
  {N.~C.}\ \bibnamefont {Rubin}}, \bibinfo {author} {\bibfnamefont
  {D.}~\bibnamefont {Sank}}, \bibinfo {author} {\bibfnamefont {K.~J.}\
  \bibnamefont {Satzinger}}, \bibinfo {author} {\bibfnamefont {V.}~\bibnamefont
  {Smelyanskiy}}, \bibinfo {author} {\bibfnamefont {K.~J.}\ \bibnamefont
  {Sung}}, \bibinfo {author} {\bibfnamefont {M.~D.}\ \bibnamefont
  {Trevithick}}, \bibinfo {author} {\bibfnamefont {A.}~\bibnamefont
  {Vainsencher}}, \bibinfo {author} {\bibfnamefont {B.}~\bibnamefont
  {Villalonga}}, \bibinfo {author} {\bibfnamefont {T.}~\bibnamefont {White}},
  \bibinfo {author} {\bibfnamefont {Z.~J.}\ \bibnamefont {Yao}}, \bibinfo
  {author} {\bibfnamefont {P.}~\bibnamefont {Yeh}}, \bibinfo {author}
  {\bibfnamefont {A.}~\bibnamefont {Zalcman}}, \bibinfo {author} {\bibfnamefont
  {H.}~\bibnamefont {Neven}}, \ and\ \bibinfo {author} {\bibfnamefont {J.~M.}\
  \bibnamefont {Martinis}},\ }\href {\doibase 10.1038/s41586-019-1666-5}
  {\bibfield  {journal} {\bibinfo  {journal} {Nature}\ }\textbf {\bibinfo
  {volume} {574}},\ \bibinfo {pages} {505} (\bibinfo {year}
  {2019})}\BibitemShut {NoStop}%
\bibitem [{\citenamefont {Hempel}\ \emph {et~al.}(2018)\citenamefont {Hempel},
  \citenamefont {Maier}, \citenamefont {Romero}, \citenamefont {McClean},
  \citenamefont {Monz}, \citenamefont {Shen}, \citenamefont {Jurcevic},
  \citenamefont {Lanyon}, \citenamefont {Love}, \citenamefont {Babbush},
  \citenamefont {Aspuru-Guzik}, \citenamefont {Blatt},\ and\ \citenamefont
  {Roos}}]{PhysRevX.8.031022}%
  \BibitemOpen
  \bibfield  {author} {\bibinfo {author} {\bibfnamefont {C.}~\bibnamefont
  {Hempel}}, \bibinfo {author} {\bibfnamefont {C.}~\bibnamefont {Maier}},
  \bibinfo {author} {\bibfnamefont {J.}~\bibnamefont {Romero}}, \bibinfo
  {author} {\bibfnamefont {J.}~\bibnamefont {McClean}}, \bibinfo {author}
  {\bibfnamefont {T.}~\bibnamefont {Monz}}, \bibinfo {author} {\bibfnamefont
  {H.}~\bibnamefont {Shen}}, \bibinfo {author} {\bibfnamefont {P.}~\bibnamefont
  {Jurcevic}}, \bibinfo {author} {\bibfnamefont {B.~P.}\ \bibnamefont
  {Lanyon}}, \bibinfo {author} {\bibfnamefont {P.}~\bibnamefont {Love}},
  \bibinfo {author} {\bibfnamefont {R.}~\bibnamefont {Babbush}}, \bibinfo
  {author} {\bibfnamefont {A.}~\bibnamefont {Aspuru-Guzik}}, \bibinfo {author}
  {\bibfnamefont {R.}~\bibnamefont {Blatt}}, \ and\ \bibinfo {author}
  {\bibfnamefont {C.~F.}\ \bibnamefont {Roos}},\ }\href {\doibase
  10.1103/PhysRevX.8.031022} {\bibfield  {journal} {\bibinfo  {journal} {Phys.
  Rev. X}\ }\textbf {\bibinfo {volume} {8}},\ \bibinfo {pages} {031022}
  (\bibinfo {year} {2018})}\BibitemShut {NoStop}%
\bibitem [{\citenamefont {Kandala}\ \emph {et~al.}(2019)\citenamefont
  {Kandala}, \citenamefont {Temme}, \citenamefont {C{\'o}rcoles}, \citenamefont
  {Mezzacapo}, \citenamefont {Chow},\ and\ \citenamefont
  {Gambetta}}]{kandala2019error}%
  \BibitemOpen
  \bibfield  {author} {\bibinfo {author} {\bibfnamefont {A.}~\bibnamefont
  {Kandala}}, \bibinfo {author} {\bibfnamefont {K.}~\bibnamefont {Temme}},
  \bibinfo {author} {\bibfnamefont {A.~D.}\ \bibnamefont {C{\'o}rcoles}},
  \bibinfo {author} {\bibfnamefont {A.}~\bibnamefont {Mezzacapo}}, \bibinfo
  {author} {\bibfnamefont {J.~M.}\ \bibnamefont {Chow}}, \ and\ \bibinfo
  {author} {\bibfnamefont {J.~M.}\ \bibnamefont {Gambetta}},\ }\href
  {https://www.nature.com/articles/s41586-019-1040-7} {\bibfield  {journal}
  {\bibinfo  {journal} {Nature}\ }\textbf {\bibinfo {volume} {567}},\ \bibinfo
  {pages} {491} (\bibinfo {year} {2019})}\BibitemShut {NoStop}%
\bibitem [{\citenamefont {Colless}\ \emph {et~al.}(2018)\citenamefont
  {Colless}, \citenamefont {Ramasesh}, \citenamefont {Dahlen}, \citenamefont
  {Blok}, \citenamefont {McClean}, \citenamefont {Carter}, \citenamefont
  {de~Jong},\ and\ \citenamefont {Siddiqi}}]{Siddiqi2017}%
  \BibitemOpen
  \bibfield  {author} {\bibinfo {author} {\bibfnamefont {J.~I.}\ \bibnamefont
  {Colless}}, \bibinfo {author} {\bibfnamefont {V.~V.}\ \bibnamefont
  {Ramasesh}}, \bibinfo {author} {\bibfnamefont {D.}~\bibnamefont {Dahlen}},
  \bibinfo {author} {\bibfnamefont {M.~S.}\ \bibnamefont {Blok}}, \bibinfo
  {author} {\bibfnamefont {J.~R.}\ \bibnamefont {McClean}}, \bibinfo {author}
  {\bibfnamefont {J.}~\bibnamefont {Carter}}, \bibinfo {author} {\bibfnamefont
  {W.~A.}\ \bibnamefont {de~Jong}}, \ and\ \bibinfo {author} {\bibfnamefont
  {I.}~\bibnamefont {Siddiqi}},\ }\href {\doibase 10.1103/PhysRevX.8.011021}
  {\bibfield  {journal} {\bibinfo  {journal} {Phys. Rev. X}\ }\textbf {\bibinfo
  {volume} {8}},\ \bibinfo {pages} {011021} (\bibinfo {year}
  {2018})}\BibitemShut {NoStop}%
\bibitem [{\citenamefont {Smart}\ and\ \citenamefont
  {Mazziotti}(2019)}]{PhysRevA.100.022517}%
  \BibitemOpen
  \bibfield  {author} {\bibinfo {author} {\bibfnamefont {S.~E.}\ \bibnamefont
  {Smart}}\ and\ \bibinfo {author} {\bibfnamefont {D.~A.}\ \bibnamefont
  {Mazziotti}},\ }\href {\doibase 10.1103/PhysRevA.100.022517} {\bibfield
  {journal} {\bibinfo  {journal} {Phys. Rev. A}\ }\textbf {\bibinfo {volume}
  {100}},\ \bibinfo {pages} {022517} (\bibinfo {year} {2019})}\BibitemShut
  {NoStop}%
\bibitem [{\citenamefont {Sagastizabal}\ \emph {et~al.}(2019)\citenamefont
  {Sagastizabal}, \citenamefont {Bonet-Monroig}, \citenamefont {Singh},
  \citenamefont {Rol}, \citenamefont {Bultink}, \citenamefont {Fu},
  \citenamefont {Price}, \citenamefont {Ostroukh}, \citenamefont
  {Muthusubramanian}, \citenamefont {Bruno}, \citenamefont {Beekman},
  \citenamefont {Haider}, \citenamefont {O'Brien},\ and\ \citenamefont
  {DiCarlo}}]{PhysRevA.100.010302}%
  \BibitemOpen
  \bibfield  {author} {\bibinfo {author} {\bibfnamefont {R.}~\bibnamefont
  {Sagastizabal}}, \bibinfo {author} {\bibfnamefont {X.}~\bibnamefont
  {Bonet-Monroig}}, \bibinfo {author} {\bibfnamefont {M.}~\bibnamefont
  {Singh}}, \bibinfo {author} {\bibfnamefont {M.~A.}\ \bibnamefont {Rol}},
  \bibinfo {author} {\bibfnamefont {C.~C.}\ \bibnamefont {Bultink}}, \bibinfo
  {author} {\bibfnamefont {X.}~\bibnamefont {Fu}}, \bibinfo {author}
  {\bibfnamefont {C.~H.}\ \bibnamefont {Price}}, \bibinfo {author}
  {\bibfnamefont {V.~P.}\ \bibnamefont {Ostroukh}}, \bibinfo {author}
  {\bibfnamefont {N.}~\bibnamefont {Muthusubramanian}}, \bibinfo {author}
  {\bibfnamefont {A.}~\bibnamefont {Bruno}}, \bibinfo {author} {\bibfnamefont
  {M.}~\bibnamefont {Beekman}}, \bibinfo {author} {\bibfnamefont
  {N.}~\bibnamefont {Haider}}, \bibinfo {author} {\bibfnamefont {T.~E.}\
  \bibnamefont {O'Brien}}, \ and\ \bibinfo {author} {\bibfnamefont
  {L.}~\bibnamefont {DiCarlo}},\ }\href {\doibase 10.1103/PhysRevA.100.010302}
  {\bibfield  {journal} {\bibinfo  {journal} {Phys. Rev. A}\ }\textbf {\bibinfo
  {volume} {100}},\ \bibinfo {pages} {010302} (\bibinfo {year}
  {2019})}\BibitemShut {NoStop}%
\bibitem [{\citenamefont {Quantum}\ \emph {et~al.}(2020)\citenamefont {Quantum}
  \emph {et~al.}}]{google2020hartree}%
  \BibitemOpen
  \bibfield  {author} {\bibinfo {author} {\bibfnamefont {G.~A.}\ \bibnamefont
  {Quantum}} \emph {et~al.},\ }\href
  {https://science.sciencemag.org/content/369/6507/1084} {\bibfield  {journal}
  {\bibinfo  {journal} {Science}\ }\textbf {\bibinfo {volume} {369}},\ \bibinfo
  {pages} {1084} (\bibinfo {year} {2020})}\BibitemShut {NoStop}%
\bibitem [{\citenamefont {Otterbach}\ \emph {et~al.}(2017)\citenamefont
  {Otterbach}, \citenamefont {Manenti}, \citenamefont {Alidoust}, \citenamefont
  {Bestwick}, \citenamefont {Block}, \citenamefont {Bloom}, \citenamefont
  {Caldwell}, \citenamefont {Didier}, \citenamefont {Fried}, \citenamefont
  {Hong} \emph {et~al.}}]{otterbach2017unsupervised}%
  \BibitemOpen
  \bibfield  {author} {\bibinfo {author} {\bibfnamefont {J.}~\bibnamefont
  {Otterbach}}, \bibinfo {author} {\bibfnamefont {R.}~\bibnamefont {Manenti}},
  \bibinfo {author} {\bibfnamefont {N.}~\bibnamefont {Alidoust}}, \bibinfo
  {author} {\bibfnamefont {A.}~\bibnamefont {Bestwick}}, \bibinfo {author}
  {\bibfnamefont {M.}~\bibnamefont {Block}}, \bibinfo {author} {\bibfnamefont
  {B.}~\bibnamefont {Bloom}}, \bibinfo {author} {\bibfnamefont
  {S.}~\bibnamefont {Caldwell}}, \bibinfo {author} {\bibfnamefont
  {N.}~\bibnamefont {Didier}}, \bibinfo {author} {\bibfnamefont {E.~S.}\
  \bibnamefont {Fried}}, \bibinfo {author} {\bibfnamefont {S.}~\bibnamefont
  {Hong}},  \emph {et~al.},\ }\href {https://arxiv.org/abs/1712.05771}
  {\bibfield  {journal} {\bibinfo  {journal} {arXiv preprint arXiv:1712.05771}\
  } (\bibinfo {year} {2017})}\BibitemShut {NoStop}%
\bibitem [{\citenamefont {Willsch}\ \emph {et~al.}(2020)\citenamefont
  {Willsch}, \citenamefont {Willsch}, \citenamefont {Jin}, \citenamefont
  {De~Raedt},\ and\ \citenamefont {Michielsen}}]{willsch2020benchmarking}%
  \BibitemOpen
  \bibfield  {author} {\bibinfo {author} {\bibfnamefont {M.}~\bibnamefont
  {Willsch}}, \bibinfo {author} {\bibfnamefont {D.}~\bibnamefont {Willsch}},
  \bibinfo {author} {\bibfnamefont {F.}~\bibnamefont {Jin}}, \bibinfo {author}
  {\bibfnamefont {H.}~\bibnamefont {De~Raedt}}, \ and\ \bibinfo {author}
  {\bibfnamefont {K.}~\bibnamefont {Michielsen}},\ }\href
  {https://link.springer.com/article/10.1007/s11128-020-02692-8} {\bibfield
  {journal} {\bibinfo  {journal} {Quantum Inf. Process.}\ }\textbf {\bibinfo
  {volume} {19}},\ \bibinfo {pages} {197} (\bibinfo {year} {2020})}\BibitemShut
  {NoStop}%
\bibitem [{\citenamefont {Abrams}\ \emph {et~al.}(2019)\citenamefont {Abrams},
  \citenamefont {Didier}, \citenamefont {Johnson}, \citenamefont {da~Silva},\
  and\ \citenamefont {Ryan}}]{abrams2019implementation}%
  \BibitemOpen
  \bibfield  {author} {\bibinfo {author} {\bibfnamefont {D.~M.}\ \bibnamefont
  {Abrams}}, \bibinfo {author} {\bibfnamefont {N.}~\bibnamefont {Didier}},
  \bibinfo {author} {\bibfnamefont {B.~R.}\ \bibnamefont {Johnson}}, \bibinfo
  {author} {\bibfnamefont {M.~P.}\ \bibnamefont {da~Silva}}, \ and\ \bibinfo
  {author} {\bibfnamefont {C.~A.}\ \bibnamefont {Ryan}},\ }\href
  {https://arxiv.org/abs/1912.04424} {\bibfield  {journal} {\bibinfo  {journal}
  {arXiv preprint arXiv:1912.04424}\ } (\bibinfo {year} {2019})}\BibitemShut
  {NoStop}%
\bibitem [{\citenamefont {Bengtsson}\ \emph {et~al.}(2019)\citenamefont
  {Bengtsson}, \citenamefont {Vikst{\aa}l}, \citenamefont {Warren},
  \citenamefont {Svensson}, \citenamefont {Gu}, \citenamefont {Kockum},
  \citenamefont {Krantz}, \citenamefont {Kri{\v{z}}an}, \citenamefont {Shiri},
  \citenamefont {Svensson} \emph {et~al.}}]{bengtsson2019quantum}%
  \BibitemOpen
  \bibfield  {author} {\bibinfo {author} {\bibfnamefont {A.}~\bibnamefont
  {Bengtsson}}, \bibinfo {author} {\bibfnamefont {P.}~\bibnamefont
  {Vikst{\aa}l}}, \bibinfo {author} {\bibfnamefont {C.}~\bibnamefont {Warren}},
  \bibinfo {author} {\bibfnamefont {M.}~\bibnamefont {Svensson}}, \bibinfo
  {author} {\bibfnamefont {X.}~\bibnamefont {Gu}}, \bibinfo {author}
  {\bibfnamefont {A.~F.}\ \bibnamefont {Kockum}}, \bibinfo {author}
  {\bibfnamefont {P.}~\bibnamefont {Krantz}}, \bibinfo {author} {\bibfnamefont
  {C.}~\bibnamefont {Kri{\v{z}}an}}, \bibinfo {author} {\bibfnamefont
  {D.}~\bibnamefont {Shiri}}, \bibinfo {author} {\bibfnamefont {I.-M.}\
  \bibnamefont {Svensson}},  \emph {et~al.},\ }\href
  {https://arxiv.org/abs/1912.10495v2} {\bibfield  {journal} {\bibinfo
  {journal} {arXiv preprint arXiv:1912.10495}\ } (\bibinfo {year}
  {2019})}\BibitemShut {NoStop}%
\bibitem [{\citenamefont {Arute}\ \emph {et~al.}(2020)\citenamefont {Arute},
  \citenamefont {Arya}, \citenamefont {Babbush}, \citenamefont {Bacon},
  \citenamefont {Bardin}, \citenamefont {Barends}, \citenamefont {Boixo},
  \citenamefont {Broughton}, \citenamefont {Buckley}, \citenamefont {Buell}
  \emph {et~al.}}]{arute2020quantum}%
  \BibitemOpen
  \bibfield  {author} {\bibinfo {author} {\bibfnamefont {F.}~\bibnamefont
  {Arute}}, \bibinfo {author} {\bibfnamefont {K.}~\bibnamefont {Arya}},
  \bibinfo {author} {\bibfnamefont {R.}~\bibnamefont {Babbush}}, \bibinfo
  {author} {\bibfnamefont {D.}~\bibnamefont {Bacon}}, \bibinfo {author}
  {\bibfnamefont {J.~C.}\ \bibnamefont {Bardin}}, \bibinfo {author}
  {\bibfnamefont {R.}~\bibnamefont {Barends}}, \bibinfo {author} {\bibfnamefont
  {S.}~\bibnamefont {Boixo}}, \bibinfo {author} {\bibfnamefont
  {M.}~\bibnamefont {Broughton}}, \bibinfo {author} {\bibfnamefont {B.~B.}\
  \bibnamefont {Buckley}}, \bibinfo {author} {\bibfnamefont {D.~A.}\
  \bibnamefont {Buell}},  \emph {et~al.},\ }\href
  {https://arxiv.org/abs/2004.04197} {\bibfield  {journal} {\bibinfo  {journal}
  {arXiv preprint arXiv:2004.04197}\ } (\bibinfo {year} {2020})}\BibitemShut
  {NoStop}%
\bibitem [{\citenamefont {Qiang}\ \emph {et~al.}(2018)\citenamefont {Qiang},
  \citenamefont {Zhou}, \citenamefont {Wang}, \citenamefont {Wilkes},
  \citenamefont {Loke}, \citenamefont {O’Gara}, \citenamefont {Kling},
  \citenamefont {Marshall}, \citenamefont {Santagati}, \citenamefont {Ralph}
  \emph {et~al.}}]{qiang2018large}%
  \BibitemOpen
  \bibfield  {author} {\bibinfo {author} {\bibfnamefont {X.}~\bibnamefont
  {Qiang}}, \bibinfo {author} {\bibfnamefont {X.}~\bibnamefont {Zhou}},
  \bibinfo {author} {\bibfnamefont {J.}~\bibnamefont {Wang}}, \bibinfo {author}
  {\bibfnamefont {C.~M.}\ \bibnamefont {Wilkes}}, \bibinfo {author}
  {\bibfnamefont {T.}~\bibnamefont {Loke}}, \bibinfo {author} {\bibfnamefont
  {S.}~\bibnamefont {O’Gara}}, \bibinfo {author} {\bibfnamefont
  {L.}~\bibnamefont {Kling}}, \bibinfo {author} {\bibfnamefont {G.~D.}\
  \bibnamefont {Marshall}}, \bibinfo {author} {\bibfnamefont {R.}~\bibnamefont
  {Santagati}}, \bibinfo {author} {\bibfnamefont {T.~C.}\ \bibnamefont
  {Ralph}},  \emph {et~al.},\ }\href
  {https://www.nature.com/articles/s41566-018-0236-y} {\bibfield  {journal}
  {\bibinfo  {journal} {Nat. Photonics}\ }\textbf {\bibinfo {volume} {12}},\
  \bibinfo {pages} {534} (\bibinfo {year} {2018})}\BibitemShut {NoStop}%
\bibitem [{\citenamefont {Pagano}\ \emph {et~al.}(2019)\citenamefont {Pagano},
  \citenamefont {Bapat}, \citenamefont {Becker}, \citenamefont {Collins},
  \citenamefont {De}, \citenamefont {Hess}, \citenamefont {Kaplan},
  \citenamefont {Kyprianidis}, \citenamefont {Tan}, \citenamefont {Baldwin}
  \emph {et~al.}}]{pagano2019quantum}%
  \BibitemOpen
  \bibfield  {author} {\bibinfo {author} {\bibfnamefont {G.}~\bibnamefont
  {Pagano}}, \bibinfo {author} {\bibfnamefont {A.}~\bibnamefont {Bapat}},
  \bibinfo {author} {\bibfnamefont {P.}~\bibnamefont {Becker}}, \bibinfo
  {author} {\bibfnamefont {K.}~\bibnamefont {Collins}}, \bibinfo {author}
  {\bibfnamefont {A.}~\bibnamefont {De}}, \bibinfo {author} {\bibfnamefont
  {P.}~\bibnamefont {Hess}}, \bibinfo {author} {\bibfnamefont {H.}~\bibnamefont
  {Kaplan}}, \bibinfo {author} {\bibfnamefont {A.}~\bibnamefont {Kyprianidis}},
  \bibinfo {author} {\bibfnamefont {W.}~\bibnamefont {Tan}}, \bibinfo {author}
  {\bibfnamefont {C.}~\bibnamefont {Baldwin}},  \emph {et~al.},\ }\href
  {https://arxiv.org/abs/1906.02700} {\bibfield  {journal} {\bibinfo  {journal}
  {arXiv preprint arXiv:1906.02700}\ } (\bibinfo {year} {2019})}\BibitemShut
  {NoStop}%
\bibitem [{\citenamefont {McClean}\ \emph {et~al.}(2020)\citenamefont
  {McClean}, \citenamefont {Harrigan}, \citenamefont {Mohseni}, \citenamefont
  {Rubin}, \citenamefont {Jiang}, \citenamefont {Boixo}, \citenamefont
  {Smelyanskiy}, \citenamefont {Babbush},\ and\ \citenamefont
  {Neven}}]{mcclean2020low}%
  \BibitemOpen
  \bibfield  {author} {\bibinfo {author} {\bibfnamefont {J.~R.}\ \bibnamefont
  {McClean}}, \bibinfo {author} {\bibfnamefont {M.~P.}\ \bibnamefont
  {Harrigan}}, \bibinfo {author} {\bibfnamefont {M.}~\bibnamefont {Mohseni}},
  \bibinfo {author} {\bibfnamefont {N.~C.}\ \bibnamefont {Rubin}}, \bibinfo
  {author} {\bibfnamefont {Z.}~\bibnamefont {Jiang}}, \bibinfo {author}
  {\bibfnamefont {S.}~\bibnamefont {Boixo}}, \bibinfo {author} {\bibfnamefont
  {V.~N.}\ \bibnamefont {Smelyanskiy}}, \bibinfo {author} {\bibfnamefont
  {R.}~\bibnamefont {Babbush}}, \ and\ \bibinfo {author} {\bibfnamefont
  {H.}~\bibnamefont {Neven}},\ }\href {https://arxiv.org/abs/2008.08615}
  {\bibfield  {journal} {\bibinfo  {journal} {arXiv preprint arXiv:2008.08615}\
  } (\bibinfo {year} {2020})}\BibitemShut {NoStop}%
\bibitem [{\citenamefont {Kosloff}\ \emph {et~al.}(1989)\citenamefont
  {Kosloff}, \citenamefont {Rice}, \citenamefont {Gaspard}, \citenamefont
  {Tersigni},\ and\ \citenamefont {Tannor}}]{kosloff1989wavepacket}%
  \BibitemOpen
  \bibfield  {author} {\bibinfo {author} {\bibfnamefont {R.}~\bibnamefont
  {Kosloff}}, \bibinfo {author} {\bibfnamefont {S.~A.}\ \bibnamefont {Rice}},
  \bibinfo {author} {\bibfnamefont {P.}~\bibnamefont {Gaspard}}, \bibinfo
  {author} {\bibfnamefont {S.}~\bibnamefont {Tersigni}}, \ and\ \bibinfo
  {author} {\bibfnamefont {D.}~\bibnamefont {Tannor}},\ }\href
  {https://www.sciencedirect.com/science/article/abs/pii/0301010489900128}
  {\bibfield  {journal} {\bibinfo  {journal} {Chem. Phys.}\ }\textbf {\bibinfo
  {volume} {139}},\ \bibinfo {pages} {201} (\bibinfo {year}
  {1989})}\BibitemShut {NoStop}%
\bibitem [{\citenamefont {Shi}\ and\ \citenamefont
  {Rabitz}(1990)}]{shi1990quantum}%
  \BibitemOpen
  \bibfield  {author} {\bibinfo {author} {\bibfnamefont {S.}~\bibnamefont
  {Shi}}\ and\ \bibinfo {author} {\bibfnamefont {H.}~\bibnamefont {Rabitz}},\
  }\href {https://aip.scitation.org/doi/10.1063/1.458438} {\bibfield  {journal}
  {\bibinfo  {journal} {J. Chem. Phys.}\ }\textbf {\bibinfo {volume} {92}},\
  \bibinfo {pages} {364} (\bibinfo {year} {1990})}\BibitemShut {NoStop}%
\bibitem [{\citenamefont {Sundermann}\ and\ \citenamefont
  {de~Vivie-Riedle}(1999)}]{sundermann1999extensions}%
  \BibitemOpen
  \bibfield  {author} {\bibinfo {author} {\bibfnamefont {K.}~\bibnamefont
  {Sundermann}}\ and\ \bibinfo {author} {\bibfnamefont {R.}~\bibnamefont
  {de~Vivie-Riedle}},\ }\href
  {https://aip.scitation.org/doi/abs/10.1063/1.477856?journalCode=jcp}
  {\bibfield  {journal} {\bibinfo  {journal} {J. Chem. Phys.}\ }\textbf
  {\bibinfo {volume} {110}},\ \bibinfo {pages} {1896} (\bibinfo {year}
  {1999})}\BibitemShut {NoStop}%
\bibitem [{\citenamefont {Dong}\ and\ \citenamefont
  {Petersen}(2010)}]{dong2010quantum}%
  \BibitemOpen
  \bibfield  {author} {\bibinfo {author} {\bibfnamefont {D.}~\bibnamefont
  {Dong}}\ and\ \bibinfo {author} {\bibfnamefont {I.~R.}\ \bibnamefont
  {Petersen}},\ }\href {https://ieeexplore.ieee.org/document/5676677}
  {\bibfield  {journal} {\bibinfo  {journal} {IET Control. Theory Appl.}\
  }\textbf {\bibinfo {volume} {4}},\ \bibinfo {pages} {2651} (\bibinfo {year}
  {2010})}\BibitemShut {NoStop}%
\bibitem [{\citenamefont {Lloyd}\ and\ \citenamefont
  {Montangero}(2014)}]{lloyd2014information}%
  \BibitemOpen
  \bibfield  {author} {\bibinfo {author} {\bibfnamefont {S.}~\bibnamefont
  {Lloyd}}\ and\ \bibinfo {author} {\bibfnamefont {S.}~\bibnamefont
  {Montangero}},\ }\href
  {https://journals.aps.org/prl/abstract/10.1103/PhysRevLett.113.010502}
  {\bibfield  {journal} {\bibinfo  {journal} {Phys. Rev. Lett.}\ }\textbf
  {\bibinfo {volume} {113}},\ \bibinfo {pages} {010502} (\bibinfo {year}
  {2014})}\BibitemShut {NoStop}%
\bibitem [{\citenamefont {Peirce}\ \emph {et~al.}(1988)\citenamefont {Peirce},
  \citenamefont {Dahleh},\ and\ \citenamefont {Rabitz}}]{PhysRevA.37.4950}%
  \BibitemOpen
  \bibfield  {author} {\bibinfo {author} {\bibfnamefont {A.~P.}\ \bibnamefont
  {Peirce}}, \bibinfo {author} {\bibfnamefont {M.~A.}\ \bibnamefont {Dahleh}},
  \ and\ \bibinfo {author} {\bibfnamefont {H.}~\bibnamefont {Rabitz}},\ }\href
  {\doibase 10.1103/PhysRevA.37.4950} {\bibfield  {journal} {\bibinfo
  {journal} {Phys. Rev. A}\ }\textbf {\bibinfo {volume} {37}},\ \bibinfo
  {pages} {4950} (\bibinfo {year} {1988})}\BibitemShut {NoStop}%
\bibitem [{\citenamefont {Brif}\ \emph {et~al.}(2010)\citenamefont {Brif},
  \citenamefont {Chakrabarti},\ and\ \citenamefont {Rabitz}}]{Brif2011}%
  \BibitemOpen
  \bibfield  {author} {\bibinfo {author} {\bibfnamefont {C.}~\bibnamefont
  {Brif}}, \bibinfo {author} {\bibfnamefont {R.}~\bibnamefont {Chakrabarti}}, \
  and\ \bibinfo {author} {\bibfnamefont {H.}~\bibnamefont {Rabitz}},\ }\href
  {http://stacks.iop.org/1367-2630/12/i=7/a=075008} {\bibfield  {journal}
  {\bibinfo  {journal} {New J. Phys.}\ }\textbf {\bibinfo {volume} {12}},\
  \bibinfo {pages} {075008} (\bibinfo {year} {2010})}\BibitemShut {NoStop}%
\bibitem [{\citenamefont {Elliott}(2009)}]{elliott2009bilinear}%
  \BibitemOpen
  \bibfield  {author} {\bibinfo {author} {\bibfnamefont {D.}~\bibnamefont
  {Elliott}},\ }\href@noop {} {\emph {\bibinfo {title} {Bilinear control
  systems: matrices in action}}},\ Vol.\ \bibinfo {volume} {169}\ (\bibinfo
  {publisher} {Springer Science \& Business Media},\ \bibinfo {year}
  {2009})\BibitemShut {NoStop}%
\bibitem [{\citenamefont {d'Alessandro}(2007)}]{d2007introduction}%
  \BibitemOpen
  \bibfield  {author} {\bibinfo {author} {\bibfnamefont {D.}~\bibnamefont
  {d'Alessandro}},\ }\href@noop {} {\emph {\bibinfo {title} {Introduction to
  quantum control and dynamics}}}\ (\bibinfo  {publisher} {CRC press},\
  \bibinfo {year} {2007})\BibitemShut {NoStop}%
\bibitem [{\citenamefont {Jurdjevic}\ and\ \citenamefont
  {Sussmann}(1972)}]{jurdjevic1972control}%
  \BibitemOpen
  \bibfield  {author} {\bibinfo {author} {\bibfnamefont {V.}~\bibnamefont
  {Jurdjevic}}\ and\ \bibinfo {author} {\bibfnamefont {H.~J.}\ \bibnamefont
  {Sussmann}},\ }\href
  {https://www.sciencedirect.com/science/article/pii/0022039672900356}
  {\bibfield  {journal} {\bibinfo  {journal} {J. Differ. Equ.}\ }\textbf
  {\bibinfo {volume} {12}},\ \bibinfo {pages} {313} (\bibinfo {year}
  {1972})}\BibitemShut {NoStop}%
\bibitem [{\citenamefont {Stengel}(1994)}]{stengel_optimal_1994}%
  \BibitemOpen
  \bibfield  {author} {\bibinfo {author} {\bibfnamefont {R.~F.}\ \bibnamefont
  {Stengel}},\ }\href@noop {} {\emph {\bibinfo {title} {Optimal control and
  estimation}}},\ Book\ (\bibinfo  {publisher} {Springer},\ \bibinfo {year}
  {1994})\BibitemShut {NoStop}%
\bibitem [{\citenamefont {Khaneja}\ \emph {et~al.}(2005)\citenamefont
  {Khaneja}, \citenamefont {Reiss}, \citenamefont {Kehlet}, \citenamefont
  {Schulte-Herbr{\"u}ggen},\ and\ \citenamefont {Glaser}}]{khaneja2005optimal}%
  \BibitemOpen
  \bibfield  {author} {\bibinfo {author} {\bibfnamefont {N.}~\bibnamefont
  {Khaneja}}, \bibinfo {author} {\bibfnamefont {T.}~\bibnamefont {Reiss}},
  \bibinfo {author} {\bibfnamefont {C.}~\bibnamefont {Kehlet}}, \bibinfo
  {author} {\bibfnamefont {T.}~\bibnamefont {Schulte-Herbr{\"u}ggen}}, \ and\
  \bibinfo {author} {\bibfnamefont {S.~J.}\ \bibnamefont {Glaser}},\ }\href
  {https://www.sciencedirect.com/science/article/abs/pii/S1090780704003696}
  {\bibfield  {journal} {\bibinfo  {journal} {J. Magn. Reson.}\ }\textbf
  {\bibinfo {volume} {172}},\ \bibinfo {pages} {296} (\bibinfo {year}
  {2005})}\BibitemShut {NoStop}%
\bibitem [{\citenamefont {Palao}\ and\ \citenamefont
  {Kosloff}(2003)}]{palao2003optimal}%
  \BibitemOpen
  \bibfield  {author} {\bibinfo {author} {\bibfnamefont {J.~P.}\ \bibnamefont
  {Palao}}\ and\ \bibinfo {author} {\bibfnamefont {R.}~\bibnamefont
  {Kosloff}},\ }\href
  {https://journals.aps.org/pra/abstract/10.1103/PhysRevA.68.062308} {\bibfield
   {journal} {\bibinfo  {journal} {Phys. Rev. A}\ }\textbf {\bibinfo {volume}
  {68}},\ \bibinfo {pages} {062308} (\bibinfo {year} {2003})}\BibitemShut
  {NoStop}%
\bibitem [{\citenamefont {Reich}\ \emph {et~al.}(2012)\citenamefont {Reich},
  \citenamefont {Ndong},\ and\ \citenamefont {Koch}}]{reich2012monotonically}%
  \BibitemOpen
  \bibfield  {author} {\bibinfo {author} {\bibfnamefont {D.~M.}\ \bibnamefont
  {Reich}}, \bibinfo {author} {\bibfnamefont {M.}~\bibnamefont {Ndong}}, \ and\
  \bibinfo {author} {\bibfnamefont {C.~P.}\ \bibnamefont {Koch}},\ }\href
  {https://aip.scitation.org/doi/10.1063/1.3691827} {\bibfield  {journal}
  {\bibinfo  {journal} {J. Chem. Phys.}\ }\textbf {\bibinfo {volume} {136}},\
  \bibinfo {pages} {104103} (\bibinfo {year} {2012})}\BibitemShut {NoStop}%
\bibitem [{\citenamefont {Goerz}\ \emph {et~al.}(2019)\citenamefont {Goerz},
  \citenamefont {Basilewitsch}, \citenamefont {Gago-Encinas}, \citenamefont
  {Krauss}, \citenamefont {Horn}, \citenamefont {Reich},\ and\ \citenamefont
  {Koch}}]{goerz2019krotov}%
  \BibitemOpen
  \bibfield  {author} {\bibinfo {author} {\bibfnamefont {M.~H.}\ \bibnamefont
  {Goerz}}, \bibinfo {author} {\bibfnamefont {D.}~\bibnamefont {Basilewitsch}},
  \bibinfo {author} {\bibfnamefont {F.}~\bibnamefont {Gago-Encinas}}, \bibinfo
  {author} {\bibfnamefont {M.~G.}\ \bibnamefont {Krauss}}, \bibinfo {author}
  {\bibfnamefont {K.~P.}\ \bibnamefont {Horn}}, \bibinfo {author}
  {\bibfnamefont {D.~M.}\ \bibnamefont {Reich}}, \ and\ \bibinfo {author}
  {\bibfnamefont {C.~P.}\ \bibnamefont {Koch}},\ }\href
  {https://scipost.org/SciPostPhys.7.6.080} {\bibfield  {journal} {\bibinfo
  {journal} {SciPost Phys.}\ }\textbf {\bibinfo {volume} {7}} (\bibinfo {year}
  {2019})}\BibitemShut {NoStop}%
\bibitem [{\citenamefont {Ho}\ and\ \citenamefont
  {Rabitz}(2010)}]{ho2010accelerated}%
  \BibitemOpen
  \bibfield  {author} {\bibinfo {author} {\bibfnamefont {T.-S.}\ \bibnamefont
  {Ho}}\ and\ \bibinfo {author} {\bibfnamefont {H.}~\bibnamefont {Rabitz}},\
  }\href {https://journals.aps.org/pre/abstract/10.1103/PhysRevE.82.026703}
  {\bibfield  {journal} {\bibinfo  {journal} {Phys. Rev. E}\ }\textbf {\bibinfo
  {volume} {82}},\ \bibinfo {pages} {026703} (\bibinfo {year}
  {2010})}\BibitemShut {NoStop}%
\bibitem [{\citenamefont {Ho}\ \emph {et~al.}(2011)\citenamefont {Ho},
  \citenamefont {Rabitz},\ and\ \citenamefont {Chu}}]{ho2011general}%
  \BibitemOpen
  \bibfield  {author} {\bibinfo {author} {\bibfnamefont {T.-S.}\ \bibnamefont
  {Ho}}, \bibinfo {author} {\bibfnamefont {H.}~\bibnamefont {Rabitz}}, \ and\
  \bibinfo {author} {\bibfnamefont {S.-I.}\ \bibnamefont {Chu}},\ }\href
  {https://www.sciencedirect.com/science/article/pii/S0010465510002833}
  {\bibfield  {journal} {\bibinfo  {journal} {Comput. Phys. Commun.}\ }\textbf
  {\bibinfo {volume} {182}},\ \bibinfo {pages} {14} (\bibinfo {year}
  {2011})}\BibitemShut {NoStop}%
\bibitem [{\citenamefont {Liao}\ \emph {et~al.}(2011)\citenamefont {Liao},
  \citenamefont {Ho}, \citenamefont {Chu},\ and\ \citenamefont
  {Rabitz}}]{liao2011fast}%
  \BibitemOpen
  \bibfield  {author} {\bibinfo {author} {\bibfnamefont {S.-L.}\ \bibnamefont
  {Liao}}, \bibinfo {author} {\bibfnamefont {T.-S.}\ \bibnamefont {Ho}},
  \bibinfo {author} {\bibfnamefont {S.-I.}\ \bibnamefont {Chu}}, \ and\
  \bibinfo {author} {\bibfnamefont {H.}~\bibnamefont {Rabitz}},\ }\href
  {https://journals.aps.org/pra/abstract/10.1103/PhysRevA.84.031401} {\bibfield
   {journal} {\bibinfo  {journal} {Phys. Rev. A}\ }\textbf {\bibinfo {volume}
  {84}},\ \bibinfo {pages} {031401} (\bibinfo {year} {2011})}\BibitemShut
  {NoStop}%
\bibitem [{\citenamefont {Rothman}\ \emph
  {et~al.}(2005{\natexlab{a}})\citenamefont {Rothman}, \citenamefont {Ho},\
  and\ \citenamefont {Rabitz}}]{rothman2005observable}%
  \BibitemOpen
  \bibfield  {author} {\bibinfo {author} {\bibfnamefont {A.}~\bibnamefont
  {Rothman}}, \bibinfo {author} {\bibfnamefont {T.-S.}\ \bibnamefont {Ho}}, \
  and\ \bibinfo {author} {\bibfnamefont {H.}~\bibnamefont {Rabitz}},\ }\href
  {https://journals.aps.org/pra/abstract/10.1103/PhysRevA.72.023416} {\bibfield
   {journal} {\bibinfo  {journal} {Phys. Rev. A}\ }\textbf {\bibinfo {volume}
  {72}},\ \bibinfo {pages} {023416} (\bibinfo {year}
  {2005}{\natexlab{a}})}\BibitemShut {NoStop}%
\bibitem [{\citenamefont {Rothman}\ \emph
  {et~al.}(2005{\natexlab{b}})\citenamefont {Rothman}, \citenamefont {Ho},\
  and\ \citenamefont {Rabitz}}]{rothman2005quantum}%
  \BibitemOpen
  \bibfield  {author} {\bibinfo {author} {\bibfnamefont {A.}~\bibnamefont
  {Rothman}}, \bibinfo {author} {\bibfnamefont {T.-S.}\ \bibnamefont {Ho}}, \
  and\ \bibinfo {author} {\bibfnamefont {H.}~\bibnamefont {Rabitz}},\ }\href
  {https://aip.scitation.org/doi/abs/10.1063/1.2042456?journalCode=jcp}
  {\bibfield  {journal} {\bibinfo  {journal} {J. Chem. Phys.}\ }\textbf
  {\bibinfo {volume} {123}},\ \bibinfo {pages} {134104} (\bibinfo {year}
  {2005}{\natexlab{b}})}\BibitemShut {NoStop}%
\bibitem [{\citenamefont {Rothman}\ \emph {et~al.}(2006)\citenamefont
  {Rothman}, \citenamefont {Ho},\ and\ \citenamefont
  {Rabitz}}]{rothman2006exploring}%
  \BibitemOpen
  \bibfield  {author} {\bibinfo {author} {\bibfnamefont {A.}~\bibnamefont
  {Rothman}}, \bibinfo {author} {\bibfnamefont {T.-S.}\ \bibnamefont {Ho}}, \
  and\ \bibinfo {author} {\bibfnamefont {H.}~\bibnamefont {Rabitz}},\ }\href
  {https://journals.aps.org/pra/abstract/10.1103/PhysRevA.73.053401} {\bibfield
   {journal} {\bibinfo  {journal} {Phys. Rev. A}\ }\textbf {\bibinfo {volume}
  {73}},\ \bibinfo {pages} {053401} (\bibinfo {year} {2006})}\BibitemShut
  {NoStop}%
\bibitem [{\citenamefont {Kosut}\ \emph {et~al.}(2013)\citenamefont {Kosut},
  \citenamefont {Grace},\ and\ \citenamefont {Brif}}]{Kosut2013_Robust}%
  \BibitemOpen
  \bibfield  {author} {\bibinfo {author} {\bibfnamefont {R.~L.}\ \bibnamefont
  {Kosut}}, \bibinfo {author} {\bibfnamefont {M.~D.}\ \bibnamefont {Grace}}, \
  and\ \bibinfo {author} {\bibfnamefont {C.}~\bibnamefont {Brif}},\ }\href
  {\doibase 10.1103/PhysRevA.88.052326} {\bibfield  {journal} {\bibinfo
  {journal} {Phys. Rev. A}\ }\textbf {\bibinfo {volume} {88}},\ \bibinfo
  {pages} {052326} (\bibinfo {year} {2013})}\BibitemShut {NoStop}%
\bibitem [{\citenamefont {Judson}\ and\ \citenamefont
  {Rabitz}(1992)}]{PhysRevLett.68.1500}%
  \BibitemOpen
  \bibfield  {author} {\bibinfo {author} {\bibfnamefont {R.~S.}\ \bibnamefont
  {Judson}}\ and\ \bibinfo {author} {\bibfnamefont {H.}~\bibnamefont
  {Rabitz}},\ }\href {\doibase 10.1103/PhysRevLett.68.1500} {\bibfield
  {journal} {\bibinfo  {journal} {Phys. Rev. Lett.}\ }\textbf {\bibinfo
  {volume} {68}},\ \bibinfo {pages} {1500} (\bibinfo {year}
  {1992})}\BibitemShut {NoStop}%
\bibitem [{\citenamefont {Egger}\ and\ \citenamefont
  {Wilhelm}(2014)}]{egger2014adaptive}%
  \BibitemOpen
  \bibfield  {author} {\bibinfo {author} {\bibfnamefont {D.~J.}\ \bibnamefont
  {Egger}}\ and\ \bibinfo {author} {\bibfnamefont {F.~K.}\ \bibnamefont
  {Wilhelm}},\ }\href
  {https://journals.aps.org/prl/abstract/10.1103/PhysRevLett.112.240503}
  {\bibfield  {journal} {\bibinfo  {journal} {Phys. Rev. Lett.}\ }\textbf
  {\bibinfo {volume} {112}},\ \bibinfo {pages} {240503} (\bibinfo {year}
  {2014})}\BibitemShut {NoStop}%
\bibitem [{\citenamefont {Lu}\ \emph {et~al.}(2017)\citenamefont {Lu},
  \citenamefont {Li}, \citenamefont {Li}, \citenamefont {Katiyar},
  \citenamefont {Park}, \citenamefont {Feng}, \citenamefont {Xin},
  \citenamefont {Li}, \citenamefont {Long}, \citenamefont {Brodutch} \emph
  {et~al.}}]{lu2017enhancing}%
  \BibitemOpen
  \bibfield  {author} {\bibinfo {author} {\bibfnamefont {D.}~\bibnamefont
  {Lu}}, \bibinfo {author} {\bibfnamefont {K.}~\bibnamefont {Li}}, \bibinfo
  {author} {\bibfnamefont {J.}~\bibnamefont {Li}}, \bibinfo {author}
  {\bibfnamefont {H.}~\bibnamefont {Katiyar}}, \bibinfo {author} {\bibfnamefont
  {A.~J.}\ \bibnamefont {Park}}, \bibinfo {author} {\bibfnamefont
  {G.}~\bibnamefont {Feng}}, \bibinfo {author} {\bibfnamefont {T.}~\bibnamefont
  {Xin}}, \bibinfo {author} {\bibfnamefont {H.}~\bibnamefont {Li}}, \bibinfo
  {author} {\bibfnamefont {G.}~\bibnamefont {Long}}, \bibinfo {author}
  {\bibfnamefont {A.}~\bibnamefont {Brodutch}},  \emph {et~al.},\ }\href
  {https://www.nature.com/articles/s41534-017-0045-z} {\bibfield  {journal}
  {\bibinfo  {journal} {NPJ Quantum Inf.}\ }\textbf {\bibinfo {volume} {3}},\
  \bibinfo {pages} {45} (\bibinfo {year} {2017})}\BibitemShut {NoStop}%
\bibitem [{\citenamefont {Li}\ \emph {et~al.}(2017)\citenamefont {Li},
  \citenamefont {Yang}, \citenamefont {Peng},\ and\ \citenamefont
  {Sun}}]{li2017hybrid}%
  \BibitemOpen
  \bibfield  {author} {\bibinfo {author} {\bibfnamefont {J.}~\bibnamefont
  {Li}}, \bibinfo {author} {\bibfnamefont {X.}~\bibnamefont {Yang}}, \bibinfo
  {author} {\bibfnamefont {X.}~\bibnamefont {Peng}}, \ and\ \bibinfo {author}
  {\bibfnamefont {C.-P.}\ \bibnamefont {Sun}},\ }\href
  {https://journals.aps.org/prl/abstract/10.1103/PhysRevLett.118.150503}
  {\bibfield  {journal} {\bibinfo  {journal} {Phys. Rev. Lett.}\ }\textbf
  {\bibinfo {volume} {118}},\ \bibinfo {pages} {150503} (\bibinfo {year}
  {2017})}\BibitemShut {NoStop}%
\bibitem [{\citenamefont {Chen}\ \emph {et~al.}(2020)\citenamefont {Chen},
  \citenamefont {Yang}, \citenamefont {Arenz}, \citenamefont {Wu},
  \citenamefont {Peng}, \citenamefont {Pelczer},\ and\ \citenamefont
  {Rabitz}}]{chen2020combining}%
  \BibitemOpen
  \bibfield  {author} {\bibinfo {author} {\bibfnamefont {Q.-M.}\ \bibnamefont
  {Chen}}, \bibinfo {author} {\bibfnamefont {X.}~\bibnamefont {Yang}}, \bibinfo
  {author} {\bibfnamefont {C.}~\bibnamefont {Arenz}}, \bibinfo {author}
  {\bibfnamefont {R.-B.}\ \bibnamefont {Wu}}, \bibinfo {author} {\bibfnamefont
  {X.}~\bibnamefont {Peng}}, \bibinfo {author} {\bibfnamefont {I.}~\bibnamefont
  {Pelczer}}, \ and\ \bibinfo {author} {\bibfnamefont {H.}~\bibnamefont
  {Rabitz}},\ }\href
  {https://journals.aps.org/pra/abstract/10.1103/PhysRevA.101.032313}
  {\bibfield  {journal} {\bibinfo  {journal} {Phys. Rev. A}\ }\textbf {\bibinfo
  {volume} {101}},\ \bibinfo {pages} {032313} (\bibinfo {year}
  {2020})}\BibitemShut {NoStop}%
\bibitem [{\citenamefont {{Yang}}\ \emph {et~al.}(2020)\citenamefont {{Yang}},
  \citenamefont {{Arenz}}, \citenamefont {{Pelczer}}, \citenamefont {{Chen}},
  \citenamefont {{Wu}}, \citenamefont {{Peng}},\ and\ \citenamefont
  {{Rabitz}}}]{2020arXiv200803874Y}%
  \BibitemOpen
  \bibfield  {author} {\bibinfo {author} {\bibfnamefont {X.-d.}\ \bibnamefont
  {{Yang}}}, \bibinfo {author} {\bibfnamefont {C.}~\bibnamefont {{Arenz}}},
  \bibinfo {author} {\bibfnamefont {I.}~\bibnamefont {{Pelczer}}}, \bibinfo
  {author} {\bibfnamefont {Q.-M.}\ \bibnamefont {{Chen}}}, \bibinfo {author}
  {\bibfnamefont {R.-B.}\ \bibnamefont {{Wu}}}, \bibinfo {author}
  {\bibfnamefont {X.-h.}\ \bibnamefont {{Peng}}}, \ and\ \bibinfo {author}
  {\bibfnamefont {H.}~\bibnamefont {{Rabitz}}},\ }\href@noop {} {\bibfield
  {journal} {\bibinfo  {journal} {arXiv e-prints}\ ,\ \bibinfo {eid}
  {arXiv:2008.03874}} (\bibinfo {year} {2020})},\ \Eprint
  {http://arxiv.org/abs/2008.03874} {arXiv:2008.03874 [quant-ph]} \BibitemShut
  {NoStop}%
\bibitem [{\citenamefont {Yang}\ \emph {et~al.}(2017)\citenamefont {Yang},
  \citenamefont {Rahmani}, \citenamefont {Shabani}, \citenamefont {Neven},\
  and\ \citenamefont {Chamon}}]{PhysRevX.7.021027}%
  \BibitemOpen
  \bibfield  {author} {\bibinfo {author} {\bibfnamefont {Z.-C.}\ \bibnamefont
  {Yang}}, \bibinfo {author} {\bibfnamefont {A.}~\bibnamefont {Rahmani}},
  \bibinfo {author} {\bibfnamefont {A.}~\bibnamefont {Shabani}}, \bibinfo
  {author} {\bibfnamefont {H.}~\bibnamefont {Neven}}, \ and\ \bibinfo {author}
  {\bibfnamefont {C.}~\bibnamefont {Chamon}},\ }\href {\doibase
  10.1103/PhysRevX.7.021027} {\bibfield  {journal} {\bibinfo  {journal} {Phys.
  Rev. X}\ }\textbf {\bibinfo {volume} {7}},\ \bibinfo {pages} {021027}
  (\bibinfo {year} {2017})}\BibitemShut {NoStop}%
\bibitem [{\citenamefont {Lin}\ \emph {et~al.}(2019)\citenamefont {Lin},
  \citenamefont {Wang}, \citenamefont {Kolesov},\ and\ \citenamefont
  {Kalabi\ifmmode~\acute{c}\else \'{c}\fi{}}}]{PhysRevA.100.022327}%
  \BibitemOpen
  \bibfield  {author} {\bibinfo {author} {\bibfnamefont {C.}~\bibnamefont
  {Lin}}, \bibinfo {author} {\bibfnamefont {Y.}~\bibnamefont {Wang}}, \bibinfo
  {author} {\bibfnamefont {G.}~\bibnamefont {Kolesov}}, \ and\ \bibinfo
  {author} {\bibfnamefont {U.~c.~v.}\ \bibnamefont
  {Kalabi\ifmmode~\acute{c}\else \'{c}\fi{}}},\ }\href {\doibase
  10.1103/PhysRevA.100.022327} {\bibfield  {journal} {\bibinfo  {journal}
  {Phys. Rev. A}\ }\textbf {\bibinfo {volume} {100}},\ \bibinfo {pages}
  {022327} (\bibinfo {year} {2019})}\BibitemShut {NoStop}%
\bibitem [{\citenamefont {Ramakrishna}\ and\ \citenamefont
  {Rabitz}(1996)}]{Ramakrishna96a}%
  \BibitemOpen
  \bibfield  {author} {\bibinfo {author} {\bibfnamefont {V.}~\bibnamefont
  {Ramakrishna}}\ and\ \bibinfo {author} {\bibfnamefont {H.}~\bibnamefont
  {Rabitz}},\ }\href {\doibase 10.1103/PhysRevA.54.1715} {\bibfield  {journal}
  {\bibinfo  {journal} {Phys. Rev. A}\ }\textbf {\bibinfo {volume} {54}},\
  \bibinfo {pages} {1715} (\bibinfo {year} {1996})}\BibitemShut {NoStop}%
\bibitem [{\citenamefont {Altafini}(2002)}]{altafini2002controllability}%
  \BibitemOpen
  \bibfield  {author} {\bibinfo {author} {\bibfnamefont {C.}~\bibnamefont
  {Altafini}},\ }\href {https://aip.scitation.org/doi/10.1063/1.1467611}
  {\bibfield  {journal} {\bibinfo  {journal} {J. Math. Phys.}\ }\textbf
  {\bibinfo {volume} {43}},\ \bibinfo {pages} {2051} (\bibinfo {year}
  {2002})}\BibitemShut {NoStop}%
\bibitem [{\citenamefont {Albertini}\ and\ \citenamefont
  {D'Alessandro}(2002)}]{albertini2002lie}%
  \BibitemOpen
  \bibfield  {author} {\bibinfo {author} {\bibfnamefont {F.}~\bibnamefont
  {Albertini}}\ and\ \bibinfo {author} {\bibfnamefont {D.}~\bibnamefont
  {D'Alessandro}},\ }\href
  {https://www.sciencedirect.com/science/article/pii/S0024379502002902}
  {\bibfield  {journal} {\bibinfo  {journal} {Linear Algebra Appl.}\ }\textbf
  {\bibinfo {volume} {350}},\ \bibinfo {pages} {213} (\bibinfo {year}
  {2002})}\BibitemShut {NoStop}%
\bibitem [{\citenamefont {Schirmer}\ \emph {et~al.}(2001)\citenamefont
  {Schirmer}, \citenamefont {Fu},\ and\ \citenamefont
  {Solomon}}]{schirmer2001complete}%
  \BibitemOpen
  \bibfield  {author} {\bibinfo {author} {\bibfnamefont {S.~G.}\ \bibnamefont
  {Schirmer}}, \bibinfo {author} {\bibfnamefont {H.}~\bibnamefont {Fu}}, \ and\
  \bibinfo {author} {\bibfnamefont {A.~I.}\ \bibnamefont {Solomon}},\ }\href
  {https://journals.aps.org/pra/abstract/10.1103/PhysRevA.63.063410} {\bibfield
   {journal} {\bibinfo  {journal} {Phys. Rev. A}\ }\textbf {\bibinfo {volume}
  {63}},\ \bibinfo {pages} {063410} (\bibinfo {year} {2001})}\BibitemShut
  {NoStop}%
\bibitem [{\citenamefont {Turinici}(2000)}]{turinici2000controllability}%
  \BibitemOpen
  \bibfield  {author} {\bibinfo {author} {\bibfnamefont {G.}~\bibnamefont
  {Turinici}},\ }in\ \href@noop {} {\emph {\bibinfo {booktitle} {Mathematical
  models and methods for ab initio Quantum Chemistry}}}\ (\bibinfo  {publisher}
  {Springer},\ \bibinfo {year} {2000})\ pp.\ \bibinfo {pages}
  {75--92}\BibitemShut {NoStop}%
\bibitem [{\citenamefont {Fu}\ \emph {et~al.}(2001)\citenamefont {Fu},
  \citenamefont {Schirmer},\ and\ \citenamefont {Solomon}}]{fu2001complete}%
  \BibitemOpen
  \bibfield  {author} {\bibinfo {author} {\bibfnamefont {H.}~\bibnamefont
  {Fu}}, \bibinfo {author} {\bibfnamefont {S.~G.}\ \bibnamefont {Schirmer}}, \
  and\ \bibinfo {author} {\bibfnamefont {A.~I.}\ \bibnamefont {Solomon}},\
  }\href {https://iopscience.iop.org/article/10.1088/0305-4470/34/8/313/meta}
  {\bibfield  {journal} {\bibinfo  {journal} {J. Phys. A}\ }\textbf {\bibinfo
  {volume} {34}},\ \bibinfo {pages} {1679} (\bibinfo {year}
  {2001})}\BibitemShut {NoStop}%
\bibitem [{\citenamefont {Turinici}\ and\ \citenamefont
  {Rabitz}(2003)}]{turinici2003wavefunction}%
  \BibitemOpen
  \bibfield  {author} {\bibinfo {author} {\bibfnamefont {G.}~\bibnamefont
  {Turinici}}\ and\ \bibinfo {author} {\bibfnamefont {H.}~\bibnamefont
  {Rabitz}},\ }\href
  {https://iopscience.iop.org/article/10.1088/0305-4470/36/10/316/meta}
  {\bibfield  {journal} {\bibinfo  {journal} {J. Phys. A}\ }\textbf {\bibinfo
  {volume} {36}},\ \bibinfo {pages} {2565} (\bibinfo {year}
  {2003})}\BibitemShut {NoStop}%
\bibitem [{\citenamefont {Burgarth}\ \emph {et~al.}(2013)\citenamefont
  {Burgarth}, \citenamefont {D'Alessandro}, \citenamefont {Hogben},
  \citenamefont {Severini},\ and\ \citenamefont {Young}}]{burgarth2013zero}%
  \BibitemOpen
  \bibfield  {author} {\bibinfo {author} {\bibfnamefont {D.}~\bibnamefont
  {Burgarth}}, \bibinfo {author} {\bibfnamefont {D.}~\bibnamefont
  {D'Alessandro}}, \bibinfo {author} {\bibfnamefont {L.}~\bibnamefont
  {Hogben}}, \bibinfo {author} {\bibfnamefont {S.}~\bibnamefont {Severini}}, \
  and\ \bibinfo {author} {\bibfnamefont {M.}~\bibnamefont {Young}},\ }\href
  {https://ieeexplore.ieee.org/document/6471752} {\bibfield  {journal}
  {\bibinfo  {journal} {IEEE Trans. Automat. Contr.}\ }\textbf {\bibinfo
  {volume} {58}},\ \bibinfo {pages} {2349} (\bibinfo {year}
  {2013})}\BibitemShut {NoStop}%
\bibitem [{\citenamefont {Heule}\ \emph {et~al.}(2010)\citenamefont {Heule},
  \citenamefont {Bruder}, \citenamefont {Burgarth},\ and\ \citenamefont
  {Stojanovi{\'c}}}]{heule2010local}%
  \BibitemOpen
  \bibfield  {author} {\bibinfo {author} {\bibfnamefont {R.}~\bibnamefont
  {Heule}}, \bibinfo {author} {\bibfnamefont {C.}~\bibnamefont {Bruder}},
  \bibinfo {author} {\bibfnamefont {D.}~\bibnamefont {Burgarth}}, \ and\
  \bibinfo {author} {\bibfnamefont {V.~M.}\ \bibnamefont {Stojanovi{\'c}}},\
  }\href {https://journals.aps.org/pra/abstract/10.1103/PhysRevA.82.052333}
  {\bibfield  {journal} {\bibinfo  {journal} {Phys. Rev. A}\ }\textbf {\bibinfo
  {volume} {82}},\ \bibinfo {pages} {052333} (\bibinfo {year}
  {2010})}\BibitemShut {NoStop}%
\bibitem [{\citenamefont {Zimbor{\'a}s}\ \emph {et~al.}(2014)\citenamefont
  {Zimbor{\'a}s}, \citenamefont {Zeier}, \citenamefont {Keyl},\ and\
  \citenamefont {Schulte-Herbr{\"u}ggen}}]{zimboras2014dynamic}%
  \BibitemOpen
  \bibfield  {author} {\bibinfo {author} {\bibfnamefont {Z.}~\bibnamefont
  {Zimbor{\'a}s}}, \bibinfo {author} {\bibfnamefont {R.}~\bibnamefont {Zeier}},
  \bibinfo {author} {\bibfnamefont {M.}~\bibnamefont {Keyl}}, \ and\ \bibinfo
  {author} {\bibfnamefont {T.}~\bibnamefont {Schulte-Herbr{\"u}ggen}},\ }\href
  {https://epjquantumtechnology.springeropen.com/articles/10.1140/epjqt11}
  {\bibfield  {journal} {\bibinfo  {journal} {EPJ Quantum Technol.}\ }\textbf
  {\bibinfo {volume} {1}},\ \bibinfo {pages} {11} (\bibinfo {year}
  {2014})}\BibitemShut {NoStop}%
\bibitem [{\citenamefont {Genoni}\ \emph {et~al.}(2012)\citenamefont {Genoni},
  \citenamefont {Serafini}, \citenamefont {Kim},\ and\ \citenamefont
  {Burgarth}}]{genoni2012dynamical}%
  \BibitemOpen
  \bibfield  {author} {\bibinfo {author} {\bibfnamefont {M.~G.}\ \bibnamefont
  {Genoni}}, \bibinfo {author} {\bibfnamefont {A.}~\bibnamefont {Serafini}},
  \bibinfo {author} {\bibfnamefont {M.}~\bibnamefont {Kim}}, \ and\ \bibinfo
  {author} {\bibfnamefont {D.}~\bibnamefont {Burgarth}},\ }\href
  {https://journals.aps.org/prl/abstract/10.1103/PhysRevLett.108.150501}
  {\bibfield  {journal} {\bibinfo  {journal} {Phys. Rev. Lett.}\ }\textbf
  {\bibinfo {volume} {108}},\ \bibinfo {pages} {150501} (\bibinfo {year}
  {2012})}\BibitemShut {NoStop}%
\bibitem [{\citenamefont {Dirr}\ \emph {et~al.}(2009)\citenamefont {Dirr},
  \citenamefont {Helmke}, \citenamefont {Kurniawan},\ and\ \citenamefont
  {Schulte-Herbr{\"u}ggen}}]{dirr2009lie}%
  \BibitemOpen
  \bibfield  {author} {\bibinfo {author} {\bibfnamefont {G.}~\bibnamefont
  {Dirr}}, \bibinfo {author} {\bibfnamefont {U.}~\bibnamefont {Helmke}},
  \bibinfo {author} {\bibfnamefont {I.}~\bibnamefont {Kurniawan}}, \ and\
  \bibinfo {author} {\bibfnamefont {T.}~\bibnamefont
  {Schulte-Herbr{\"u}ggen}},\ }\href
  {https://www.sciencedirect.com/science/article/pii/S0034487709900222}
  {\bibfield  {journal} {\bibinfo  {journal} {Rep. Math. Phys.}\ }\textbf
  {\bibinfo {volume} {64}},\ \bibinfo {pages} {93} (\bibinfo {year}
  {2009})}\BibitemShut {NoStop}%
\bibitem [{\citenamefont {Burgarth}\ \emph {et~al.}(2014)\citenamefont
  {Burgarth}, \citenamefont {Facchi}, \citenamefont {Giovannetti},
  \citenamefont {Nakazato}, \citenamefont {Pascazio},\ and\ \citenamefont
  {Yuasa}}]{burgarth2014exponential}%
  \BibitemOpen
  \bibfield  {author} {\bibinfo {author} {\bibfnamefont {D.~K.}\ \bibnamefont
  {Burgarth}}, \bibinfo {author} {\bibfnamefont {P.}~\bibnamefont {Facchi}},
  \bibinfo {author} {\bibfnamefont {V.}~\bibnamefont {Giovannetti}}, \bibinfo
  {author} {\bibfnamefont {H.}~\bibnamefont {Nakazato}}, \bibinfo {author}
  {\bibfnamefont {S.}~\bibnamefont {Pascazio}}, \ and\ \bibinfo {author}
  {\bibfnamefont {K.}~\bibnamefont {Yuasa}},\ }\href
  {https://www.nature.com/articles/ncomms6173} {\bibfield  {journal} {\bibinfo
  {journal} {Nat. Commun.}\ }\textbf {\bibinfo {volume} {5}},\ \bibinfo {pages}
  {1} (\bibinfo {year} {2014})}\BibitemShut {NoStop}%
\bibitem [{\citenamefont {Arenz}\ \emph {et~al.}(2016)\citenamefont {Arenz},
  \citenamefont {Burgarth}, \citenamefont {Facchi}, \citenamefont
  {Giovannetti}, \citenamefont {Nakazato}, \citenamefont {Pascazio},\ and\
  \citenamefont {Yuasa}}]{arenz2016universal}%
  \BibitemOpen
  \bibfield  {author} {\bibinfo {author} {\bibfnamefont {C.}~\bibnamefont
  {Arenz}}, \bibinfo {author} {\bibfnamefont {D.}~\bibnamefont {Burgarth}},
  \bibinfo {author} {\bibfnamefont {P.}~\bibnamefont {Facchi}}, \bibinfo
  {author} {\bibfnamefont {V.}~\bibnamefont {Giovannetti}}, \bibinfo {author}
  {\bibfnamefont {H.}~\bibnamefont {Nakazato}}, \bibinfo {author}
  {\bibfnamefont {S.}~\bibnamefont {Pascazio}}, \ and\ \bibinfo {author}
  {\bibfnamefont {K.}~\bibnamefont {Yuasa}},\ }\href@noop {} {\bibfield
  {journal} {\bibinfo  {journal} {Phys. Rev. A}\ }\textbf {\bibinfo {volume}
  {93}},\ \bibinfo {pages} {062308} (\bibinfo {year} {2016})}\BibitemShut
  {NoStop}%
\bibitem [{\citenamefont {{Lloyd}}(2018)}]{2018arXiv181211075L}%
  \BibitemOpen
  \bibfield  {author} {\bibinfo {author} {\bibfnamefont {S.}~\bibnamefont
  {{Lloyd}}},\ }\href@noop {} {\bibfield  {journal} {\bibinfo  {journal} {arXiv
  e-prints}\ ,\ \bibinfo {eid} {arXiv:1812.11075}} (\bibinfo {year} {2018})},\
  \Eprint {http://arxiv.org/abs/1812.11075} {arXiv:1812.11075 [quant-ph]}
  \BibitemShut {NoStop}%
\bibitem [{\citenamefont {Morales}\ \emph {et~al.}(2020)\citenamefont
  {Morales}, \citenamefont {Biamonte},\ and\ \citenamefont
  {Zimbor{\'a}s}}]{2019arXiv190903123M}%
  \BibitemOpen
  \bibfield  {author} {\bibinfo {author} {\bibfnamefont {M.~E.}\ \bibnamefont
  {Morales}}, \bibinfo {author} {\bibfnamefont {J.}~\bibnamefont {Biamonte}}, \
  and\ \bibinfo {author} {\bibfnamefont {Z.}~\bibnamefont {Zimbor{\'a}s}},\
  }\href {https://link.springer.com/article/10.1007/s11128-020-02748-9}
  {\bibfield  {journal} {\bibinfo  {journal} {Quantum Inf. Process.}\ }\textbf
  {\bibinfo {volume} {19}},\ \bibinfo {pages} {1} (\bibinfo {year}
  {2020})}\BibitemShut {NoStop}%
\bibitem [{\citenamefont {{Bigan Mbeng}}\ \emph {et~al.}(2019)\citenamefont
  {{Bigan Mbeng}}, \citenamefont {{Fazio}},\ and\ \citenamefont
  {{Santoro}}}]{2019arXiv190608948B}%
  \BibitemOpen
  \bibfield  {author} {\bibinfo {author} {\bibfnamefont {G.}~\bibnamefont
  {{Bigan Mbeng}}}, \bibinfo {author} {\bibfnamefont {R.}~\bibnamefont
  {{Fazio}}}, \ and\ \bibinfo {author} {\bibfnamefont {G.}~\bibnamefont
  {{Santoro}}},\ }\href@noop {} {\bibfield  {journal} {\bibinfo  {journal}
  {arXiv e-prints}\ ,\ \bibinfo {eid} {arXiv:1906.08948}} (\bibinfo {year}
  {2019})},\ \Eprint {http://arxiv.org/abs/1906.08948} {arXiv:1906.08948
  [quant-ph]} \BibitemShut {NoStop}%
\bibitem [{\citenamefont {Akshay}\ \emph {et~al.}(2020)\citenamefont {Akshay},
  \citenamefont {Philathong}, \citenamefont {Morales},\ and\ \citenamefont
  {Biamonte}}]{PhysRevLett.124.090504}%
  \BibitemOpen
  \bibfield  {author} {\bibinfo {author} {\bibfnamefont {V.}~\bibnamefont
  {Akshay}}, \bibinfo {author} {\bibfnamefont {H.}~\bibnamefont {Philathong}},
  \bibinfo {author} {\bibfnamefont {M.~E.~S.}\ \bibnamefont {Morales}}, \ and\
  \bibinfo {author} {\bibfnamefont {J.~D.}\ \bibnamefont {Biamonte}},\ }\href
  {\doibase 10.1103/PhysRevLett.124.090504} {\bibfield  {journal} {\bibinfo
  {journal} {Phys. Rev. Lett.}\ }\textbf {\bibinfo {volume} {124}},\ \bibinfo
  {pages} {090504} (\bibinfo {year} {2020})}\BibitemShut {NoStop}%
\bibitem [{\citenamefont {{Dong}}\ \emph {et~al.}(2019)\citenamefont {{Dong}},
  \citenamefont {{Meng}}, \citenamefont {{Lin}}, \citenamefont {{Kosut}},\ and\
  \citenamefont {{Whaley}}}]{2019arXiv191100789D}%
  \BibitemOpen
  \bibfield  {author} {\bibinfo {author} {\bibfnamefont {Y.}~\bibnamefont
  {{Dong}}}, \bibinfo {author} {\bibfnamefont {X.}~\bibnamefont {{Meng}}},
  \bibinfo {author} {\bibfnamefont {L.}~\bibnamefont {{Lin}}}, \bibinfo
  {author} {\bibfnamefont {R.}~\bibnamefont {{Kosut}}}, \ and\ \bibinfo
  {author} {\bibfnamefont {K.~B.}\ \bibnamefont {{Whaley}}},\ }\href@noop {}
  {\bibfield  {journal} {\bibinfo  {journal} {arXiv e-prints}\ ,\ \bibinfo
  {eid} {arXiv:1911.00789}} (\bibinfo {year} {2019})},\ \Eprint
  {http://arxiv.org/abs/1911.00789} {arXiv:1911.00789 [quant-ph]} \BibitemShut
  {NoStop}%
\bibitem [{\citenamefont {Niu}\ \emph {et~al.}(2019)\citenamefont {Niu},
  \citenamefont {Lu},\ and\ \citenamefont {Chuang}}]{Niu2019_Optimizing}%
  \BibitemOpen
  \bibfield  {author} {\bibinfo {author} {\bibfnamefont {M.~Y.}\ \bibnamefont
  {Niu}}, \bibinfo {author} {\bibfnamefont {S.}~\bibnamefont {Lu}}, \ and\
  \bibinfo {author} {\bibfnamefont {I.~L.}\ \bibnamefont {Chuang}},\ }\href
  {https://arxiv.org/abs/1905.12134} {\bibfield  {journal} {\bibinfo  {journal}
  {arXiv:1905.12134 [quant-ph]}\ } (\bibinfo {year} {2019})},\ \bibinfo {note}
  {arXiv: 1905.12134}\BibitemShut {NoStop}%
\bibitem [{\citenamefont {{Yao}}\ \emph {et~al.}(2020)\citenamefont {{Yao}},
  \citenamefont {{Bukov}},\ and\ \citenamefont {{Lin}}}]{2020arXiv200201068Y}%
  \BibitemOpen
  \bibfield  {author} {\bibinfo {author} {\bibfnamefont {J.}~\bibnamefont
  {{Yao}}}, \bibinfo {author} {\bibfnamefont {M.}~\bibnamefont {{Bukov}}}, \
  and\ \bibinfo {author} {\bibfnamefont {L.}~\bibnamefont {{Lin}}},\
  }\href@noop {} {\bibfield  {journal} {\bibinfo  {journal} {arXiv e-prints}\
  ,\ \bibinfo {eid} {arXiv:2002.01068}} (\bibinfo {year} {2020})},\ \Eprint
  {http://arxiv.org/abs/2002.01068} {arXiv:2002.01068 [quant-ph]} \BibitemShut
  {NoStop}%
\bibitem [{\citenamefont {Bapat}\ and\ \citenamefont
  {Jordan}(2019)}]{Bapat2019_Bang}%
  \BibitemOpen
  \bibfield  {author} {\bibinfo {author} {\bibfnamefont {A.}~\bibnamefont
  {Bapat}}\ and\ \bibinfo {author} {\bibfnamefont {S.}~\bibnamefont {Jordan}},\
  }\href {https://dl.acm.org/doi/10.5555/3370251.3370255} {\bibfield  {journal}
  {\bibinfo  {journal} {Quantum Inf. Comput.}\ }\textbf {\bibinfo {volume}
  {19}},\ \bibinfo {pages} {424} (\bibinfo {year} {2019})}\BibitemShut
  {NoStop}%
\bibitem [{\citenamefont {{Brady}}\ \emph {et~al.}(2020)\citenamefont
  {{Brady}}, \citenamefont {{Baldwin}}, \citenamefont {{Bapat}}, \citenamefont
  {{Kharkov}},\ and\ \citenamefont {{Gorshkov}}}]{2020arXiv200308952B}%
  \BibitemOpen
  \bibfield  {author} {\bibinfo {author} {\bibfnamefont {L.~T.}\ \bibnamefont
  {{Brady}}}, \bibinfo {author} {\bibfnamefont {C.~L.}\ \bibnamefont
  {{Baldwin}}}, \bibinfo {author} {\bibfnamefont {A.}~\bibnamefont {{Bapat}}},
  \bibinfo {author} {\bibfnamefont {Y.}~\bibnamefont {{Kharkov}}}, \ and\
  \bibinfo {author} {\bibfnamefont {A.~V.}\ \bibnamefont {{Gorshkov}}},\
  }\href@noop {} {\bibfield  {journal} {\bibinfo  {journal} {arXiv e-prints}\
  ,\ \bibinfo {eid} {arXiv:2003.08952}} (\bibinfo {year} {2020})},\ \Eprint
  {http://arxiv.org/abs/2003.08952} {arXiv:2003.08952 [quant-ph]} \BibitemShut
  {NoStop}%
\bibitem [{\citenamefont {Wang}\ \emph {et~al.}(2018)\citenamefont {Wang},
  \citenamefont {Hadfield}, \citenamefont {Jiang},\ and\ \citenamefont
  {Rieffel}}]{PhysRevA.97.022304}%
  \BibitemOpen
  \bibfield  {author} {\bibinfo {author} {\bibfnamefont {Z.}~\bibnamefont
  {Wang}}, \bibinfo {author} {\bibfnamefont {S.}~\bibnamefont {Hadfield}},
  \bibinfo {author} {\bibfnamefont {Z.}~\bibnamefont {Jiang}}, \ and\ \bibinfo
  {author} {\bibfnamefont {E.~G.}\ \bibnamefont {Rieffel}},\ }\href {\doibase
  10.1103/PhysRevA.97.022304} {\bibfield  {journal} {\bibinfo  {journal} {Phys.
  Rev. A}\ }\textbf {\bibinfo {volume} {97}},\ \bibinfo {pages} {022304}
  (\bibinfo {year} {2018})}\BibitemShut {NoStop}%
\bibitem [{\citenamefont {{Wu}}\ \emph {et~al.}(2020)\citenamefont {{Wu}},
  \citenamefont {{Cao}}, \citenamefont {{Xie}},\ and\ \citenamefont
  {{Liu}}}]{2020arXiv200313658W}%
  \BibitemOpen
  \bibfield  {author} {\bibinfo {author} {\bibfnamefont {R.-B.}\ \bibnamefont
  {{Wu}}}, \bibinfo {author} {\bibfnamefont {X.}~\bibnamefont {{Cao}}},
  \bibinfo {author} {\bibfnamefont {P.}~\bibnamefont {{Xie}}}, \ and\ \bibinfo
  {author} {\bibfnamefont {Y.-x.}\ \bibnamefont {{Liu}}},\ }\href@noop {}
  {\bibfield  {journal} {\bibinfo  {journal} {arXiv e-prints}\ ,\ \bibinfo
  {eid} {arXiv:2003.13658}} (\bibinfo {year} {2020})},\ \Eprint
  {http://arxiv.org/abs/2003.13658} {arXiv:2003.13658 [quant-ph]} \BibitemShut
  {NoStop}%
\bibitem [{\citenamefont {Grimsley}\ \emph
  {et~al.}(2019{\natexlab{b}})\citenamefont {Grimsley}, \citenamefont
  {Economou}, \citenamefont {Barnes},\ and\ \citenamefont
  {Mayhall}}]{Grimsley2019_Adaptive}%
  \BibitemOpen
  \bibfield  {author} {\bibinfo {author} {\bibfnamefont {H.~R.}\ \bibnamefont
  {Grimsley}}, \bibinfo {author} {\bibfnamefont {S.~E.}\ \bibnamefont
  {Economou}}, \bibinfo {author} {\bibfnamefont {E.}~\bibnamefont {Barnes}}, \
  and\ \bibinfo {author} {\bibfnamefont {N.~J.}\ \bibnamefont {Mayhall}},\
  }\href {\doibase 10.1038/s41467-019-10988-2} {\bibfield  {journal} {\bibinfo
  {journal} {Nat. Commun.}\ }\textbf {\bibinfo {volume} {10}},\ \bibinfo
  {pages} {3007} (\bibinfo {year} {2019}{\natexlab{b}})}\BibitemShut {NoStop}%
\bibitem [{\citenamefont {Tang}\ \emph {et~al.}(2020)\citenamefont {Tang},
  \citenamefont {Shkolnikov}, \citenamefont {Barron}, \citenamefont {Grimsley},
  \citenamefont {Mayhall}, \citenamefont {Barnes},\ and\ \citenamefont
  {Economou}}]{Tang2020_Qubit}%
  \BibitemOpen
  \bibfield  {author} {\bibinfo {author} {\bibfnamefont {H.~L.}\ \bibnamefont
  {Tang}}, \bibinfo {author} {\bibfnamefont {V.~O.}\ \bibnamefont
  {Shkolnikov}}, \bibinfo {author} {\bibfnamefont {G.~S.}\ \bibnamefont
  {Barron}}, \bibinfo {author} {\bibfnamefont {H.~R.}\ \bibnamefont
  {Grimsley}}, \bibinfo {author} {\bibfnamefont {N.~J.}\ \bibnamefont
  {Mayhall}}, \bibinfo {author} {\bibfnamefont {E.}~\bibnamefont {Barnes}}, \
  and\ \bibinfo {author} {\bibfnamefont {S.~E.}\ \bibnamefont {Economou}},\
  }\href {http://arxiv.org/abs/1911.10205} {\bibfield  {journal} {\bibinfo
  {journal} {arXiv:1911.10205 [quant-ph]}\ } (\bibinfo {year} {2020})},\
  \bibinfo {note} {arXiv: 1911.10205}\BibitemShut {NoStop}%
\bibitem [{\citenamefont {Zhu}\ \emph {et~al.}(2020)\citenamefont {Zhu},
  \citenamefont {Tang}, \citenamefont {Barron}, \citenamefont {Mayhall},
  \citenamefont {Barnes},\ and\ \citenamefont {Economou}}]{Zhu2020_Adaptive}%
  \BibitemOpen
  \bibfield  {author} {\bibinfo {author} {\bibfnamefont {L.}~\bibnamefont
  {Zhu}}, \bibinfo {author} {\bibfnamefont {H.~L.}\ \bibnamefont {Tang}},
  \bibinfo {author} {\bibfnamefont {G.~S.}\ \bibnamefont {Barron}}, \bibinfo
  {author} {\bibfnamefont {N.~J.}\ \bibnamefont {Mayhall}}, \bibinfo {author}
  {\bibfnamefont {E.}~\bibnamefont {Barnes}}, \ and\ \bibinfo {author}
  {\bibfnamefont {S.~E.}\ \bibnamefont {Economou}},\ }\href
  {http://arxiv.org/abs/2005.10258} {\bibfield  {journal} {\bibinfo  {journal}
  {arXiv:2005.10258 [quant-ph]}\ } (\bibinfo {year} {2020})},\ \bibinfo {note}
  {arXiv: 2005.10258}\BibitemShut {NoStop}%
\bibitem [{\citenamefont {Choquette}\ \emph {et~al.}(2020)\citenamefont
  {Choquette}, \citenamefont {Di~Paolo}, \citenamefont {Barkoutsos},
  \citenamefont {S\'{e}n\'{e}chal}, \citenamefont {Tavernelli},\ and\
  \citenamefont {Blais}}]{Choquette2020_QOCA}%
  \BibitemOpen
  \bibfield  {author} {\bibinfo {author} {\bibfnamefont {A.}~\bibnamefont
  {Choquette}}, \bibinfo {author} {\bibfnamefont {A.}~\bibnamefont {Di~Paolo}},
  \bibinfo {author} {\bibfnamefont {P.~K.}\ \bibnamefont {Barkoutsos}},
  \bibinfo {author} {\bibfnamefont {D.}~\bibnamefont {S\'{e}n\'{e}chal}},
  \bibinfo {author} {\bibfnamefont {I.}~\bibnamefont {Tavernelli}}, \ and\
  \bibinfo {author} {\bibfnamefont {A.}~\bibnamefont {Blais}},\ }\href
  {http://arxiv.org/abs/2008.01098} {\bibfield  {journal} {\bibinfo  {journal}
  {arXiv:2008.01098 [quant-ph]}\ } (\bibinfo {year} {2020})}\BibitemShut
  {NoStop}%
\bibitem [{\citenamefont {Meitei}\ \emph {et~al.}(2020)\citenamefont {Meitei},
  \citenamefont {Gard}, \citenamefont {Barron}, \citenamefont {Pappas},
  \citenamefont {Economou}, \citenamefont {Barnes},\ and\ \citenamefont
  {Mayhall}}]{Meitei2020_Gate}%
  \BibitemOpen
  \bibfield  {author} {\bibinfo {author} {\bibfnamefont {O.~R.}\ \bibnamefont
  {Meitei}}, \bibinfo {author} {\bibfnamefont {B.~T.}\ \bibnamefont {Gard}},
  \bibinfo {author} {\bibfnamefont {G.~S.}\ \bibnamefont {Barron}}, \bibinfo
  {author} {\bibfnamefont {D.~P.}\ \bibnamefont {Pappas}}, \bibinfo {author}
  {\bibfnamefont {S.~E.}\ \bibnamefont {Economou}}, \bibinfo {author}
  {\bibfnamefont {E.}~\bibnamefont {Barnes}}, \ and\ \bibinfo {author}
  {\bibfnamefont {N.~J.}\ \bibnamefont {Mayhall}},\ }\href
  {http://arxiv.org/abs/2008.04302} {\bibfield  {journal} {\bibinfo  {journal}
  {arXiv:2008.04302}\ } (\bibinfo {year} {2020})},\ \bibinfo {note} {arXiv:
  2008.04302}\BibitemShut {NoStop}%
\bibitem [{\citenamefont {Zhang}\ \emph {et~al.}(2020)\citenamefont {Zhang},
  \citenamefont {Kyaw}, \citenamefont {Kottmann}, \citenamefont {Degroote},\
  and\ \citenamefont {Aspuru-Guzik}}]{Zhang2020_Mutual}%
  \BibitemOpen
  \bibfield  {author} {\bibinfo {author} {\bibfnamefont {Z.-J.}\ \bibnamefont
  {Zhang}}, \bibinfo {author} {\bibfnamefont {T.~H.}\ \bibnamefont {Kyaw}},
  \bibinfo {author} {\bibfnamefont {J.~S.}\ \bibnamefont {Kottmann}}, \bibinfo
  {author} {\bibfnamefont {M.}~\bibnamefont {Degroote}}, \ and\ \bibinfo
  {author} {\bibfnamefont {A.}~\bibnamefont {Aspuru-Guzik}},\ }\href
  {http://arxiv.org/abs/2008.07553} {\bibfield  {journal} {\bibinfo  {journal}
  {arXiv:2008.07553 [quant-ph]}\ } (\bibinfo {year} {2020})},\ \bibinfo {note}
  {arXiv: 2008.07553}\BibitemShut {NoStop}%
\bibitem [{\citenamefont {{M. Reimpell and R. F.
  Werner}}(2005)}]{ReimpellW:05}%
  \BibitemOpen
  \bibfield  {author} {\bibinfo {author} {\bibnamefont {{M. Reimpell and R. F.
  Werner}}},\ }\href
  {https://journals.aps.org/prl/abstract/10.1103/PhysRevLett.94.080501}
  {\bibfield  {journal} {\bibinfo  {journal} {Phys. Rev. Lett.}\ }\textbf
  {\bibinfo {volume} {94}},\ \bibinfo {pages} {080501} (\bibinfo {year}
  {2005})}\BibitemShut {NoStop}%
\bibitem [{\citenamefont {Fletcher}\ \emph {et~al.}(2007)\citenamefont
  {Fletcher}, \citenamefont {Shor},\ and\ \citenamefont {Win}}]{FletcherSW:06}%
  \BibitemOpen
  \bibfield  {author} {\bibinfo {author} {\bibfnamefont {A.~S.}\ \bibnamefont
  {Fletcher}}, \bibinfo {author} {\bibfnamefont {P.~W.}\ \bibnamefont {Shor}},
  \ and\ \bibinfo {author} {\bibfnamefont {M.~Z.}\ \bibnamefont {Win}},\ }\href
  {https://journals.aps.org/pra/abstract/10.1103/PhysRevA.75.012338} {\bibfield
   {journal} {\bibinfo  {journal} {Phys. Rev. A}\ }\textbf {\bibinfo {volume}
  {75}},\ \bibinfo {pages} {012338} (\bibinfo {year} {2007})}\BibitemShut
  {NoStop}%
\bibitem [{\citenamefont {Kosut}\ \emph {et~al.}(2008)\citenamefont {Kosut},
  \citenamefont {Shabani},\ and\ \citenamefont {Lidar}}]{KosutSL:08a}%
  \BibitemOpen
  \bibfield  {author} {\bibinfo {author} {\bibfnamefont {R.~L.}\ \bibnamefont
  {Kosut}}, \bibinfo {author} {\bibfnamefont {A.}~\bibnamefont {Shabani}}, \
  and\ \bibinfo {author} {\bibfnamefont {D.~A.}\ \bibnamefont {Lidar}},\ }\href
  {https://journals.aps.org/prl/abstract/10.1103/PhysRevLett.100.020502}
  {\bibfield  {journal} {\bibinfo  {journal} {Phys. Rev. Lett.}\ }\textbf
  {\bibinfo {volume} {100}},\ \bibinfo {pages} {020502} (\bibinfo {year}
  {2008})}\BibitemShut {NoStop}%
\bibitem [{\citenamefont {{R. L. Kosut and D. A. Lidar}}(2009)}]{KosutL:06}%
  \BibitemOpen
  \bibfield  {author} {\bibinfo {author} {\bibnamefont {{R. L. Kosut and D. A.
  Lidar}}},\ }\href
  {https://link.springer.com/article/10.1007/s11128-009-0120-2} {\bibfield
  {journal} {\bibinfo  {journal} {Quantum Inf. Process.}\ }\textbf {\bibinfo
  {volume} {8}},\ \bibinfo {pages} {443} (\bibinfo {year} {2009})}\BibitemShut
  {NoStop}%
\bibitem [{\citenamefont {Taghavi}\ \emph {et~al.}(2010)\citenamefont
  {Taghavi}, \citenamefont {Kosut},\ and\ \citenamefont
  {Lidar}}]{TaghaviKL:10}%
  \BibitemOpen
  \bibfield  {author} {\bibinfo {author} {\bibfnamefont {S.}~\bibnamefont
  {Taghavi}}, \bibinfo {author} {\bibfnamefont {R.~L.}\ \bibnamefont {Kosut}},
  \ and\ \bibinfo {author} {\bibfnamefont {D.~A.}\ \bibnamefont {Lidar}},\
  }\href {https://ieeexplore.ieee.org/abstract/document/5429112} {\bibfield
  {journal} {\bibinfo  {journal} {IEEE Trans. Inf. Theory}\ }\textbf {\bibinfo
  {volume} {56}},\ \bibinfo {pages} {1461} (\bibinfo {year}
  {2010})}\BibitemShut {NoStop}%
\bibitem [{\citenamefont {Suzuki}(1991)}]{suzuki1991general}%
  \BibitemOpen
  \bibfield  {author} {\bibinfo {author} {\bibfnamefont {M.}~\bibnamefont
  {Suzuki}},\ }\href {https://aip.scitation.org/doi/10.1063/1.529425}
  {\bibfield  {journal} {\bibinfo  {journal} {J. Math. Phys.}\ }\textbf
  {\bibinfo {volume} {32}},\ \bibinfo {pages} {400} (\bibinfo {year}
  {1991})}\BibitemShut {NoStop}%
\bibitem [{\citenamefont {Berry}\ and\ \citenamefont
  {Childs}(2009)}]{berry2009black}%
  \BibitemOpen
  \bibfield  {author} {\bibinfo {author} {\bibfnamefont {D.~W.}\ \bibnamefont
  {Berry}}\ and\ \bibinfo {author} {\bibfnamefont {A.~M.}\ \bibnamefont
  {Childs}},\ }\href {https://arxiv.org/abs/0910.4157} {\bibfield  {journal}
  {\bibinfo  {journal} {arXiv preprint arXiv:0910.4157}\ } (\bibinfo {year}
  {2009})}\BibitemShut {NoStop}%
\bibitem [{\citenamefont {Berry}\ \emph {et~al.}(2015)\citenamefont {Berry},
  \citenamefont {Childs}, \citenamefont {Cleve}, \citenamefont {Kothari},\ and\
  \citenamefont {Somma}}]{berry2015simulating}%
  \BibitemOpen
  \bibfield  {author} {\bibinfo {author} {\bibfnamefont {D.~W.}\ \bibnamefont
  {Berry}}, \bibinfo {author} {\bibfnamefont {A.~M.}\ \bibnamefont {Childs}},
  \bibinfo {author} {\bibfnamefont {R.}~\bibnamefont {Cleve}}, \bibinfo
  {author} {\bibfnamefont {R.}~\bibnamefont {Kothari}}, \ and\ \bibinfo
  {author} {\bibfnamefont {R.~D.}\ \bibnamefont {Somma}},\ }\href
  {https://journals.aps.org/prl/abstract/10.1103/PhysRevLett.114.090502}
  {\bibfield  {journal} {\bibinfo  {journal} {Phys. Rev. Lett.}\ }\textbf
  {\bibinfo {volume} {114}},\ \bibinfo {pages} {090502} (\bibinfo {year}
  {2015})}\BibitemShut {NoStop}%
\bibitem [{\citenamefont {Low}\ and\ \citenamefont
  {Chuang}(2017)}]{low2017optimal}%
  \BibitemOpen
  \bibfield  {author} {\bibinfo {author} {\bibfnamefont {G.~H.}\ \bibnamefont
  {Low}}\ and\ \bibinfo {author} {\bibfnamefont {I.~L.}\ \bibnamefont
  {Chuang}},\ }\href
  {https://journals.aps.org/prl/abstract/10.1103/PhysRevLett.118.010501}
  {\bibfield  {journal} {\bibinfo  {journal} {Phys. Rev. Lett.}\ }\textbf
  {\bibinfo {volume} {118}},\ \bibinfo {pages} {010501} (\bibinfo {year}
  {2017})}\BibitemShut {NoStop}%
\bibitem [{\citenamefont {Magann}\ \emph {et~al.}(2020)\citenamefont {Magann},
  \citenamefont {Grace}, \citenamefont {Rabitz},\ and\ \citenamefont
  {Sarovar}}]{Magann2020_Digital}%
  \BibitemOpen
  \bibfield  {author} {\bibinfo {author} {\bibfnamefont {A.~B.}\ \bibnamefont
  {Magann}}, \bibinfo {author} {\bibfnamefont {M.~D.}\ \bibnamefont {Grace}},
  \bibinfo {author} {\bibfnamefont {H.~A.}\ \bibnamefont {Rabitz}}, \ and\
  \bibinfo {author} {\bibfnamefont {M.}~\bibnamefont {Sarovar}},\ }\href
  {http://arxiv.org/abs/2002.12497} {\bibfield  {journal} {\bibinfo  {journal}
  {arXiv:2002.12497 [quant-ph]}\ } (\bibinfo {year} {2020})},\ \bibinfo {note}
  {arXiv: 2002.12497}\BibitemShut {NoStop}%
\bibitem [{\citenamefont {Rabitz}\ \emph {et~al.}(2004)\citenamefont {Rabitz},
  \citenamefont {Hsieh},\ and\ \citenamefont {Rosenthal}}]{rabitz2004quantum}%
  \BibitemOpen
  \bibfield  {author} {\bibinfo {author} {\bibfnamefont {H.~A.}\ \bibnamefont
  {Rabitz}}, \bibinfo {author} {\bibfnamefont {M.~M.}\ \bibnamefont {Hsieh}}, \
  and\ \bibinfo {author} {\bibfnamefont {C.~M.}\ \bibnamefont {Rosenthal}},\
  }\href {https://science.sciencemag.org/content/303/5666/1998} {\bibfield
  {journal} {\bibinfo  {journal} {Science}\ }\textbf {\bibinfo {volume}
  {303}},\ \bibinfo {pages} {1998} (\bibinfo {year} {2004})}\BibitemShut
  {NoStop}%
\bibitem [{\citenamefont {Rabitz}\ \emph
  {et~al.}(2006{\natexlab{a}})\citenamefont {Rabitz}, \citenamefont {Ho},
  \citenamefont {Hsieh}, \citenamefont {Kosut},\ and\ \citenamefont
  {Demiralp}}]{rabitz2006topology}%
  \BibitemOpen
  \bibfield  {author} {\bibinfo {author} {\bibfnamefont {H.}~\bibnamefont
  {Rabitz}}, \bibinfo {author} {\bibfnamefont {T.-S.}\ \bibnamefont {Ho}},
  \bibinfo {author} {\bibfnamefont {M.}~\bibnamefont {Hsieh}}, \bibinfo
  {author} {\bibfnamefont {R.}~\bibnamefont {Kosut}}, \ and\ \bibinfo {author}
  {\bibfnamefont {M.}~\bibnamefont {Demiralp}},\ }\href
  {https://journals.aps.org/pra/abstract/10.1103/PhysRevA.74.012721} {\bibfield
   {journal} {\bibinfo  {journal} {Phys. Rev. A}\ }\textbf {\bibinfo {volume}
  {74}},\ \bibinfo {pages} {012721} (\bibinfo {year}
  {2006}{\natexlab{a}})}\BibitemShut {NoStop}%
\bibitem [{\citenamefont {Ho}\ and\ \citenamefont {Rabitz}(2006)}]{HO2006226}%
  \BibitemOpen
  \bibfield  {author} {\bibinfo {author} {\bibfnamefont {T.-S.}\ \bibnamefont
  {Ho}}\ and\ \bibinfo {author} {\bibfnamefont {H.}~\bibnamefont {Rabitz}},\
  }\href {\doibase https://doi.org/10.1016/j.jphotochem.2006.03.038} {\bibfield
   {journal} {\bibinfo  {journal} {J. Photochem. Photobiol. A}\ }\textbf
  {\bibinfo {volume} {180}},\ \bibinfo {pages} {226 } (\bibinfo {year}
  {2006})}\BibitemShut {NoStop}%
\bibitem [{\citenamefont {Ho}\ \emph {et~al.}(2009)\citenamefont {Ho},
  \citenamefont {Dominy},\ and\ \citenamefont {Rabitz}}]{ho2009landscape}%
  \BibitemOpen
  \bibfield  {author} {\bibinfo {author} {\bibfnamefont {T.-S.}\ \bibnamefont
  {Ho}}, \bibinfo {author} {\bibfnamefont {J.}~\bibnamefont {Dominy}}, \ and\
  \bibinfo {author} {\bibfnamefont {H.}~\bibnamefont {Rabitz}},\ }\href
  {https://journals.aps.org/pra/abstract/10.1103/PhysRevA.79.013422} {\bibfield
   {journal} {\bibinfo  {journal} {Phys. Rev. A}\ }\textbf {\bibinfo {volume}
  {79}},\ \bibinfo {pages} {013422} (\bibinfo {year} {2009})}\BibitemShut
  {NoStop}%
\bibitem [{\citenamefont {Pechen}\ and\ \citenamefont
  {Tannor}(2011)}]{pechen2011there}%
  \BibitemOpen
  \bibfield  {author} {\bibinfo {author} {\bibfnamefont {A.~N.}\ \bibnamefont
  {Pechen}}\ and\ \bibinfo {author} {\bibfnamefont {D.~J.}\ \bibnamefont
  {Tannor}},\ }\href
  {https://journals.aps.org/prl/abstract/10.1103/PhysRevLett.106.120402}
  {\bibfield  {journal} {\bibinfo  {journal} {Phys. Rev. Lett.}\ }\textbf
  {\bibinfo {volume} {106}},\ \bibinfo {pages} {120402} (\bibinfo {year}
  {2011})}\BibitemShut {NoStop}%
\bibitem [{\citenamefont {De~Fouquieres}\ and\ \citenamefont
  {Schirmer}(2013)}]{de2013closer}%
  \BibitemOpen
  \bibfield  {author} {\bibinfo {author} {\bibfnamefont {P.}~\bibnamefont
  {De~Fouquieres}}\ and\ \bibinfo {author} {\bibfnamefont {S.~G.}\ \bibnamefont
  {Schirmer}},\ }\href
  {https://www.worldscientific.com/doi/abs/10.1142/S0219025713500215}
  {\bibfield  {journal} {\bibinfo  {journal} {Infin. Dimens. Anal. Quantum
  Probab. Relat. Top.}\ }\textbf {\bibinfo {volume} {16}},\ \bibinfo {pages}
  {1350021} (\bibinfo {year} {2013})}\BibitemShut {NoStop}%
\bibitem [{\citenamefont {Chakrabarti}\ and\ \citenamefont
  {Rabitz}(2007)}]{chakrabarti2007quantum}%
  \BibitemOpen
  \bibfield  {author} {\bibinfo {author} {\bibfnamefont {R.}~\bibnamefont
  {Chakrabarti}}\ and\ \bibinfo {author} {\bibfnamefont {H.}~\bibnamefont
  {Rabitz}},\ }\href {\doibase 10.1080/01442350701633300} {\bibfield  {journal}
  {\bibinfo  {journal} {Int. Rev. Phys. Chem.}\ }\textbf {\bibinfo {volume}
  {26}},\ \bibinfo {pages} {671} (\bibinfo {year} {2007})}\BibitemShut
  {NoStop}%
\bibitem [{\citenamefont {Rabitz}\ \emph
  {et~al.}(2006{\natexlab{b}})\citenamefont {Rabitz}, \citenamefont {Hsieh},\
  and\ \citenamefont {Rosenthal}}]{rabitz2006optimal}%
  \BibitemOpen
  \bibfield  {author} {\bibinfo {author} {\bibfnamefont {H.}~\bibnamefont
  {Rabitz}}, \bibinfo {author} {\bibfnamefont {M.}~\bibnamefont {Hsieh}}, \
  and\ \bibinfo {author} {\bibfnamefont {C.}~\bibnamefont {Rosenthal}},\ }\href
  {\doibase 10.1063/1.2198837} {\bibfield  {journal} {\bibinfo  {journal} {J.
  Chem. Phys.}\ }\textbf {\bibinfo {volume} {124}},\ \bibinfo {pages} {204107}
  (\bibinfo {year} {2006}{\natexlab{b}})}\BibitemShut {NoStop}%
\bibitem [{\citenamefont {Hsieh}\ and\ \citenamefont
  {Rabitz}(2008)}]{hsieh2008optimal}%
  \BibitemOpen
  \bibfield  {author} {\bibinfo {author} {\bibfnamefont {M.}~\bibnamefont
  {Hsieh}}\ and\ \bibinfo {author} {\bibfnamefont {H.}~\bibnamefont {Rabitz}},\
  }\href {https://journals.aps.org/pra/abstract/10.1103/PhysRevA.77.042306}
  {\bibfield  {journal} {\bibinfo  {journal} {Phys. Rev. A}\ }\textbf {\bibinfo
  {volume} {77}},\ \bibinfo {pages} {042306} (\bibinfo {year}
  {2008})}\BibitemShut {NoStop}%
\bibitem [{\citenamefont {Rabitz}\ \emph {et~al.}(2012)\citenamefont {Rabitz},
  \citenamefont {Ho}, \citenamefont {Long}, \citenamefont {Wu},\ and\
  \citenamefont {Brif}}]{PhysRevLett.108.198901}%
  \BibitemOpen
  \bibfield  {author} {\bibinfo {author} {\bibfnamefont {H.}~\bibnamefont
  {Rabitz}}, \bibinfo {author} {\bibfnamefont {T.-S.}\ \bibnamefont {Ho}},
  \bibinfo {author} {\bibfnamefont {R.}~\bibnamefont {Long}}, \bibinfo {author}
  {\bibfnamefont {R.}~\bibnamefont {Wu}}, \ and\ \bibinfo {author}
  {\bibfnamefont {C.}~\bibnamefont {Brif}},\ }\href
  {https://link.aps.org/doi/10.1103/PhysRevLett.108.198901} {\bibfield
  {journal} {\bibinfo  {journal} {Phys. Rev. Lett.}\ }\textbf {\bibinfo
  {volume} {108}},\ \bibinfo {pages} {198901} (\bibinfo {year}
  {2012})}\BibitemShut {NoStop}%
\bibitem [{\citenamefont {Pechen}\ and\ \citenamefont
  {Tannor}(2012)}]{PhysRevLett.108.198902}%
  \BibitemOpen
  \bibfield  {author} {\bibinfo {author} {\bibfnamefont {A.~N.}\ \bibnamefont
  {Pechen}}\ and\ \bibinfo {author} {\bibfnamefont {D.~J.}\ \bibnamefont
  {Tannor}},\ }\href {https://link.aps.org/doi/10.1103/PhysRevLett.108.198902}
  {\bibfield  {journal} {\bibinfo  {journal} {Phys. Rev. Lett.}\ }\textbf
  {\bibinfo {volume} {108}},\ \bibinfo {pages} {198902} (\bibinfo {year}
  {2012})}\BibitemShut {NoStop}%
\bibitem [{\citenamefont {Moore~Tibbetts}\ \emph {et~al.}(2012)\citenamefont
  {Moore~Tibbetts}, \citenamefont {Brif}, \citenamefont {Grace}, \citenamefont
  {Donovan}, \citenamefont {Hocker}, \citenamefont {Ho}, \citenamefont {Wu},\
  and\ \citenamefont {Rabitz}}]{Moore-Tibbetts2012Exploring}%
  \BibitemOpen
  \bibfield  {author} {\bibinfo {author} {\bibfnamefont {K.~W.}\ \bibnamefont
  {Moore~Tibbetts}}, \bibinfo {author} {\bibfnamefont {C.}~\bibnamefont
  {Brif}}, \bibinfo {author} {\bibfnamefont {M.~D.}\ \bibnamefont {Grace}},
  \bibinfo {author} {\bibfnamefont {A.}~\bibnamefont {Donovan}}, \bibinfo
  {author} {\bibfnamefont {D.~L.}\ \bibnamefont {Hocker}}, \bibinfo {author}
  {\bibfnamefont {T.-S.}\ \bibnamefont {Ho}}, \bibinfo {author} {\bibfnamefont
  {R.-B.}\ \bibnamefont {Wu}}, \ and\ \bibinfo {author} {\bibfnamefont
  {H.}~\bibnamefont {Rabitz}},\ }\href {\doibase 10.1103/PhysRevA.86.062309}
  {\bibfield  {journal} {\bibinfo  {journal} {Phys. Rev. A}\ }\textbf {\bibinfo
  {volume} {86}},\ \bibinfo {pages} {062309} (\bibinfo {year}
  {2012})}\BibitemShut {NoStop}%
\bibitem [{\citenamefont {Riviello}\ \emph {et~al.}(2014)\citenamefont
  {Riviello}, \citenamefont {Brif}, \citenamefont {Long}, \citenamefont {Wu},
  \citenamefont {Tibbetts}, \citenamefont {Ho},\ and\ \citenamefont
  {Rabitz}}]{riviello2014searching}%
  \BibitemOpen
  \bibfield  {author} {\bibinfo {author} {\bibfnamefont {G.}~\bibnamefont
  {Riviello}}, \bibinfo {author} {\bibfnamefont {C.}~\bibnamefont {Brif}},
  \bibinfo {author} {\bibfnamefont {R.}~\bibnamefont {Long}}, \bibinfo {author}
  {\bibfnamefont {R.-B.}\ \bibnamefont {Wu}}, \bibinfo {author} {\bibfnamefont
  {K.~M.}\ \bibnamefont {Tibbetts}}, \bibinfo {author} {\bibfnamefont {T.-S.}\
  \bibnamefont {Ho}}, \ and\ \bibinfo {author} {\bibfnamefont {H.}~\bibnamefont
  {Rabitz}},\ }\href
  {https://journals.aps.org/pra/abstract/10.1103/PhysRevA.90.013404} {\bibfield
   {journal} {\bibinfo  {journal} {Phys. Rev. A}\ }\textbf {\bibinfo {volume}
  {90}},\ \bibinfo {pages} {013404} (\bibinfo {year} {2014})}\BibitemShut
  {NoStop}%
\bibitem [{\citenamefont {Russell}\ \emph {et~al.}(2017)\citenamefont
  {Russell}, \citenamefont {Rabitz},\ and\ \citenamefont
  {Wu}}]{russell2017control}%
  \BibitemOpen
  \bibfield  {author} {\bibinfo {author} {\bibfnamefont {B.}~\bibnamefont
  {Russell}}, \bibinfo {author} {\bibfnamefont {H.}~\bibnamefont {Rabitz}}, \
  and\ \bibinfo {author} {\bibfnamefont {R.-B.}\ \bibnamefont {Wu}},\ }\href
  {https://iopscience.iop.org/article/10.1088/1751-8121/aa6b77} {\bibfield
  {journal} {\bibinfo  {journal} {J. Phys. A}\ }\textbf {\bibinfo {volume}
  {50}},\ \bibinfo {pages} {205302} (\bibinfo {year} {2017})}\BibitemShut
  {NoStop}%
\bibitem [{\citenamefont {Zhdanov}(2018)}]{Zhdanov_2018}%
  \BibitemOpen
  \bibfield  {author} {\bibinfo {author} {\bibfnamefont {D.~V.}\ \bibnamefont
  {Zhdanov}},\ }\href
  {https://iopscience.iop.org/article/10.1088/1751-8121/aaecf6/meta} {\bibfield
   {journal} {\bibinfo  {journal} {J. Phys. A: Math. Theor.}\ }\textbf
  {\bibinfo {volume} {51}},\ \bibinfo {pages} {508001} (\bibinfo {year}
  {2018})}\BibitemShut {NoStop}%
\bibitem [{\citenamefont {Russell}\ \emph {et~al.}(2018)\citenamefont
  {Russell}, \citenamefont {Wu},\ and\ \citenamefont {Rabitz}}]{Russell_2018}%
  \BibitemOpen
  \bibfield  {author} {\bibinfo {author} {\bibfnamefont {B.}~\bibnamefont
  {Russell}}, \bibinfo {author} {\bibfnamefont {R.}~\bibnamefont {Wu}}, \ and\
  \bibinfo {author} {\bibfnamefont {H.}~\bibnamefont {Rabitz}},\ }\href
  {https://iopscience.iop.org/article/10.1088/1751-8121/aaecf2} {\bibfield
  {journal} {\bibinfo  {journal} {J. Phys. A: Math. Theor.}\ }\textbf {\bibinfo
  {volume} {51}},\ \bibinfo {pages} {508002} (\bibinfo {year}
  {2018})}\BibitemShut {NoStop}%
\bibitem [{\citenamefont {{Arenz}}\ and\ \citenamefont
  {{Rabitz}}(2020)}]{2020arXiv200402729A}%
  \BibitemOpen
  \bibfield  {author} {\bibinfo {author} {\bibfnamefont {C.}~\bibnamefont
  {{Arenz}}}\ and\ \bibinfo {author} {\bibfnamefont {H.}~\bibnamefont
  {{Rabitz}}},\ }\href@noop {} {\bibfield  {journal} {\bibinfo  {journal}
  {arXiv e-prints}\ ,\ \bibinfo {eid} {arXiv:2004.02729}} (\bibinfo {year}
  {2020})},\ \Eprint {http://arxiv.org/abs/2004.02729} {arXiv:2004.02729
  [quant-ph]} \BibitemShut {NoStop}%
\bibitem [{\citenamefont {Banchi}\ \emph {et~al.}(2017)\citenamefont {Banchi},
  \citenamefont {Burgarth},\ and\ \citenamefont
  {Kastoryano}}]{banchi2017driven}%
  \BibitemOpen
  \bibfield  {author} {\bibinfo {author} {\bibfnamefont {L.}~\bibnamefont
  {Banchi}}, \bibinfo {author} {\bibfnamefont {D.}~\bibnamefont {Burgarth}}, \
  and\ \bibinfo {author} {\bibfnamefont {M.~J.}\ \bibnamefont {Kastoryano}},\
  }\href {https://journals.aps.org/prx/abstract/10.1103/PhysRevX.7.041015}
  {\bibfield  {journal} {\bibinfo  {journal} {Phys. Rev. X}\ }\textbf {\bibinfo
  {volume} {7}},\ \bibinfo {pages} {041015} (\bibinfo {year}
  {2017})}\BibitemShut {NoStop}%
\bibitem [{\citenamefont {Moore}\ and\ \citenamefont
  {Rabitz}(2011)}]{moore2011exploring}%
  \BibitemOpen
  \bibfield  {author} {\bibinfo {author} {\bibfnamefont {K.~W.}\ \bibnamefont
  {Moore}}\ and\ \bibinfo {author} {\bibfnamefont {H.}~\bibnamefont {Rabitz}},\
  }\href {https://journals.aps.org/pra/abstract/10.1103/PhysRevA.84.012109}
  {\bibfield  {journal} {\bibinfo  {journal} {Phys. Rev. A}\ }\textbf {\bibinfo
  {volume} {84}},\ \bibinfo {pages} {012109} (\bibinfo {year}
  {2011})}\BibitemShut {NoStop}%
\bibitem [{\citenamefont {Hocker}\ \emph {et~al.}(2014)\citenamefont {Hocker},
  \citenamefont {Brif}, \citenamefont {Grace}, \citenamefont {Donovan},
  \citenamefont {Ho}, \citenamefont {Tibbetts}, \citenamefont {Wu},\ and\
  \citenamefont {Rabitz}}]{Hocker2014_Characterization}%
  \BibitemOpen
  \bibfield  {author} {\bibinfo {author} {\bibfnamefont {D.}~\bibnamefont
  {Hocker}}, \bibinfo {author} {\bibfnamefont {C.}~\bibnamefont {Brif}},
  \bibinfo {author} {\bibfnamefont {M.~D.}\ \bibnamefont {Grace}}, \bibinfo
  {author} {\bibfnamefont {A.}~\bibnamefont {Donovan}}, \bibinfo {author}
  {\bibfnamefont {T.-S.}\ \bibnamefont {Ho}}, \bibinfo {author} {\bibfnamefont
  {K.~M.}\ \bibnamefont {Tibbetts}}, \bibinfo {author} {\bibfnamefont
  {R.}~\bibnamefont {Wu}}, \ and\ \bibinfo {author} {\bibfnamefont
  {H.}~\bibnamefont {Rabitz}},\ }\href {\doibase 10.1103/PhysRevA.90.062309}
  {\bibfield  {journal} {\bibinfo  {journal} {Phys. Rev. A}\ }\textbf {\bibinfo
  {volume} {90}},\ \bibinfo {pages} {062309} (\bibinfo {year}
  {2014})}\BibitemShut {NoStop}%
\bibitem [{\citenamefont {Kerschke}(2017)}]{Kerschke2017_FLACCO}%
  \BibitemOpen
  \bibfield  {author} {\bibinfo {author} {\bibfnamefont {P.}~\bibnamefont
  {Kerschke}},\ }\href {http://arxiv.org/abs/1708.05258} {\bibfield  {journal}
  {\bibinfo  {journal} {arXiv:1708.05258 [stat.ML]}\ } (\bibinfo {year}
  {2017})}\BibitemShut {NoStop}%
\bibitem [{\citenamefont {Riviello}\ \emph {et~al.}(2015)\citenamefont
  {Riviello}, \citenamefont {Moore-Tibbetts}, \citenamefont {Brif},
  \citenamefont {Long}, \citenamefont {Wu}, \citenamefont {Ho},\ and\
  \citenamefont {Rabitz}}]{Riviello2015_Searching}%
  \BibitemOpen
  \bibfield  {author} {\bibinfo {author} {\bibfnamefont {G.}~\bibnamefont
  {Riviello}}, \bibinfo {author} {\bibfnamefont {K.}~\bibnamefont
  {Moore-Tibbetts}}, \bibinfo {author} {\bibfnamefont {C.}~\bibnamefont
  {Brif}}, \bibinfo {author} {\bibfnamefont {R.}~\bibnamefont {Long}}, \bibinfo
  {author} {\bibfnamefont {R.-B.}\ \bibnamefont {Wu}}, \bibinfo {author}
  {\bibfnamefont {T.-S.}\ \bibnamefont {Ho}}, \ and\ \bibinfo {author}
  {\bibfnamefont {H.}~\bibnamefont {Rabitz}},\ }\href {\doibase
  10.1103/PhysRevA.91.043401} {\bibfield  {journal} {\bibinfo  {journal} {Phys.
  Rev. A}\ }\textbf {\bibinfo {volume} {91}},\ \bibinfo {pages} {043401}
  (\bibinfo {year} {2015})}\BibitemShut {NoStop}%
\bibitem [{\citenamefont {Volkoff}\ and\ \citenamefont
  {Coles}(2020)}]{Volkoff2020_Large}%
  \BibitemOpen
  \bibfield  {author} {\bibinfo {author} {\bibfnamefont {T.}~\bibnamefont
  {Volkoff}}\ and\ \bibinfo {author} {\bibfnamefont {P.~J.}\ \bibnamefont
  {Coles}},\ }\href {http://arxiv.org/abs/2005.12200} {\bibfield  {journal}
  {\bibinfo  {journal} {arXiv:2005.12200 [quant-ph]}\ } (\bibinfo {year}
  {2020})},\ \bibinfo {note} {arXiv: 2005.12200}\BibitemShut {NoStop}%
\bibitem [{\citenamefont {Carlini}\ \emph {et~al.}(2006)\citenamefont
  {Carlini}, \citenamefont {Hosoya}, \citenamefont {Koike},\ and\ \citenamefont
  {Okudaira}}]{carlini2006time}%
  \BibitemOpen
  \bibfield  {author} {\bibinfo {author} {\bibfnamefont {A.}~\bibnamefont
  {Carlini}}, \bibinfo {author} {\bibfnamefont {A.}~\bibnamefont {Hosoya}},
  \bibinfo {author} {\bibfnamefont {T.}~\bibnamefont {Koike}}, \ and\ \bibinfo
  {author} {\bibfnamefont {Y.}~\bibnamefont {Okudaira}},\ }\href
  {https://journals.aps.org/prl/abstract/10.1103/PhysRevLett.96.060503}
  {\bibfield  {journal} {\bibinfo  {journal} {Phys. Rev. Lett.}\ }\textbf
  {\bibinfo {volume} {96}},\ \bibinfo {pages} {060503} (\bibinfo {year}
  {2006})}\BibitemShut {NoStop}%
\bibitem [{\citenamefont {Nielsen}\ \emph
  {et~al.}(2006{\natexlab{a}})\citenamefont {Nielsen}, \citenamefont {Dowling},
  \citenamefont {Gu},\ and\ \citenamefont {Doherty}}]{nielsen2006quantum}%
  \BibitemOpen
  \bibfield  {author} {\bibinfo {author} {\bibfnamefont {M.~A.}\ \bibnamefont
  {Nielsen}}, \bibinfo {author} {\bibfnamefont {M.~R.}\ \bibnamefont
  {Dowling}}, \bibinfo {author} {\bibfnamefont {M.}~\bibnamefont {Gu}}, \ and\
  \bibinfo {author} {\bibfnamefont {A.~C.}\ \bibnamefont {Doherty}},\ }\href
  {https://science.sciencemag.org/content/311/5764/1133} {\bibfield  {journal}
  {\bibinfo  {journal} {Science}\ }\textbf {\bibinfo {volume} {311}},\ \bibinfo
  {pages} {1133} (\bibinfo {year} {2006}{\natexlab{a}})}\BibitemShut {NoStop}%
\bibitem [{\citenamefont {Nielsen}\ \emph
  {et~al.}(2006{\natexlab{b}})\citenamefont {Nielsen}, \citenamefont {Dowling},
  \citenamefont {Gu},\ and\ \citenamefont {Doherty}}]{nielsen2006optimal}%
  \BibitemOpen
  \bibfield  {author} {\bibinfo {author} {\bibfnamefont {M.~A.}\ \bibnamefont
  {Nielsen}}, \bibinfo {author} {\bibfnamefont {M.~R.}\ \bibnamefont
  {Dowling}}, \bibinfo {author} {\bibfnamefont {M.}~\bibnamefont {Gu}}, \ and\
  \bibinfo {author} {\bibfnamefont {A.~C.}\ \bibnamefont {Doherty}},\ }\href
  {https://journals.aps.org/pra/abstract/10.1103/PhysRevA.73.062323} {\bibfield
   {journal} {\bibinfo  {journal} {Phys. Rev. A}\ }\textbf {\bibinfo {volume}
  {73}},\ \bibinfo {pages} {062323} (\bibinfo {year}
  {2006}{\natexlab{b}})}\BibitemShut {NoStop}%
\bibitem [{\citenamefont {Carlini}\ \emph {et~al.}(2007)\citenamefont
  {Carlini}, \citenamefont {Hosoya}, \citenamefont {Koike},\ and\ \citenamefont
  {Okudaira}}]{carlini2007time}%
  \BibitemOpen
  \bibfield  {author} {\bibinfo {author} {\bibfnamefont {A.}~\bibnamefont
  {Carlini}}, \bibinfo {author} {\bibfnamefont {A.}~\bibnamefont {Hosoya}},
  \bibinfo {author} {\bibfnamefont {T.}~\bibnamefont {Koike}}, \ and\ \bibinfo
  {author} {\bibfnamefont {Y.}~\bibnamefont {Okudaira}},\ }\href
  {https://journals.aps.org/pra/abstract/10.1103/PhysRevA.75.042308} {\bibfield
   {journal} {\bibinfo  {journal} {Phys. Rev. A}\ }\textbf {\bibinfo {volume}
  {75}},\ \bibinfo {pages} {042308} (\bibinfo {year} {2007})}\BibitemShut
  {NoStop}%
\bibitem [{\citenamefont {Wang}\ \emph {et~al.}(2015)\citenamefont {Wang},
  \citenamefont {Allegra}, \citenamefont {Jacobs}, \citenamefont {Lloyd},
  \citenamefont {Lupo},\ and\ \citenamefont {Mohseni}}]{wang2015quantum}%
  \BibitemOpen
  \bibfield  {author} {\bibinfo {author} {\bibfnamefont {X.}~\bibnamefont
  {Wang}}, \bibinfo {author} {\bibfnamefont {M.}~\bibnamefont {Allegra}},
  \bibinfo {author} {\bibfnamefont {K.}~\bibnamefont {Jacobs}}, \bibinfo
  {author} {\bibfnamefont {S.}~\bibnamefont {Lloyd}}, \bibinfo {author}
  {\bibfnamefont {C.}~\bibnamefont {Lupo}}, \ and\ \bibinfo {author}
  {\bibfnamefont {M.}~\bibnamefont {Mohseni}},\ }\href
  {https://journals.aps.org/prl/abstract/10.1103/PhysRevLett.114.170501}
  {\bibfield  {journal} {\bibinfo  {journal} {Phys. Rev. Lett.}\ }\textbf
  {\bibinfo {volume} {114}},\ \bibinfo {pages} {170501} (\bibinfo {year}
  {2015})}\BibitemShut {NoStop}%
\bibitem [{\citenamefont {Albertini}\ and\ \citenamefont
  {D'Alessandro}(2015)}]{doi:10.1063/1.4906137}%
  \BibitemOpen
  \bibfield  {author} {\bibinfo {author} {\bibfnamefont {F.}~\bibnamefont
  {Albertini}}\ and\ \bibinfo {author} {\bibfnamefont {D.}~\bibnamefont
  {D'Alessandro}},\ }\href {https://doi.org/10.1063/1.4906137} {\bibfield
  {journal} {\bibinfo  {journal} {J. Math. Phys.}\ }\textbf {\bibinfo {volume}
  {56}},\ \bibinfo {pages} {012106} (\bibinfo {year} {2015})}\BibitemShut
  {NoStop}%
\bibitem [{\citenamefont {Romano}(2014)}]{PhysRevA.90.062302}%
  \BibitemOpen
  \bibfield  {author} {\bibinfo {author} {\bibfnamefont {R.}~\bibnamefont
  {Romano}},\ }\href {https://link.aps.org/doi/10.1103/PhysRevA.90.062302}
  {\bibfield  {journal} {\bibinfo  {journal} {Phys. Rev. A}\ }\textbf {\bibinfo
  {volume} {90}},\ \bibinfo {pages} {062302} (\bibinfo {year}
  {2014})}\BibitemShut {NoStop}%
\bibitem [{\citenamefont {Hegerfeldt}(2013)}]{PhysRevLett.111.260501}%
  \BibitemOpen
  \bibfield  {author} {\bibinfo {author} {\bibfnamefont {G.~C.}\ \bibnamefont
  {Hegerfeldt}},\ }\href
  {https://link.aps.org/doi/10.1103/PhysRevLett.111.260501} {\bibfield
  {journal} {\bibinfo  {journal} {Phys. Rev. Lett.}\ }\textbf {\bibinfo
  {volume} {111}},\ \bibinfo {pages} {260501} (\bibinfo {year}
  {2013})}\BibitemShut {NoStop}%
\bibitem [{\citenamefont {Khaneja}\ \emph {et~al.}(2001)\citenamefont
  {Khaneja}, \citenamefont {Brockett},\ and\ \citenamefont
  {Glaser}}]{khaneja2001time}%
  \BibitemOpen
  \bibfield  {author} {\bibinfo {author} {\bibfnamefont {N.}~\bibnamefont
  {Khaneja}}, \bibinfo {author} {\bibfnamefont {R.}~\bibnamefont {Brockett}}, \
  and\ \bibinfo {author} {\bibfnamefont {S.~J.}\ \bibnamefont {Glaser}},\
  }\href {https://journals.aps.org/pra/abstract/10.1103/PhysRevA.63.032308}
  {\bibfield  {journal} {\bibinfo  {journal} {Phys. Rev. A}\ }\textbf {\bibinfo
  {volume} {63}},\ \bibinfo {pages} {032308} (\bibinfo {year}
  {2001})}\BibitemShut {NoStop}%
\bibitem [{\citenamefont {Khaneja}\ \emph {et~al.}(2002)\citenamefont
  {Khaneja}, \citenamefont {Glaser},\ and\ \citenamefont
  {Brockett}}]{khaneja2002sub}%
  \BibitemOpen
  \bibfield  {author} {\bibinfo {author} {\bibfnamefont {N.}~\bibnamefont
  {Khaneja}}, \bibinfo {author} {\bibfnamefont {S.~J.}\ \bibnamefont {Glaser}},
  \ and\ \bibinfo {author} {\bibfnamefont {R.}~\bibnamefont {Brockett}},\
  }\href {https://journals.aps.org/pra/abstract/10.1103/PhysRevA.65.032301}
  {\bibfield  {journal} {\bibinfo  {journal} {Phys. Rev. A}\ }\textbf {\bibinfo
  {volume} {65}},\ \bibinfo {pages} {032301} (\bibinfo {year}
  {2002})}\BibitemShut {NoStop}%
\bibitem [{\citenamefont {Arenz}\ \emph {et~al.}(2017)\citenamefont {Arenz},
  \citenamefont {Russell}, \citenamefont {Burgarth},\ and\ \citenamefont
  {Rabitz}}]{arenz2017roles}%
  \BibitemOpen
  \bibfield  {author} {\bibinfo {author} {\bibfnamefont {C.}~\bibnamefont
  {Arenz}}, \bibinfo {author} {\bibfnamefont {B.}~\bibnamefont {Russell}},
  \bibinfo {author} {\bibfnamefont {D.}~\bibnamefont {Burgarth}}, \ and\
  \bibinfo {author} {\bibfnamefont {H.}~\bibnamefont {Rabitz}},\ }\href
  {https://iopscience.iop.org/article/10.1088/1367-2630/aa8242} {\bibfield
  {journal} {\bibinfo  {journal} {New J. Phys.}\ }\textbf {\bibinfo {volume}
  {19}},\ \bibinfo {pages} {103015} (\bibinfo {year} {2017})}\BibitemShut
  {NoStop}%
\bibitem [{\citenamefont {Lee}\ \emph {et~al.}(2018)\citenamefont {Lee},
  \citenamefont {Arenz}, \citenamefont {Rabitz},\ and\ \citenamefont
  {Russell}}]{lee2018dependence}%
  \BibitemOpen
  \bibfield  {author} {\bibinfo {author} {\bibfnamefont {J.}~\bibnamefont
  {Lee}}, \bibinfo {author} {\bibfnamefont {C.}~\bibnamefont {Arenz}}, \bibinfo
  {author} {\bibfnamefont {H.}~\bibnamefont {Rabitz}}, \ and\ \bibinfo {author}
  {\bibfnamefont {B.}~\bibnamefont {Russell}},\ }\href
  {https://iopscience.iop.org/article/10.1088/1367-2630/aac6f3/meta} {\bibfield
   {journal} {\bibinfo  {journal} {New J. Phys.}\ }\textbf {\bibinfo {volume}
  {20}},\ \bibinfo {pages} {063002} (\bibinfo {year} {2018})}\BibitemShut
  {NoStop}%
\bibitem [{\citenamefont {Arenz}\ and\ \citenamefont
  {Rabitz}(2018)}]{arenz2018controlling}%
  \BibitemOpen
  \bibfield  {author} {\bibinfo {author} {\bibfnamefont {C.}~\bibnamefont
  {Arenz}}\ and\ \bibinfo {author} {\bibfnamefont {H.}~\bibnamefont {Rabitz}},\
  }\href {https://journals.aps.org/prl/abstract/10.1103/PhysRevLett.120.220503}
  {\bibfield  {journal} {\bibinfo  {journal} {Phys. Rev. Lett.}\ }\textbf
  {\bibinfo {volume} {120}},\ \bibinfo {pages} {220503} (\bibinfo {year}
  {2018})}\BibitemShut {NoStop}%
\bibitem [{\citenamefont {Lee}\ \emph {et~al.}(2020)\citenamefont {Lee},
  \citenamefont {Arenz}, \citenamefont {Burgarth},\ and\ \citenamefont
  {Rabitz}}]{lee2020upper}%
  \BibitemOpen
  \bibfield  {author} {\bibinfo {author} {\bibfnamefont {J.}~\bibnamefont
  {Lee}}, \bibinfo {author} {\bibfnamefont {C.}~\bibnamefont {Arenz}}, \bibinfo
  {author} {\bibfnamefont {D.}~\bibnamefont {Burgarth}}, \ and\ \bibinfo
  {author} {\bibfnamefont {H.}~\bibnamefont {Rabitz}},\ }\href
  {https://iopscience.iop.org/article/10.1088/1751-8121/ab7498} {\bibfield
  {journal} {\bibinfo  {journal} {J. Phys. A}\ }\textbf {\bibinfo {volume}
  {53}},\ \bibinfo {pages} {125304} (\bibinfo {year} {2020})}\BibitemShut
  {NoStop}%
\bibitem [{\citenamefont {Gokler}\ \emph {et~al.}(2017)\citenamefont {Gokler},
  \citenamefont {Lloyd}, \citenamefont {Shor},\ and\ \citenamefont
  {Thompson}}]{gokler2017efficiently}%
  \BibitemOpen
  \bibfield  {author} {\bibinfo {author} {\bibfnamefont {C.}~\bibnamefont
  {Gokler}}, \bibinfo {author} {\bibfnamefont {S.}~\bibnamefont {Lloyd}},
  \bibinfo {author} {\bibfnamefont {P.}~\bibnamefont {Shor}}, \ and\ \bibinfo
  {author} {\bibfnamefont {K.}~\bibnamefont {Thompson}},\ }\href
  {https://journals.aps.org/prl/abstract/10.1103/PhysRevLett.118.260501}
  {\bibfield  {journal} {\bibinfo  {journal} {Phys. Rev. Lett.}\ }\textbf
  {\bibinfo {volume} {118}},\ \bibinfo {pages} {260501} (\bibinfo {year}
  {2017})}\BibitemShut {NoStop}%
\bibitem [{\citenamefont {Viola}(2002)}]{viola2002quantum}%
  \BibitemOpen
  \bibfield  {author} {\bibinfo {author} {\bibfnamefont {L.}~\bibnamefont
  {Viola}},\ }\href
  {https://journals.aps.org/pra/abstract/10.1103/PhysRevA.66.012307} {\bibfield
   {journal} {\bibinfo  {journal} {Phys. Rev. A}\ }\textbf {\bibinfo {volume}
  {66}},\ \bibinfo {pages} {012307} (\bibinfo {year} {2002})}\BibitemShut
  {NoStop}%
\bibitem [{\citenamefont {West}\ \emph {et~al.}(2010)\citenamefont {West},
  \citenamefont {Lidar}, \citenamefont {Fong},\ and\ \citenamefont
  {Gyure}}]{west2010high}%
  \BibitemOpen
  \bibfield  {author} {\bibinfo {author} {\bibfnamefont {J.~R.}\ \bibnamefont
  {West}}, \bibinfo {author} {\bibfnamefont {D.~A.}\ \bibnamefont {Lidar}},
  \bibinfo {author} {\bibfnamefont {B.~H.}\ \bibnamefont {Fong}}, \ and\
  \bibinfo {author} {\bibfnamefont {M.~F.}\ \bibnamefont {Gyure}},\ }\href
  {https://journals.aps.org/prl/abstract/10.1103/PhysRevLett.105.230503}
  {\bibfield  {journal} {\bibinfo  {journal} {Phys. Rev. Lett.}\ }\textbf
  {\bibinfo {volume} {105}},\ \bibinfo {pages} {230503} (\bibinfo {year}
  {2010})}\BibitemShut {NoStop}%
\bibitem [{\citenamefont {Arenz}\ \emph {et~al.}(2014)\citenamefont {Arenz},
  \citenamefont {Gualdi},\ and\ \citenamefont {Burgarth}}]{arenz2014control}%
  \BibitemOpen
  \bibfield  {author} {\bibinfo {author} {\bibfnamefont {C.}~\bibnamefont
  {Arenz}}, \bibinfo {author} {\bibfnamefont {G.}~\bibnamefont {Gualdi}}, \
  and\ \bibinfo {author} {\bibfnamefont {D.}~\bibnamefont {Burgarth}},\ }\href
  {https://iopscience.iop.org/article/10.1088/1367-2630/16/6/065023} {\bibfield
   {journal} {\bibinfo  {journal} {New J. Phys.}\ }\textbf {\bibinfo {volume}
  {16}},\ \bibinfo {pages} {065023} (\bibinfo {year} {2014})}\BibitemShut
  {NoStop}%
\bibitem [{\citenamefont {Kosut}\ \emph {et~al.}(2019)\citenamefont {Kosut},
  \citenamefont {Arenz},\ and\ \citenamefont {Rabitz}}]{kosut2019quantum}%
  \BibitemOpen
  \bibfield  {author} {\bibinfo {author} {\bibfnamefont {R.~L.}\ \bibnamefont
  {Kosut}}, \bibinfo {author} {\bibfnamefont {C.}~\bibnamefont {Arenz}}, \ and\
  \bibinfo {author} {\bibfnamefont {H.}~\bibnamefont {Rabitz}},\ }\href
  {https://iopscience.iop.org/article/10.1088/1751-8121/ab0dc9} {\bibfield
  {journal} {\bibinfo  {journal} {J. Phys. A}\ }\textbf {\bibinfo {volume}
  {52}},\ \bibinfo {pages} {165305} (\bibinfo {year} {2019})}\BibitemShut
  {NoStop}%
\bibitem [{\citenamefont {Dowdall}\ \emph {et~al.}(2017)\citenamefont
  {Dowdall}, \citenamefont {Benseny}, \citenamefont {Busch},\ and\
  \citenamefont {Ruschhaupt}}]{dowdall_fast_2017}%
  \BibitemOpen
  \bibfield  {author} {\bibinfo {author} {\bibfnamefont {T.}~\bibnamefont
  {Dowdall}}, \bibinfo {author} {\bibfnamefont {A.}~\bibnamefont {Benseny}},
  \bibinfo {author} {\bibfnamefont {T.}~\bibnamefont {Busch}}, \ and\ \bibinfo
  {author} {\bibfnamefont {A.}~\bibnamefont {Ruschhaupt}},\ }\href {\doibase
  10.1103/PhysRevA.96.043601} {\bibfield  {journal} {\bibinfo  {journal} {Phys.
  Rev. A}\ }\textbf {\bibinfo {volume} {96}},\ \bibinfo {pages} {043601}
  (\bibinfo {year} {2017})}\BibitemShut {NoStop}%
\bibitem [{\citenamefont {Fogarty}\ \emph {et~al.}(2020)\citenamefont
  {Fogarty}, \citenamefont {Deffner}, \citenamefont {Busch},\ and\
  \citenamefont {Campbell}}]{fogarty_orthogonality_2020}%
  \BibitemOpen
  \bibfield  {author} {\bibinfo {author} {\bibfnamefont {T.}~\bibnamefont
  {Fogarty}}, \bibinfo {author} {\bibfnamefont {S.}~\bibnamefont {Deffner}},
  \bibinfo {author} {\bibfnamefont {T.}~\bibnamefont {Busch}}, \ and\ \bibinfo
  {author} {\bibfnamefont {S.}~\bibnamefont {Campbell}},\ }\href {\doibase
  10.1103/PhysRevLett.124.110601} {\bibfield  {journal} {\bibinfo  {journal}
  {Phys. Rev. Lett.}\ }\textbf {\bibinfo {volume} {124}},\ \bibinfo {pages}
  {110601} (\bibinfo {year} {2020})}\BibitemShut {NoStop}%
\bibitem [{\citenamefont {{Knill}}(1996)}]{knill1996quantum}%
  \BibitemOpen
  \bibfield  {author} {\bibinfo {author} {\bibfnamefont {E.}~\bibnamefont
  {{Knill}}},\ }\href@noop {} {\bibfield  {journal} {\bibinfo  {journal} {arXiv
  e-prints}\ ,\ \bibinfo {eid} {quant-ph/9610012}} (\bibinfo {year} {1996})},\
  \Eprint {http://arxiv.org/abs/quant-ph/9610012} {arXiv:quant-ph/9610012
  [quant-ph]} \BibitemShut {NoStop}%
\bibitem [{\citenamefont {Cleve}\ \emph {et~al.}(2004)\citenamefont {Cleve},
  \citenamefont {Hoyer}, \citenamefont {Toner},\ and\ \citenamefont
  {Watrous}}]{cleve2004consequences}%
  \BibitemOpen
  \bibfield  {author} {\bibinfo {author} {\bibfnamefont {R.}~\bibnamefont
  {Cleve}}, \bibinfo {author} {\bibfnamefont {P.}~\bibnamefont {Hoyer}},
  \bibinfo {author} {\bibfnamefont {B.}~\bibnamefont {Toner}}, \ and\ \bibinfo
  {author} {\bibfnamefont {J.}~\bibnamefont {Watrous}},\ }in\ \href
  {https://ieeexplore.ieee.org/document/1313847} {\emph {\bibinfo {booktitle}
  {Proceedings. 19th IEEE Annual Conference on Computational Complexity,
  2004.}}}\ (\bibinfo {organization} {IEEE},\ \bibinfo {year} {2004})\ pp.\
  \bibinfo {pages} {236--249}\BibitemShut {NoStop}%
\bibitem [{\citenamefont {Harrow}\ \emph {et~al.}()\citenamefont {Harrow},
  \citenamefont {Lin},\ and\ \citenamefont {Montanaro}}]{Harrow:2017a}%
  \BibitemOpen
  \bibfield  {author} {\bibinfo {author} {\bibfnamefont {A.~W.}\ \bibnamefont
  {Harrow}}, \bibinfo {author} {\bibfnamefont {C.~Y.-Y.}\ \bibnamefont {Lin}},
  \ and\ \bibinfo {author} {\bibfnamefont {A.}~\bibnamefont {Montanaro}},\
  }\enquote {\bibinfo {title} {Sequential measurements, disturbance and
  property testing},}\ in\ \href {\doibase 10.1137/1.9781611974782.105} {\emph
  {\bibinfo {booktitle} {Proceedings of the 2017 Annual ACM-SIAM Symposium on
  Discrete Algorithms}}},\ pp.\ \bibinfo {pages} {1598--1611}\BibitemShut
  {NoStop}%
\bibitem [{\citenamefont {Bravyi}\ \emph {et~al.}(2018)\citenamefont {Bravyi},
  \citenamefont {Gosset},\ and\ \citenamefont {K{\"o}nig}}]{bravyi2018quantum}%
  \BibitemOpen
  \bibfield  {author} {\bibinfo {author} {\bibfnamefont {S.}~\bibnamefont
  {Bravyi}}, \bibinfo {author} {\bibfnamefont {D.}~\bibnamefont {Gosset}}, \
  and\ \bibinfo {author} {\bibfnamefont {R.}~\bibnamefont {K{\"o}nig}},\ }\href
  {https://science.sciencemag.org/content/362/6412/308} {\bibfield  {journal}
  {\bibinfo  {journal} {Science}\ }\textbf {\bibinfo {volume} {362}},\ \bibinfo
  {pages} {308} (\bibinfo {year} {2018})}\BibitemShut {NoStop}%
\bibitem [{\citenamefont {Centrone}\ \emph {et~al.}(2020)\citenamefont
  {Centrone}, \citenamefont {Kumar}, \citenamefont {Diamanti},\ and\
  \citenamefont {Kerenidis}}]{centrone2020experimental}%
  \BibitemOpen
  \bibfield  {author} {\bibinfo {author} {\bibfnamefont {F.}~\bibnamefont
  {Centrone}}, \bibinfo {author} {\bibfnamefont {N.}~\bibnamefont {Kumar}},
  \bibinfo {author} {\bibfnamefont {E.}~\bibnamefont {Diamanti}}, \ and\
  \bibinfo {author} {\bibfnamefont {I.}~\bibnamefont {Kerenidis}},\ }\href
  {https://arxiv.org/abs/2007.15876} {\bibfield  {journal} {\bibinfo  {journal}
  {arXiv preprint arXiv:2007.15876}\ } (\bibinfo {year} {2020})}\BibitemShut
  {NoStop}%
\end{thebibliography}%

\end{document}